\documentclass[prd,aps,twocolumn,preprintnumbers, showpacs, nofootinbib,superscriptaddress,notitlepage]{revtex4-1}
\usepackage{mathrsfs}
\usepackage{amsfonts}
\usepackage{amsmath}
\usepackage{slashed}
\usepackage{array}
\usepackage{verbatim}
\usepackage{epsfig}
\usepackage{graphicx}
\usepackage{color}
\usepackage{bm}
\usepackage{multirow}
\usepackage[dvipsnames]{xcolor}
    \usepackage{subfigure}

\newcommand{\beq}{\begin{eqnarray}}
\newcommand{\eeq}{\end{eqnarray}}

\def\be{\begin{equation}}
\def\ee{\end{equation}}
\def\bea{\begin{eqnarray}}
\def\eea{\end{eqnarray}}

\begin{document}

\title{Two-nucleon systems at $m_{\pi}\approx292$ MeV from lattice QCD}

\author{Kuan Zhang}
\affiliation{
Institute of Modern Physics, Chinese Academy of Sciences, Lanzhou, 730000, China}

\author{Kang Yu}
\affiliation{
School of Physical Sciences, University of Chinese Academy of Sciences, Beijing 100049, China}

\author{Yiqi Geng}
\affiliation{
Institute of Modern Physics, Chinese Academy of Sciences, Lanzhou, 730000, China}
\affiliation{
Nanjing Normal University, Nanjing, Jiangsu 210023, China}

\author{Chuan Liu}
\affiliation{
School of Physics, Peking University, Beijing 100871, China}
\affiliation{
Center for High Energy Physics, Peking University, Beijing 100871, China}
\affiliation{Collaborative Innovation Center of Quantum Matter, Beijing 100871, China}

\author{Liuming Liu}
\email{Corresponding author: liuming@impcas.ac.cn}
\affiliation{
Institute of Modern Physics, Chinese Academy of Sciences, Lanzhou, 730000, China}

\author{Peng Sun}
\email{Corresponding author: pengsun@impcas.ac.cn}
\affiliation{
Institute of Modern Physics, Chinese Academy of Sciences, Lanzhou, 730000, China}

\author{Jia-Jun Wu}
\email{Corresponding author: wujiajun@ucas.ac.cn}
\affiliation{
School of Physical Sciences, University of Chinese Academy of Sciences, Beijing 100049, China}
\affiliation{
Southern Center for Nuclear-Science Theory (SCNT), Institute of Modern Physics, Chinese Academy
of Sciences, Huizhou 516000, China}

\author{Ruilin Zhu}
\affiliation{
Nanjing Normal University, Nanjing, Jiangsu 210023, China}

\date{\today}

\begin{abstract}

Nucleon-nucleon systems in the $^3S_1$ and the $^1S_0$ channels are studied in lattice quantum chromodynamics at a pion mass of approximately $m_{\pi}\approx292$ MeV, employing three $N_f = 2+1$ ensembles with the same pion mass and lattice spacing $a=0.10530(18)$ fm but different spatial volumes. 
Finite-volume energies of the nucleon-nucleon systems are determined in both the rest frame and a moving frame. 
The distillation quark smearing method is applied to improve the precision and to ensure the symmetric correlators by using the same interpolating operators at sink and source.
The scattering amplitudes are extracted from the finite-volume spectra using the Lüscher's finite-volume method. 
At the studied pion mass, both the $^3S_1$ (deuteron) and $^1S_0$(di-neutron) channels exhibit a virtual state pole, with binding energies of $6^{+5}_{-3}$ MeV and $11^{+6}_{-5}$ MeV, respectively. 
To investigate the effects of the left-hand cut, an alternative method -- the Non-Perturbative Hamiltonian framework (NPHF) --  is used for the scattering analysis and yields consistent results with those from the Lüscher method. 

\end{abstract}

\maketitle

\section{Introduction}

A central objective of nuclear physics is to understand low-energy nuclear phenomena through the underlying theory of strong interactions, known as Quantum Chromodynamics (QCD). 
The only established ab initio method to rigorously compute the low-energy properties of nuclear systems is lattice QCD. 
While current lattice studies of single-nucleon properties have achieved percent-level accuracy and full control of the systematics~\cite{BMW:2014pzb, Chang:2018uxx}, the study of multi-nucleon systems faces great challenges, mainly due to the signal-to-noise problem. 

Nucleon-nucleon interactions play an essential role in our understanding of the nuclear matter. 
Lattice QCD studies of nucleon-nucleon interactions have undergone a long history of controversy. 
Early studies using L\"uscher's finite-volume method~\cite{Luscher:1986pf,Luscher:1990ux} in the limit of SU(3) flavor symmetry with $m_{\pi}\approx800$ MeV established that both the deuteron and di-neutron become deeply bound in this regime~\cite{NPLQCD:2012mex,NPLQCD:2013bqy,Berkowitz:2015eaa,Wagman:2017tmp}.  
Studies at lower pion masses, $m_{\pi} \sim 300 - 500$MeV, also supported the existence of bound states in the two-nucleon systems at this pion mass range~\cite{NPLQCD:2011naw, NPLQCD:2020lxg, Orginos:2015aya, Yamazaki:2012hi, Yamazaki:2015asa}. 
All these studies employed asymmetric correlators, using local hexaquark {or spatially displaced two-nucleon} operators at the source and momentum-space operators at the sink. 
However, calculations that employ variational techniques, distillation, and boosted frames~\cite{Francis:2018qch,Horz:2020zvv} raised tensions with those earlier asymmetric correlator studies. 
Complementary calculations have explored the variational spectrum with large operator bases~\cite{Amarasinghe:2021lqa,Detmold:2024iwz}. 
On a parallel track, the HALQCD collaboration has developed potential-based extractions of nucleon forces~\cite{Ishii:2012ssm, Inoue:2011ai, Iritani:2017rlk,Aoki:2020bew}, reporting unbound two-nucleon channels at heavier-than-physical pion masses. 
A recent study~\cite{BaSc:2025yhy} investigated two-nucleon systems using both L\"uscher's formula and HALQCD method, employing both momentum-space operators and local hexaquark operators, concluding that di-nucleons do not form bound states at a heavy pion mass $m_\pi = 714$MeV, and that the previous identification of bound di-nucleons may have arisen from a misidentification of the spectrum with asymmetric correlators. {Another recent study computed the lowest finite-volume energy of the di-nucleons at three different pion masses between 140 and 700~MeV~\cite{Chakraborty:2026zeq}. It found that the energy shift of di-neutron changes from negative to positive as the pion mass approaches physical value. This study also observed a difference between the energies extracted from symmetric and asymmetric correlators.}

In this work, we study the nucleon-nucleon interactions at $m_{\pi}\approx292$ MeV. 
This bridges the gap between heavy pion masses employed in most of the previous studies and the physical point.
The distillation method is used in our calculation, which enables us to construct momentum-space operators at both source and sink, avoiding potential problems in spectrum determination using asymmetric correlators. 
We do not use the local hexaquark operators. 
Studies in Ref.~\cite{Amarasinghe:2021lqa,Detmold:2024iwz,BaSc:2025yhy} indicate that the hexaquark operators do not affect the low-lying near-threshold energies, which is the energy domain we will inspect in this work. 

Composite particles are more rigorously associated with pole singularities in scattering amplitudes. 
We employ Lüscher's finite volume method ~\cite{Luscher:1986pf,Luscher:1990ux}, which connects finite-volume energy levels and infinite-volume scattering amplitudes, to extract the scattering amplitudes of the two-nucleon systems. 
The original Lüscher formula for two identical spinless particles in the rest frame was generalized to more complex physical scenarios involving  particles with spin, boosted frames and coupled channels ~\cite{Rummukainen:1995vs,Beane:2003yx,Beane:2003da,Li:2003jn,Detmold:2004qn,Feng:2004ua,Christ:2005gi,Kim:2005gf,Bernard:2008ax,Bour:2011ef,Davoudi:2011md,Leskovec:2012gb,Gockeler:2012yj,Ishizuka:2009bx,Briceno:2013lba,He:2005ey,Briceno:2014oea}. 
However, these formulas do not consider the effects of the left-hand cut. 
In our study with a relatively low pion mass, the left-hand cut $k^2 = -m_\pi^2/4$ is very close to the threshold and may have non-negligible effects. 
To include the effects of the left-hand cut, we use the Non-Perturbative Hamiltonian framework(NPHF) to reanalyze the finite-volume spectra.
{{red}
The idea of NPHF is as follows.
The Hamiltonian operator is available both in the discrete momentum space and continuum momentum space.
Thus, this Hamiltonian can connect the finite volume spectrum and the observables in the infinite volume.
Then the energy levels in the lattice are the eigenvalues of the Hamiltonian matrix constructed in the discrete momentum which is exactly the same as the energy levels from the lattice calculation.
Then by fitting the lattice spectrum, we can fix the parameters in the Hamiltonian.
At last, we will search the pole of the system by using the Lippmann-Schwinger equation where the kernel is the potential in the Hamiltonian.
}
Our results from both methods indicate that the two-nucleon systems in both $^3S_1$ and the $^1S_0$ channels have virtual state poles. 

This paper is organized as follows. 
In Section~\ref{sec:setup}, we introduce the ensembles used and the distillation method for the computations of quarks propagators. 
In Section~\ref{sec:spectrum}, the construction of interpolation operators, the variation method, and energy levels derived from the correlation functions are reviewed. Section~\ref{sec:analysis} details our scattering analysis using the Lüscher formula, discussing its specific form, the parameterization of the scattering amplitude, fitting details, and exploration of singularities in the amplitudes. 
Additionally, two-nucleon systems are studied within the framework of the NPHF method in Section~\ref{sec:HEFT} before concluding in Section~\ref{sec:summary}.

\section{Numerical setup}\label{sec:setup}

In this paper, we utilize 2+1 flavor QCD ensembles generated by the CLQCD collaboration~\cite{CLQCD:2023sdb,CLQCD:2024yyn}, employing the stout-smeared clover fermion action and Symanzik gauge action. 
Three ensembles, named C24P29, C32P29 and C48P29 are used to calculate discrete spectrum in finite volume. 
The pion mass is approximately 292 MeV across all three ensembles. 
Additional details can be found in Table~\ref{tab:CLQCD}.

\begin{table}[htbp]
  \centering
  \begin{tabular}{|c|c|c|c|c|c|c|}
\hline
\hline
ID    & $(L^3 \times T)/a$ & $a(\mathrm{fm})$ & $m_\pi(\mathrm{MeV})$  & $m_{\pi} L$  &$m_N(\text{MeV})$ & $N_{\rm cfg}$ \\
\hline
C24P29 &  $24^3\times72$ & 0.1053 &292 & 3.75 & 1026(4) & 838 \\
\hline
C32P29 &  $32^3\times 64$ & 0.1053 &292  & 5.01  &1005(3) & 981\\
\hline
C48P29 &  $48^3\times 96$ & 0.1053 &292  & 7.52 &1007(1) & 731 \\
\hline
\hline
\end{tabular}
  \caption{Parameters of the ensembles, including lattice size $L^3\times T$, lattice spacing $a$, pion mass $m_\pi$, $m_{\pi}\times L$, nucleon mass $m_N$ and the number of configurations we use in this work.}
  \label{tab:CLQCD}
\end{table}

The distillation quark smearing method~\cite{HadronSpectrum:2009krc} is utilized to compute the quark propagators. 
The smearing operator is constructed from a small number($N_{ev}$) of eigenvectors associated with the $N_{ev}$ lowest eigenvalues of the three-dimensional Laplacian defined in terms of the gauge fields:
\begin{align}\label{eq:Laplacian}
-\nabla^2_{xy}=6\delta_{xy}-\sum^3_{j=1}\Big(\tilde{U}_j(x,t)\delta_{x+\hat{j},y}+\tilde{U}^{\dagger}_j(x-\hat{j},t)\delta_{x-\hat{j},y}\Big),
\end{align}
where $\tilde{U}$ denotes the HYP-smearing gauge fields. 
The smearing operator is given by $\square=V(t)V^{\dagger}(t)$, where $V(t)$ is a matrix of dimensions $3L^3\times N_{ev}$ whose rows are the $N_{ev}$ eigenvector described above. 
This approach allows for the efficient computation of correlation functions for numerous interpolating operators, and significantly improving precision with manageable computational cost. 
In this study, we employ 100 eigenvectors across all three ensembles.

\section{Finite Volume Spectrum}\label{sec:spectrum}

%\subsection{Interpolating operators}
%To obtain the finite-volume energies of the two-nucleon systems, we start with the construction of the interpolating operators. In the three-dimensional continuum space, the rotational symmetry group is $SU(2)$, its irreducible representations are denoted by the total angular momentum $J=0,1/2,1,\cdots$. While on the lattice, the continuum SU(2) group is reduced to the double covering of the cubic group $O^D$. When parity is taken into account, the $O^D$ group is promoted to $O_h^D=O^{D}\otimes Z_2$, with $Z_2$ incorporating the spatial inversion. In moving frames, the symmetry is further reduced to the little groups. 

To obtain finite volume energies in lattice simulations, it is essential to construct interpolating operators with specific quantum numbers. 
This study concentrates on the $^3S_1$ and $^1S_0$ channels in two-nucleon systems. 
In the $^3S_1$ channel, the angular momentum and parity are $J^P=1^+$ with isospin $I=0$; in the $^1S_0$ channel, $J^P=0^+$ with isospin $I=1$. 
While isospin $I$ remains a good quantum number on the lattice, handling angular momentum poses more complexity.

Lattice QCD calculations involve discretizing the theory on a four-dimensional Euclidean hypercubic grid. 
{The cubic volume inherently breaks the full three-dimensional rotational symmetry, complicating the study of states with a definite angular momentum $J$. This difficulty arises because the finite-volume spectrum—governed by the underlying geometry—suffers from a loss of angular momentum conservation. %(For extensions of finite-volume analysis to other topologies, such as a 2-sphere, see Ref.~\cite{Schuh:2022vvl}.) 
Beyond spatial geometry, spacetime discretization and the choice of hadronic operators significantly impact the extraction of the finite-volume spectrum, particularly prior to the continuum limit. In this work, as in most lattice QCD simulations, the cubic volume is discretized using a isotropic cubic lattice. Consequently, we can construct hadronic operators embedded within this reduced symmetry group, allowing us to extract a spectrum that fully respects this reduced symmetry even before taking the continuum limit.}

In the infinite volume, integer angular momentum corresponds to irreducible representations (irreps) of the rotation group $SO(3)$, and half-integer angular momentum to irreps of its double cover $SU(2)$. The reduced symmetry in this lattice simulations is described by the cubic group $O$ and its double cover $O^D$. Parity is included by extending the $O^D$ group to $O_h^D=O^{D}\otimes Z_2$, with $Z_2$ incorporating the spatial inversion. 
In moving frames, the symmetry is further reduced to little groups and parity is no longer a good quantum number.

The lattice symmetry groups are subgroups of the continuum groups. Consequently, the irreps of the continuum groups, labeled by angular momentum, become reducible when restricted to these subgroups. Using the orthogonality relation for characters of inequivalent irreps, one can decompose a given angular momentum state into irreps of the corresponding lattice group.

%{\color{red}All lattice groups discussed above are subgroups of the $O(3)$ group. Therefore, the irreducible representations of $O(3)$ labeled by orbital angular momentum become reducible when restricted to these subgroups. The same holds for their respective double covering groups. Using the orthogonality relation of characters of inequivalent irreducible representations, we can decompose a given angular momentum system into the irreducible representations of the corresponding lattice group.}
%
In this work, we compute the energies in the rest frame and a moving frame with momentum ${\bm P} \equiv\frac{2\pi}{L}\bm{d}= \frac{2\pi}{L}(0,0,1)$. 
We adopt the following irreps in the rest frame: $G_1^+$ for one nucleon, $T_1^+$ for two nucleons in the $^3S_1$ channel, and $A_1^+$ for the $^1S_0$ channel. 
In the moving frame with momentum ${\bm P} = \frac{2\pi}{L}(0,0,1)$,  we use $E_1$ for one nucleon, $E_2$ for two nucleons in the $^3S_1$ channel, and $A_1$ for the $^1S_0$ channel.

Consider a frequently used nucleon operator in lattice simulations:
\begin{align}\label{eq:nucleon}
{\cal N}=\epsilon_{abc}(d_a^{T}C\gamma_5u_b)d_c.
\end{align}
Here $C$ is the charge conjugation operator, $a$, $b$ and $c$ are the color indices. 
This operator, comprising the $ddu$ quark field, represents a neutron. 
By exchanging $d$ and $u$, it corresponds to a proton, denoted as ${\cal P}$. 
These two operators form the components of the isospin $I=1/2$ space, corresponding to $I_z=\mp1/2$, respectively.

The two-nucleon operators for the $^3S_1$ and $^1S_0$ channels are expressed as:
\begin{eqnarray}\label{eq:NNoperator}
O_{\Lambda}^{I=0}({\bm P}) =\sum_{\lambda_1,\lambda_2,{\bm p_1},{\bm p_2}} & C_{\lambda_1,\lambda_2,{\bm p_1},{\bm p_2}}  [  {\cal P}_{\lambda_1}(\bm p_1) {\cal N}_{\lambda_2}({\bm p_2})  \nonumber \\
&    - {\cal N}_{\lambda_1}(\bm p_1) {\cal P}_{\lambda_2}({\bm p_2}) ] \\
O_{\Lambda}^{I=1}({\bm P}) =\sum_{\lambda_1,\lambda_2,{\bm p_1},{\bm p_2}} & C_{\lambda_1,\lambda_2,{\bm p_1},{\bm p_2}}   {\cal P}_{\lambda_1}(\bm p_1) {\cal P}_{\lambda_2}({\bm p_2})
%& {\bm P}={\bm p_1}+{\bm p_2}.
\end{eqnarray}
with ${\bm P}={\bm p_1}+{\bm p_2}$. 
Here $\lambda_{1,2}$ label the components of the lattice irreps of the nucleons. 
The coefficients $C_{\lambda_1,\lambda_2,{\bm p_1},{\bm p_2}}$ are chosen to achieve the desired transformation properties under the irrep $\Lambda$,  {and their values can be found in the Appendix C.}
These coefficients are constructed following the method of Ref.~\cite{Yan:2025jlq} and computed using the open-source package {\bf OpTion} provided therein. 

To keep the energies near the threshold, we use only the few lowest momenta for ${\bm p_1}$ and ${\bm p_2}$ in the operators. 
For {each irrep in} the rest frame, we employ a single operator with $|{\bm p_1}|$ = $|{\bm p_2}|$ = 0 for the $L=24$ ensemble, and two operators(with $|{\bm p_1}|$ = $|{\bm p_2}|$ = 0 and $\frac{2\pi}{L}$) for the $L=32$ and 48 ensembles. Note that one can construct two linearly independent operators with $|{\bm p_1}|$ = $|{\bm p_2}|$ = $\frac{2\pi}{L}$ in the $T_1^+$ irrep. We choose the one that dominantly couples to S-wave.  
For {each irrep in} the moving frame with ${\bm P} = \frac{2\pi}{L}(0,0,1)$, we use one operator with $|{\bm p_1}| = 0, |{\bm p_2}| = \frac{2\pi}{L}$  for all three ensembles. {For the irreps with multiple dimension(e.g., $T_1^+$ and $E$), all components of each irrep in the same momentum shell are averaged to enhance the signal-to-noise ratio.}

The finite-volume energies are extracted from the correlation functions of the operators defined above:
\begin{eqnarray}
C_{ij}(t)=\sum_{t_s} \langle O_i(t+t_s)O_j^\dagger(t_s)\rangle,
\end{eqnarray}
where $i,j$ index the two-particle operators. 
The source time $t_s$ runs over all time slices to increase statistics. 

{$C$ is a $2\times2$ matrix for those irreps with two momentum shells. To disentangle them, we solve the generalized eigenvalue problem(GEVP)~\cite{Luscher:1990ck} :}
\begin{eqnarray}
\label{eq:GEVP}
C(t) v_n(t,t_0) = \lambda_n(t, t_0) C(t_0)v_n(t,t_0).
\end{eqnarray}
The value of $t_0$ is set to be $5a$ in our analysis. 
The energies are obtained by fitting the eigenvalues to a two-exponential form: 
\begin{equation}
\label{eq:two-exp fit}
\lambda_n(t, t_0) = A e^{-E (t-t_0)} + B e^{-E^\prime(t-t_0)},
\end{equation}
where the second exponential accounts for the residual excited-state contributions. 
 
For those irreps with only one operator, the energies are obtained by fitting the corresponding correlation functions to Eq.~\ref{eq:two-exp fit}. 
Further details on the fitting procedure are provided in Appendix A. {Varying $t_0$ does not affect the extracted energy levels. In Appendix B, we compare the results obtained for $t_0 = 4a, 5a$ and $6a$.}

\begin{figure}[tbph]
\centering
\includegraphics[width=8cm]{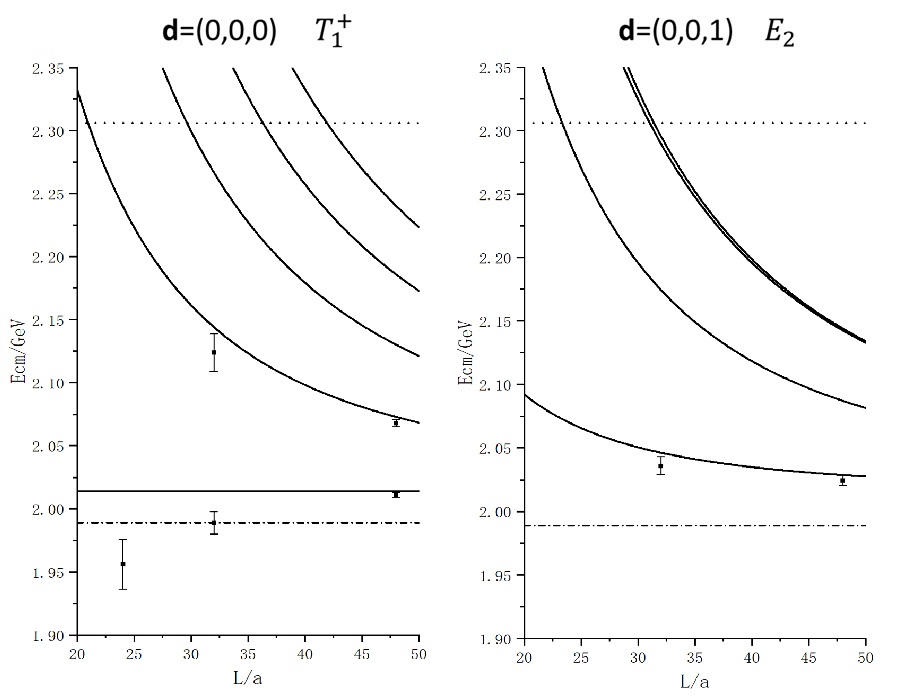}
\includegraphics[width=8cm]{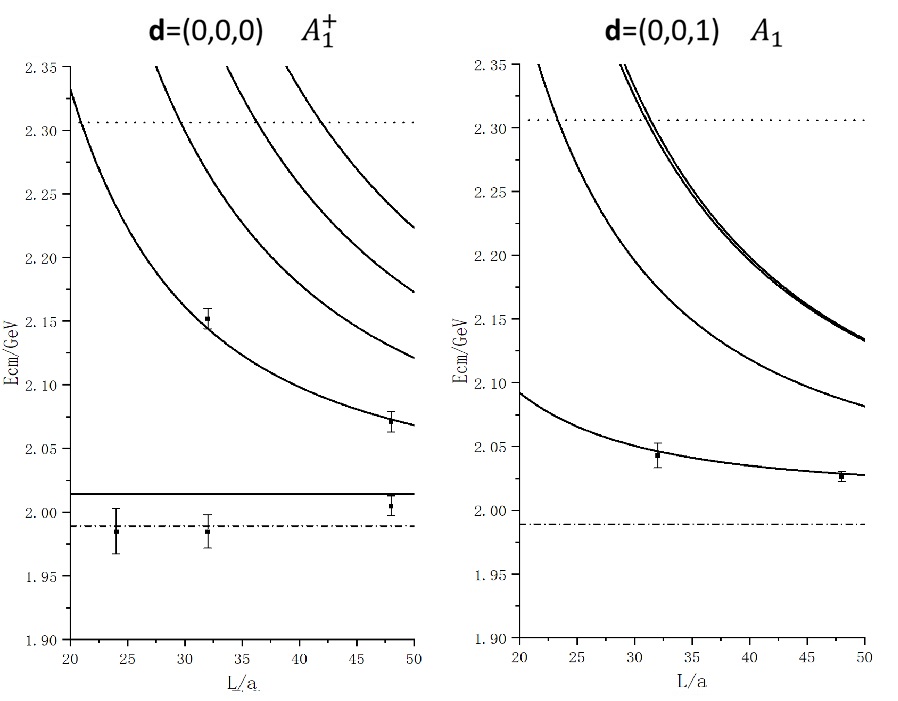}
\caption{The energy levels in the center-of-momentum frame on three ensembles, in the $^3S_1$ channel (upper section) and the $^1S_0$ channel (lower section), and in the rest frame (left section) and a moving frame (right section). In each panel, the dashed line on the top represents the $NN\pi$ threshold ($2m_N+m_{\pi}$), the dashed line on the bottom is the left hand cut in $NN$ scattering (${\bm k}^2=-\frac{1}{4}m_{\pi}^2$), and the solid curves are free energies in Eq.~\ref{eq:Efree}. All curves are plotted using the nucleon mass from ensemble C48P29. Since the nucleon mass of the ensembles C32P29 and C24P29 are slightly different with the ensemble C48P29, the energies of these two ensembles are shifted by $E^{\rm free}_{L=48}-E^{\rm free}_{L=24,32}$ to correctly show the energy shifts of the interacting energies with respect to the free energies.}
\label{fig:energy}
\end{figure}

Energies in the moving frame are converted to the center-of-momentum frame via $E_{\rm{cm}} = \sqrt{E^2-\bm{P}^2}$. 
Fig.~\ref{fig:energy} shows the resulting center-of-momentum energy levels for all three ensembles, together with the non-interacting two-particle energies:
\begin{align}\label{eq:Efree}
E^{\rm free}=\sqrt{m^2_{1}+\bm{p}_1^2}+\sqrt{m^2_{2}+\bm{p}_2^2}
,
\end{align}
where $m_1$ and $m_2$ are the masses of the two particles involved in the scattering. 
%
%Each calculated lattice level is matched with an expected nearby non-interacting level, with no lattice levels left over. 
All energy levels are below the $NN\pi$ threshold, allowing us to restrict our analysis to the elastic scattering.

We can compute the scattering momentum $\bm{k}$ using:
\begin{align}\label{eq:Ecm2}
&E_{\rm cm}=\sqrt{m^2_{1}+\bm{k}^2}+\sqrt{m^2_{2}+\bm{k}^2} \\
&\bm{k}^2=\frac{1}{4}E^2_{\rm{cm}}-\frac{1}{2}(m^2_{1}+m^2_{2})+\frac{(m^2_{1}-m^2_{2})^2}{4E_{\rm{cm}}^2}
.
\end{align}
In $NN$ scattering, the left-hand cut corresponds to ${\bm k}^2=-\frac{1}{4}m_{\pi}^2$. Since the Lüscher formula is not valid below this cut, we restrict our scattering analysis using the Lüscher formula to energy levels above this region. 

In Sec.~\ref{sec:HEFT}, we employ an alternative method, NPHF, for the scattering analysis. 
In this method, the effects of the left-hand cut is included. 
Therefore, all energy levels are included in the analysis. 

\section{Scattering analysis}\label{sec:analysis}

\subsection{Generalized Lüscher formula}

L\"ushcer's finite-volume method connects the finite-volume energy $E$ with the infinite-volume scattering S-matrix. 
The generalized Lüscher formula reads
%We work in an $L^3$ spatial volume with periodic boundary conditions. The total energy $E$ is related to the dimensionless unitary scattering S-matrix in the space of open channels via generalized Lüscher formula:
\begin{align}\label{eq:formula}
\rm{det}[T^{-1}+\delta{\cal G}(E,\bm{P};L)]=0, \quad\quad S\equiv1+iT,
\end{align}
where $T$ and $\delta{\cal G}$ are matrices in the space of scattering channels. 
In this work, we consider only single-channel scattering. 

We use an orthonormal basis of states $|Jm_J,lS\rangle$ in the continuum, where $J$ is the total angular momentum of the two particles , $m_J$ is the $z$-component of $J$, $l$ is the orbital angular momentum, and $S$ is the spin.
For $l=0$, we have $S=J$. 
In this basis, $\delta{\cal G}$ is given by ~\cite{Briceno:2014oea}

\begin{align}\label{eq:box}
&[\delta{\cal G}]_{Jm_J,lS;J^{\prime}m_J^{\prime},l^{\prime}S^{\prime}}=\delta_{SS^{\prime}}[\delta{\cal G}^S]_{Jm_Jl;J^{\prime}m_J^{\prime}l^{\prime}} \nonumber\\
&=i\delta_{SS^{\prime}}\rho\Bigg[\delta_{JJ^{\prime}}\delta_{m_Jm_J^{\prime}}\delta_{ll^{\prime}} 
+i\sum_{l^{\prime\prime},m^{\prime\prime}}\frac{(4\pi)^{3/2}}{k^{l^{\prime\prime}+1}}c^{\bm{d}}_{l^{\prime\prime}m^{\prime\prime}}(k^2;L)  \nonumber\\
&\times\sum_{m_l,m_{l^{\prime}},m_S}\langle lS,Jm_J|lm_l,Sm_S\rangle\langle l^{\prime}m_{l^{\prime}},Sm_S|l^{\prime}S,J^{\prime}m_{J^{\prime}}\rangle \nonumber\\
&\times \int d\Omega Y^*_{l,m_l}Y^*_{l^{\prime\prime},m_{l^{\prime\prime}}}Y_{l^{\prime},m_{l^{\prime}}}
\Bigg]
,
\end{align}
where $\rho$ is defined as
\begin{align}\label{eq:rho}
\rho\equiv\frac{k}{8\pi E_{\rm cm}},
\end{align}
and the function $c^{\bm{d}}_{lm}$ is defined as
\begin{align}\label{eq:funcC}
c^{\bm{d}}_{lm}\equiv\frac{\sqrt{4\pi}}{\gamma L^3}\Big(\frac{2\pi}{L}\Big)^{l-2}{\cal Z}^{\bm{d}}_{lm}[1;q^2], \quad q\equiv \frac{Lk}{2\pi}, \quad \gamma\equiv\frac{E}{E_{\rm{cm}}}
.
\end{align}

The original definition of the Lüscher zeta function ${\cal Z}^{\bm{d}}_{lm}[s;x^2]$ is,
\begin{align}\label{eq:zeta}
{\cal Z}^{\bm{d}}_{lm}[s;x^2]\equiv\sum_{\bm{r}\in {\cal P}_{\bm{d}}}\frac{|\bm{r}|^lY_{l,m}(\bm r)}{(r^2-x^2)^s}
,
\end{align}
where the sum runs over ${\cal P}_{\bm{d}}=\{\bm{r}\equiv\hat{\gamma}^{-1}(\bm{m}-\alpha\bm{d})|\bm{m}\in {\cal Z}^3\}$ with
$\alpha\equiv\frac{1}{2}\Big[1+\frac{m_{1}^2-m_{2}^2}{E_{\rm cm}^2}\Big]$. 
The operation $\hat{\gamma}^{-1}\bm{x}\equiv\gamma^{-1}\bm{x}_{\parallel}+\bm{x}_{\perp}$ is understood, where $\bm{x}_{\parallel}$ denotes the component of $\bm{x}$ parallel to the total momentum, and $\bm{x}_{\perp} \equiv \bm{x}-\bm{x}_{\parallel}$ the perpendicular component. 

The matrix $\delta{\cal G}$ becomes block-diagonal when we consider the reduced symmetries of the cube (or the boosted cube). 
Using the operation known as subduction into irreps $\Lambda$ (row $\lambda$) of the relevant symmetry group,
\begin{align}\label{eq:subduction}
&\delta_{\Lambda,\Lambda^{\prime}}\delta_{\lambda,\lambda^{\prime}}\delta{\cal G}^{S\Lambda}_{Jln;J^{\prime}l^{\prime}n^{\prime}}= \nonumber\\
&\sum_{m_J,m_J^{\prime}}\Big(S^{\Lambda\lambda n}_{Jm_J}\Big)\delta{\cal G}^S_{Jm_Jl;J^{\prime}m_J^{\prime}l^{\prime}}\Big(S^{\Lambda^{\prime}\lambda^{\prime} n^{\prime}}_{J^{\prime}m_J^{\prime}}\Big)^*
,
\end{align}
where the coefficients $S^{\Lambda\lambda n}_{Jm_J}$ can be found in~\cite{Morningstar:2017spu}. 
Here $n$ labels the occurrence of the irrep $\Lambda$ when reduced from $J$. 
In this work, the number of occurrences is always 1, so we omit this index in the following.

For the $^3S_1$ channel, we have $J=1$, $l=0$, and $S=1$. From Eq.~\ref{eq:box}, for $\bm{d}=(0,0,0)$ and $(0,0,1)$ we get

\begin{align}\label{eq:3S1_formula_1}
[\delta{\cal G}^1]_{1m_J0;1m_J^{\prime}0}=\delta_{m_Jm_J^{\prime}}\big(\rho i-\rho\frac{2}{\sqrt{\pi}Lk}\cdot\frac{1}{\gamma}{\cal Z}^{\bm{d}}_{00}[1;q^2]\big)
.
\end{align}
After subduction, 
\begin{align}\label{eq:3S1_formula_2}
[\delta{\cal G}^{1T_1^+}]_{10;10}=\rho i-\rho\frac{2}{\sqrt{\pi}Lk}\cdot\frac{1}{\gamma}{\cal Z}^{(000)}_{00}[1;q^2]
,
\end{align}
\begin{align}\label{eq:3S1_formula_3}
[\delta{\cal G}^{1E_2}]_{10;10}=\rho i-\rho\frac{2}{\sqrt{\pi}Lk}\cdot\frac{1}{\gamma}{\cal Z}^{(001)}_{00}[1;q^2]
.
\end{align}

For the $^1S_0$ channel, $J=0$, $l=0$, and $S=0$, the results are similar to those in $^3S_1$ channel. 
For $\bm{d}=(0,0,0)$ or $(0,0,1)$ we get,
\begin{align}\label{eq:1S0_formula_1}
[\delta{\cal G}^0]_{000;000}=\rho i-\rho\frac{2}{\sqrt{\pi}Lk}\cdot\frac{1}{\gamma}{\cal Z}^{\bm{d}}_{00}[1;q^2].
\end{align}
After subduction, 
\begin{align}\label{eq:1S0_formula_2}
[\delta{\cal G}^{0A_1^+}]_{00;00}=\rho i-\rho\frac{2}{\sqrt{\pi}Lk}\cdot\frac{1}{\gamma}{\cal Z}^{(000)}_{00}[1;q^2]
,
\end{align}
\begin{align}\label{eq:1S0_formula_3}
[\delta{\cal G}^{0A_1}]_{00;00}=\rho i-\rho\frac{2}{\sqrt{\pi}Lk}\cdot\frac{1}{\gamma}{\cal Z}^{(001)}_{00}[1;q^2]
.
\end{align}

\subsection{K matrix parameterization}\label{subsec:para}

For elastic scattering, the scattering amplitude can be expressed in terms of partial-wave amplitudes as $T(s,\text{cos}\theta)=\sum^{+\infty}_{l=0}(2l+1)P_l(\text{cos}\theta)T_l(s)$, where $\theta$ denotes the direction of $\bm{k}$ and $s\equiv E_{\rm cm}^2$ is the Mandelstam variable. 
Above the kinematic threshold $(E_{\rm cm}>E^{\rm thr.}=m_1+m_2)$, probability conservation is enforced by the unitarity condition:
\begin{align}\label{eq:unitary1}
\text{Im}T_l^{-1}=-\rho\Theta(E_{\rm cm}-E^{\rm thr.}).
\end{align}

Experimentally, only quantities that lead to estimates of $T(s)$ for real value of $s$ can be measured; the behavior in the complex plane is determined by analytically continuing parameterized functions of $s$.
Unitary imposes a strong constraint on the form of these parameterizations. 
This is effectively implemented using a K-matrix:
\begin{align}\label{eq:unitary3}
T^{-1}_l =K_l^{-1}-i \rho.
\end{align}
Choosing $K$ as a real function of $s$ for real energies above the kinematic threshold ensures that the unitary condition is satisfied. 

We parameterize the inverse of the $K$-matrix as a symmetric matrix of polynomials. 
Considering only S-wave elastic scattering in this work, the scattering amplitude is parameterized as:
\begin{align}\label{eq:para}
T_0^{-1}=c_0+c_1s-i\rho.
\end{align}
The generalized Lüscher formula is then given by:
\begin{align}\label{eq:Omega}
\Omega(E_{\rm cm}) = c_0+c_1s-\frac{1}{4\pi^{3/2} LE_{\rm cm}}\cdot\frac{1}{\gamma}{\cal Z}^{\bm d}_{00}[1;q^2]=0
,
\end{align}
which is applicable to both the $^3S_1$ and $^1S_0$ channels.

\subsection{Fitting details}

As discussed in Sec.~\ref{subsec:para}, we parameterize the energy dependence of the scattering matrix by a limited set of parameters $\{c_i\}$. 
For any specific values of these parameters, Eq.~(\ref{eq:Omega}) can be solved to yield a discrete spectrum of finite-volume energies $\{E_{\rm sol.}(L,{\bm P};\{c_i\})\}$. 
To quantify the resemblance between the finite-volume spectrum $\{E_{\rm latt.}(L,{\bm P})\}$ obtained in Sec.~\ref{sec:spectrum} and the spectrum $\{E_{\rm sol.}(L,{\bm P};\{c_i\})\}$
predicted by the parameterization, we introduce a residual:
\begin{align}\label{eq:chi2}
\chi^2(\{c_i\})=&\sum_{L,{\bm P}}[E_{\rm latt.}(L,{\bm P})-E_{\rm sol.}(L,{\bm P};\{c_i\})]_m~{\cal C}^{-1}_{mn} \nonumber\\
&\times[E_{\rm latt.}(L,{\bm P})-E_{\rm sol.}(L,{\bm P};\{c_i\})]_n
.
\end{align}
In this expression, the data covariance ${\cal C}$ accounts for the correlation between the determined energy level $\{E_{\rm latt.}(L,{\bm P})\}$ measured on the same ensembles, with $m$ and $n$ indexing the energy levels. 
By varying $\{c_i\}$, we minimize the $\chi^2$ to achieve the best matching description.

The continuum dispersion relation for a single nucleon is not perfectly maintained due to lattice artifacts, leading to deviations of lattice energies from their continuum counterparts. 
Given that the Lüscher formula relies on the continuum dispersion relation, we shift the two-particle finite volume energies to account for differences between the continuum and lattice single particle energies~\cite{Padmanath:2022cvl,Prelovsek:2020eiw,Piemonte:2019cbi}:
\begin{align}\label{eq:shift}
E_{\rm latt.}(L,{\bm P})\rightarrow &E_{\rm latt.}(L,{\bm P})+(M^{\rm cont.}_{N({\bm k_1})}+M^{\rm cont.}_{N({\bm k_2})}) \nonumber\\
&-(M^{\rm lat.}_{N({\bm k_1})}+M^{\rm lat.}_{N({\bm k_2})}).
\end{align}
Here, $M^{\rm cont.}_{N({\bm k_i})}$ represents the energy of a nucleon at momentum ${\bm k_i}$ as calculated using the continuum dispersion relation, and $M^{\rm lat.}_{N({\bm k_i})}$ is the corresponding energy computed on the lattice.

Unlike the straightforward process of fitting the two-point function in Sec.~\ref{sec:spectrum}, minimizing $\chi^2(\{c_i\})$ presents significant challenges.
Firstly, $\chi^2(\{c_i\})$ lacks favorable mathematical properties and is highly non-linear. 
Its gradient may not exist across the entire $\{c_i\}$ space. 
This makes traditional gradient-dependent methods, such as the Levenberg-Marquardt algorithm, inefficient and highly sensitive to the initial parameter values.
Secondly, exact solutions to Eq.~(\ref{eq:Omega}) are unattainable; instead, we obtain $\{E_{\rm sol.}(L,{\bm P};\{c_i\})\}$ through numerical methods, necessitating repeated evaluations of the Lüscher zeta function, which becomes a computational bottleneck in the fitting process.

To address the first challenge, we employ the differential evolution algorithm, which does not rely on gradients and is less prone to getting trapped in local minima. 
However, it converges more slowly than the Levenberg-Marquardt method, exacerbating the second challenge of computational intensity. 
Therefore, we must enhance the efficiency of calculating the Lüscher zeta function. 
Instead of using its original definition in Eq.~(\ref{eq:zeta}), we apply a formulation more suitable for numerical computations~\cite{Leskovec:2012gb}
\begin{align}\label{eq:zeta2}
{\cal Z}^{\bm d}_{lm}[1;x^2]&=\gamma\int^1_0dt~e^{tx^2}\sum_{{\bm n}\in Z^3,{\bm n}\neq0}(-1)^{2\alpha{\bm n}\cdot{\bm d}}(-i)^l \nonumber\\
&\times Y_{lm}(-\frac{\pi\hat{\gamma}{\bm n}}{t})(\frac{\pi}{t})^{3/2}e^{-(\pi\hat{\gamma}{\bm n})^2/t} \nonumber\\
&+\gamma\int^1_0dt~(e^{tx^2}-1)(\frac{\pi}{t})^{3/2}\frac{1}{\sqrt{4\pi}}\delta_{l0}\delta_{m0}-\gamma\pi\delta_{l0}\delta_{m0} \nonumber\\
&+\sum_{{\bm r}\in P_d}Y_{lm}({\bm r})\frac{e^{-(r^2-x^2)}}{r^2-x^2}
,
\end{align}
where the summation in the last term is taken from -5 to +5 in each of the three dimensions in this work. 
Moreover, we store the Lüscher zeta function on disk with a precision of $10^{-4}$ for $x^2$. 
The resulting precision for $E_{\rm cm}$ can be estimated by
\begin{align}\label{eq:zetaUncer}
\delta(E_{\rm cm})=\frac{\delta(E^2_{\rm cm})}{2E_{\rm cm}}=\frac{4}{2E_{\rm cm}}(\frac{2\pi}{L})^2\delta(x^2)\approx10^{-2}~{\rm MeV}
,
\end{align}
which is sufficiently high compared to the precision of $\{E_{\rm latt.}(L,{\bm P})\}$ in our analysis. 
When the fitting program initiates, we load these discretized values of the Lüscher zeta function from disk into memory and interpolate them using techniques such as Cubic Spline interpolation. 
This approach significantly enhances the efficiency of the fitting program.

\begin{figure}[tbph]
\centering
\includegraphics[width=4cm]{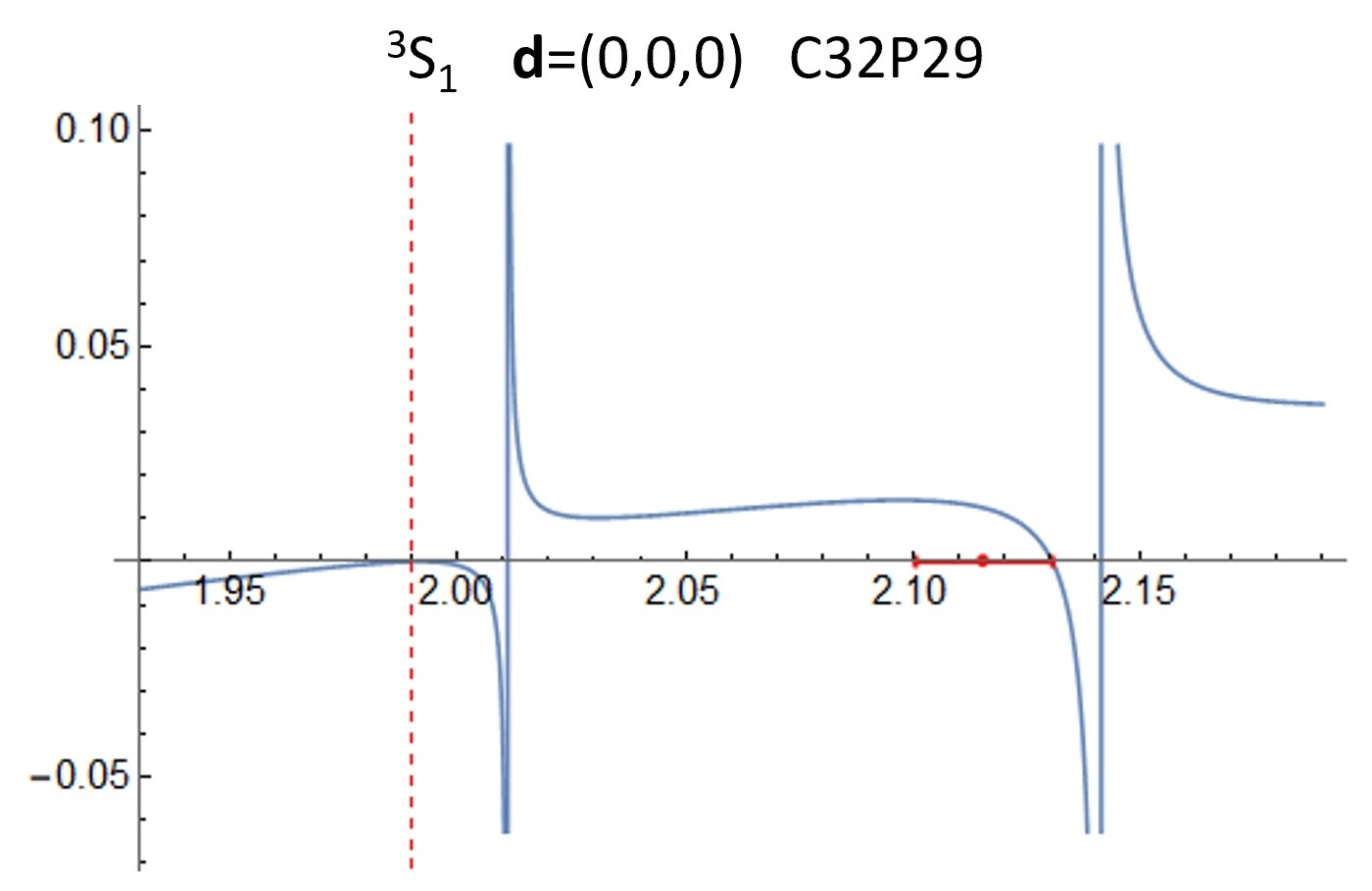}
\includegraphics[width=4cm]{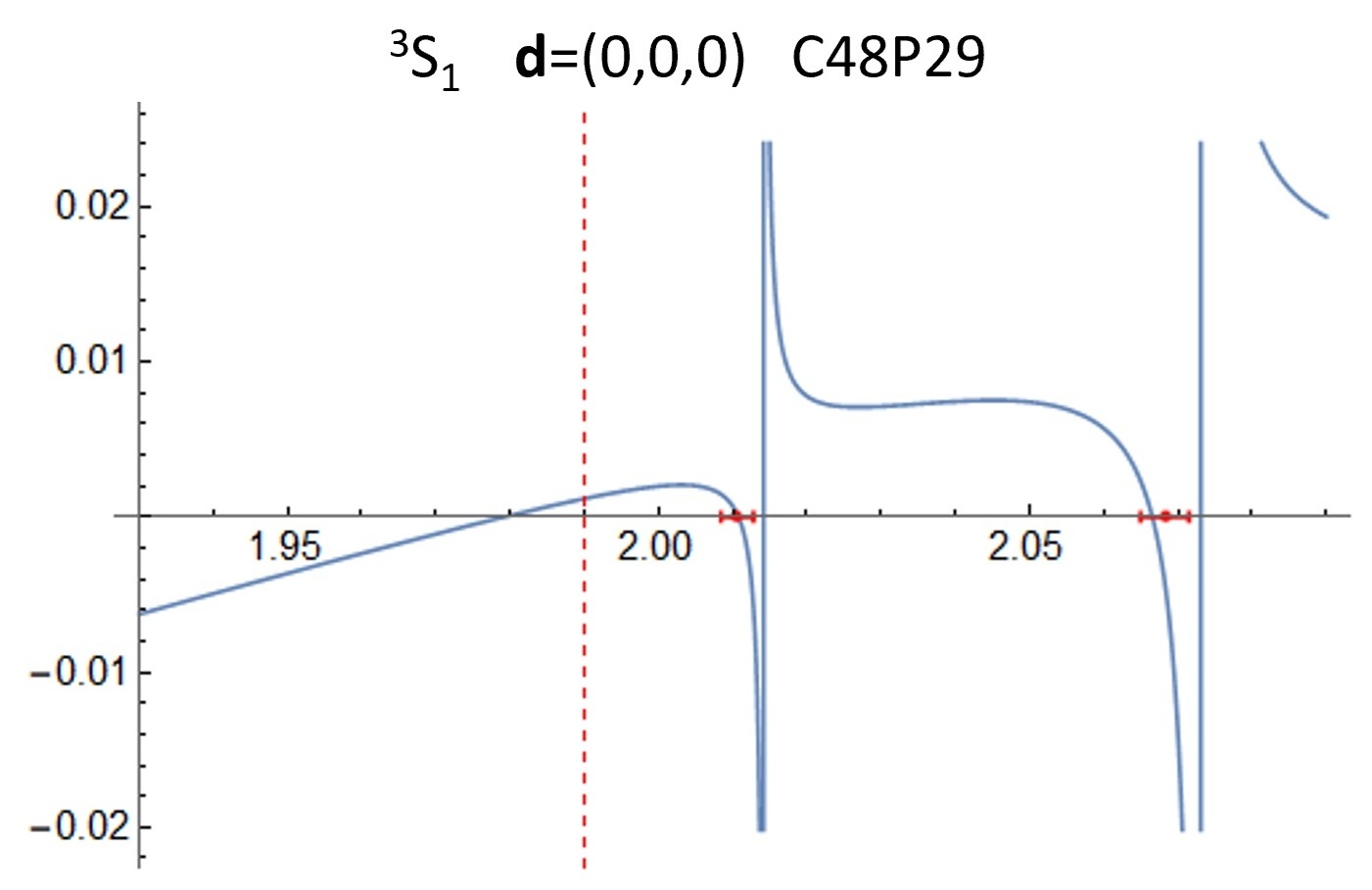}
\includegraphics[width=4cm]{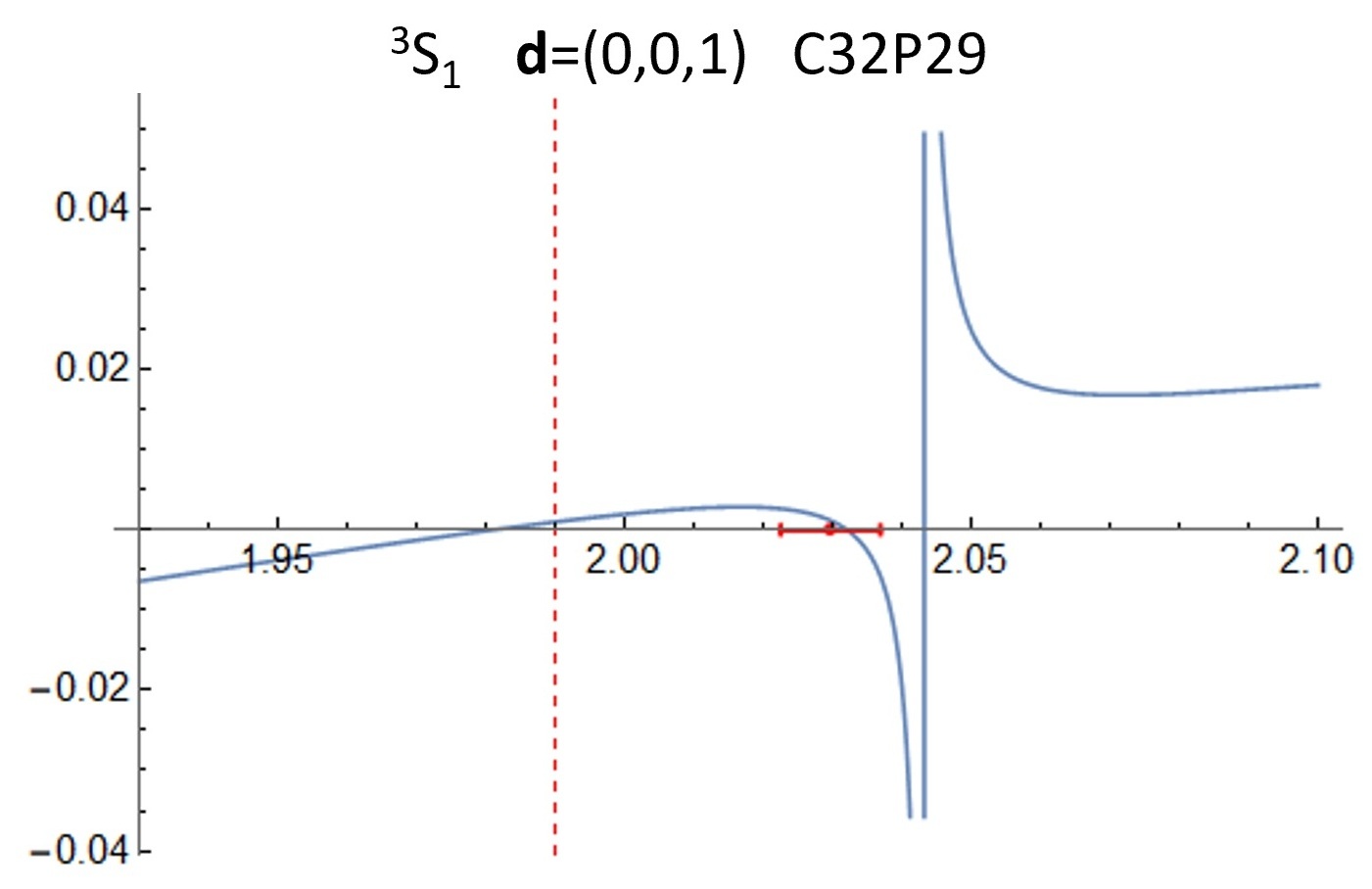}
\includegraphics[width=4cm]{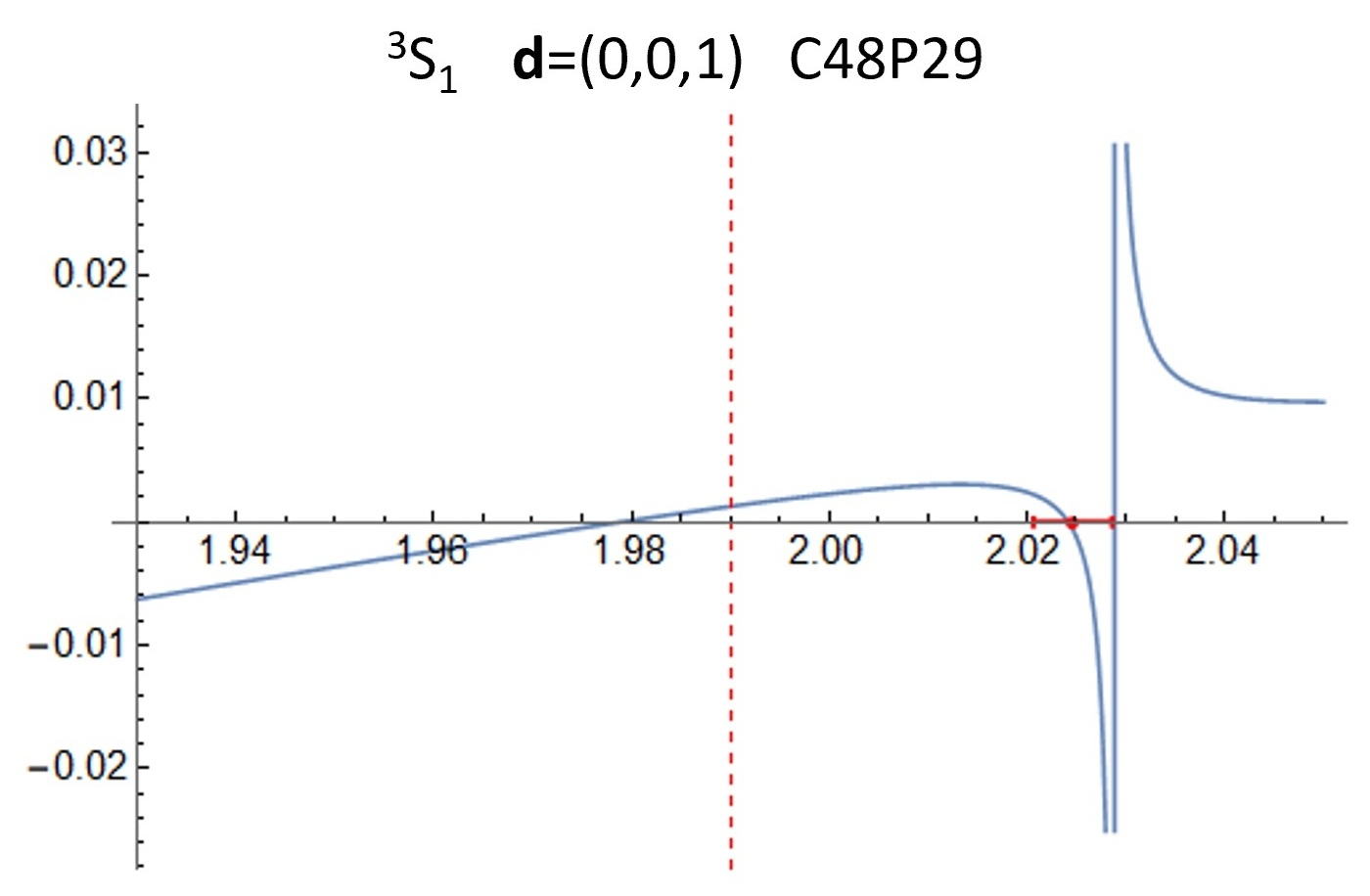}
\includegraphics[width=4cm]{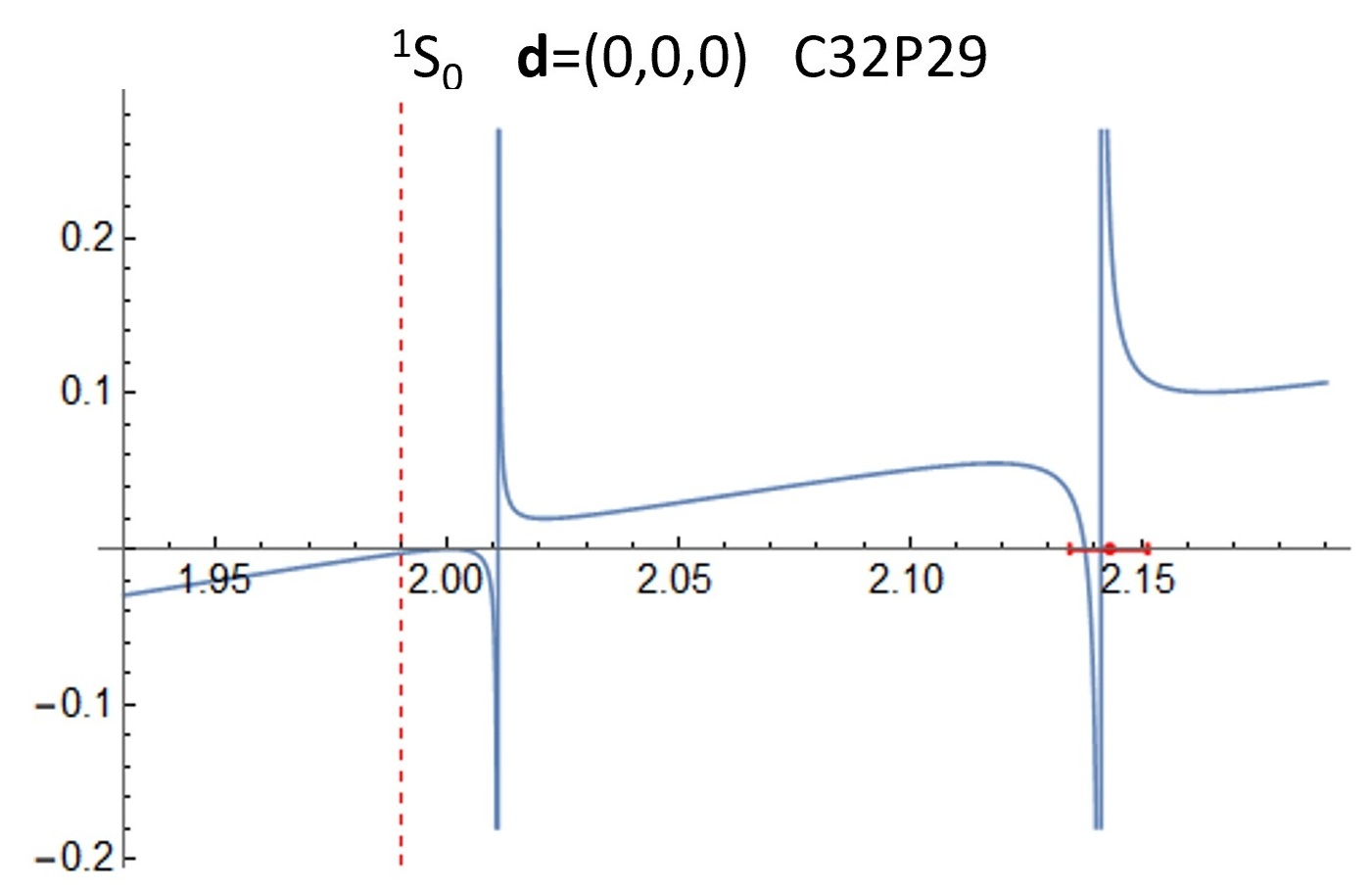}
\includegraphics[width=4cm]{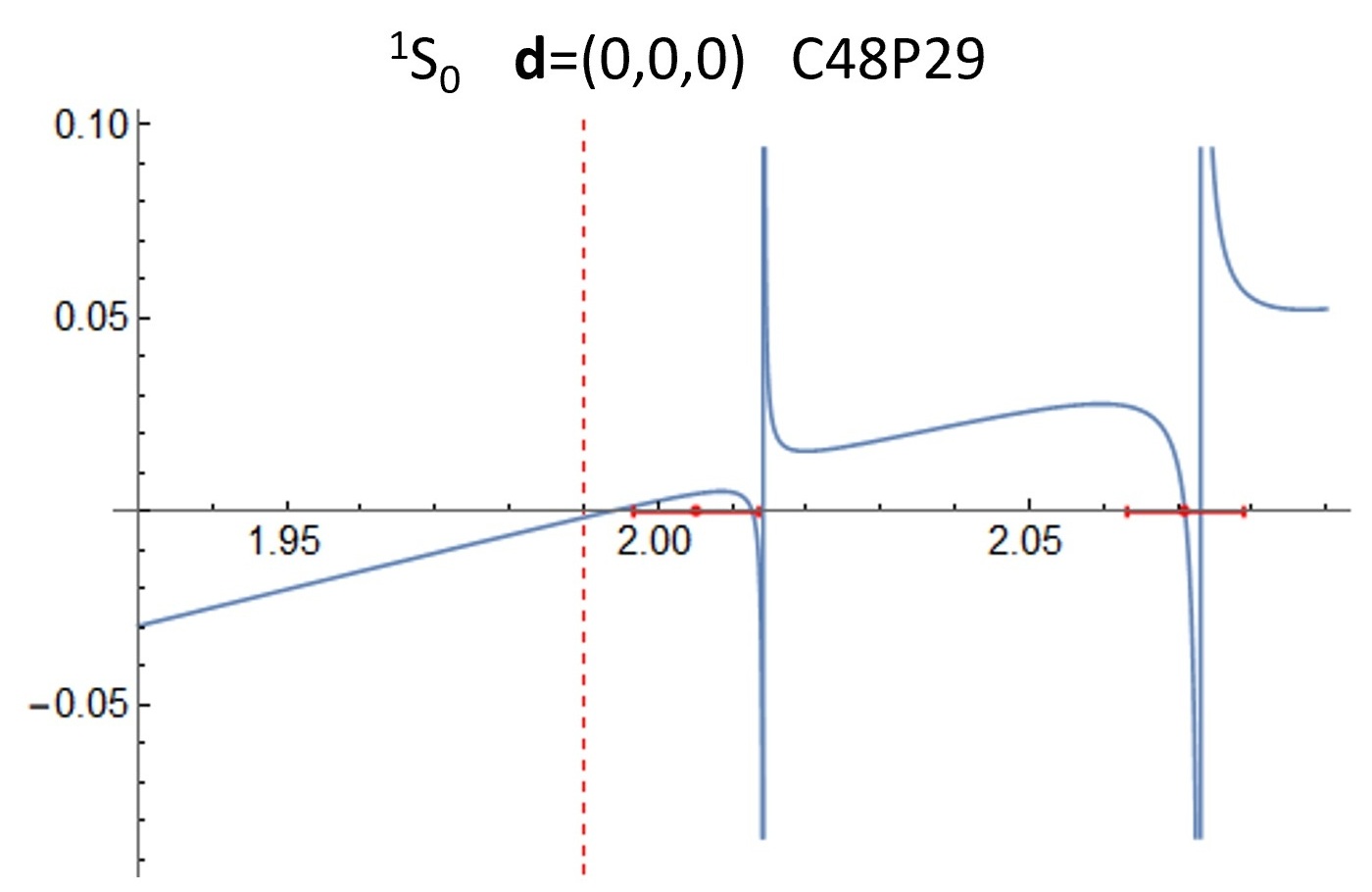}
\includegraphics[width=4cm]{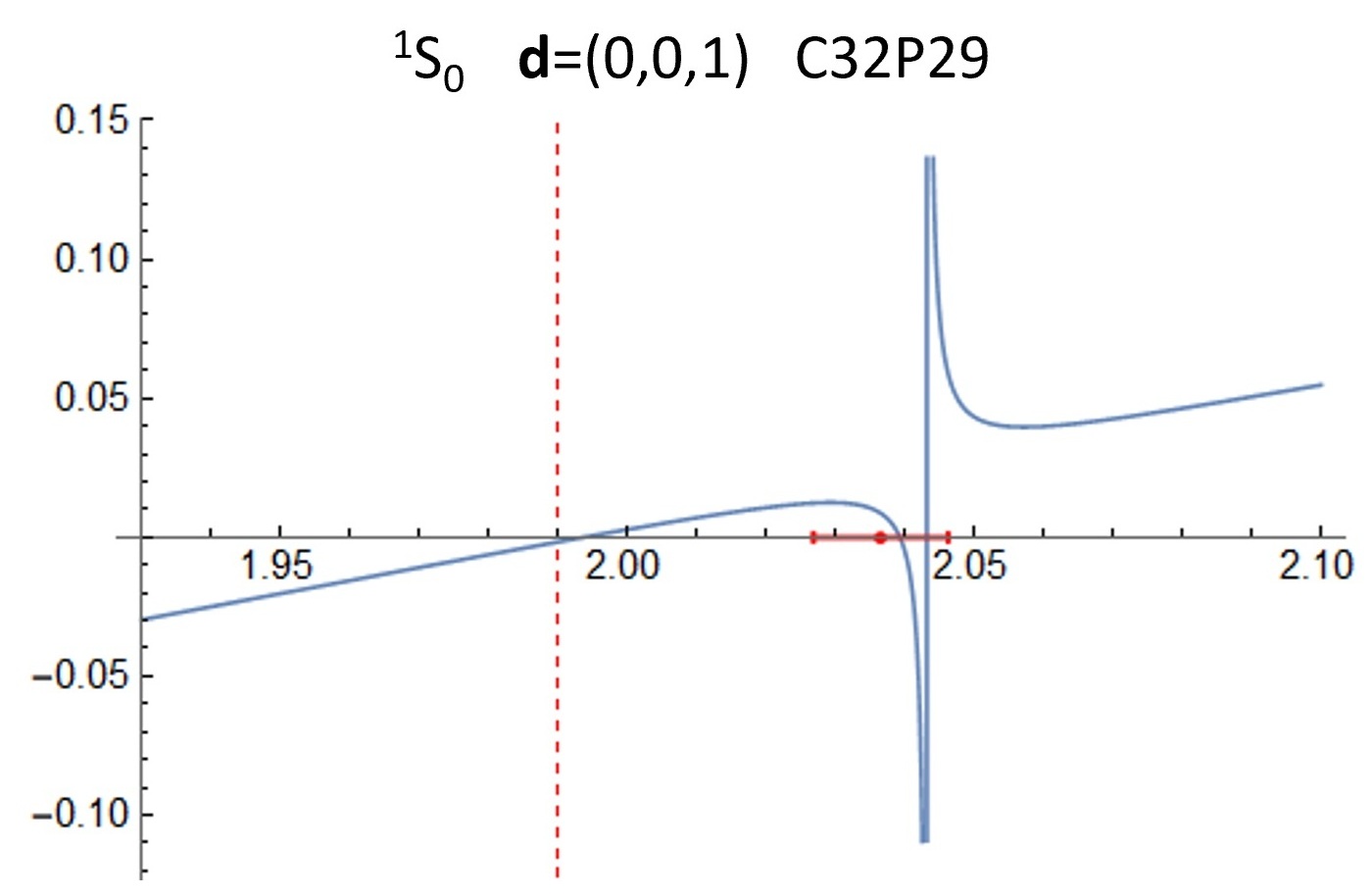}
\includegraphics[width=4cm]{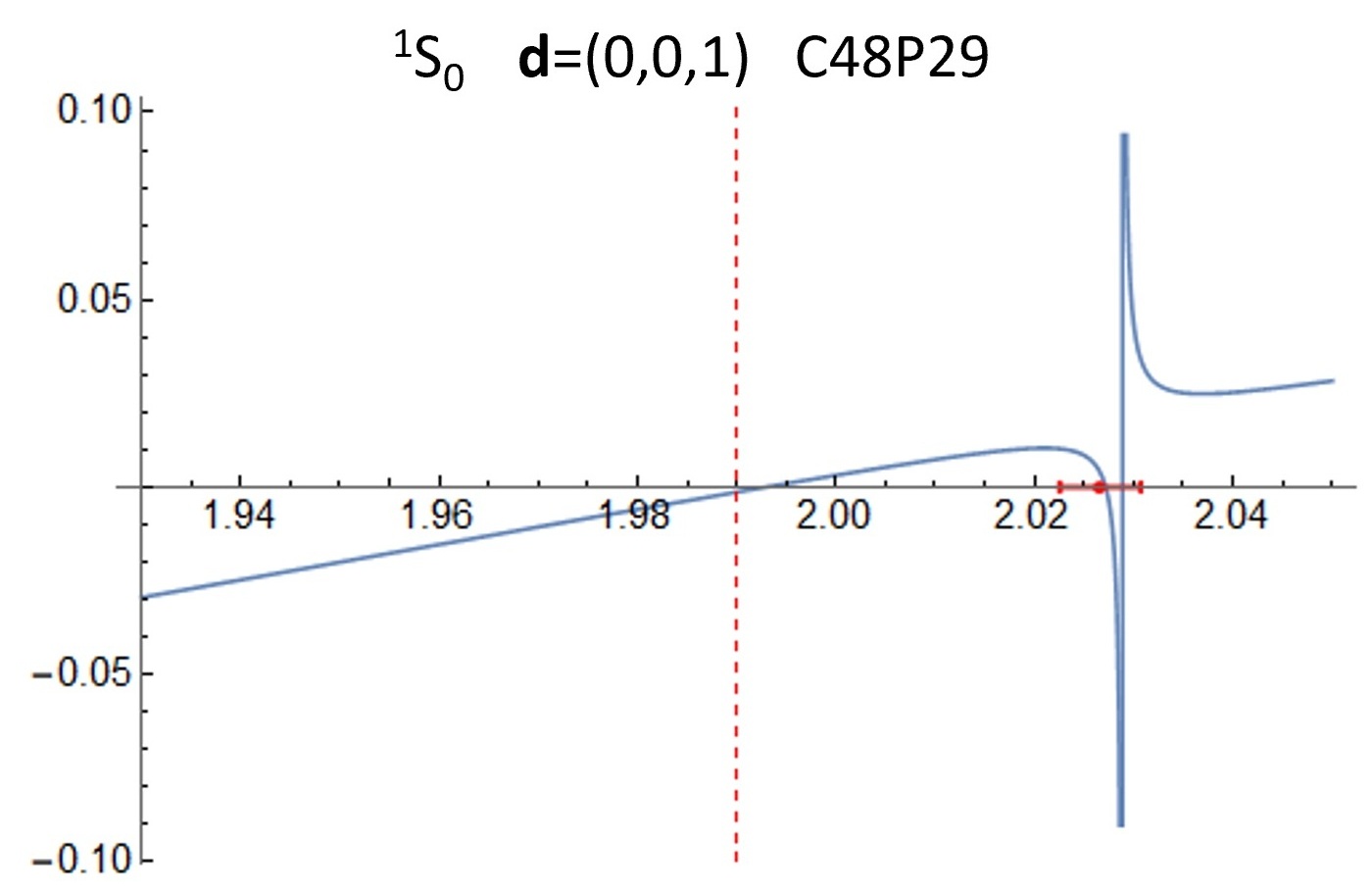}
\caption{$\Omega(E_{\rm cm})$ defined in Eq.~\ref{eq:Omega} plotted against $E_{\rm cm}$ in GeV for the parameters $c_0=-0.176698$, $c_1=0.155040$ in the $^3S_1$ channel (upper section) and $c_0=-0.526277$, $c_1=0.462596$ in the $^1S_0$ channel (lower section). The left section displays the results for C32P29 ($L=32$), while the right section for C48P29 ($L=48$). For each channel and ensemble, the upper panel corresponds to the rest frame (${\bm d}=(0,0,0)$) and the lower panel to the moving frame (${\bm d}=(0,0,1)$). The red dots are the energy levels used to constrain the parameters $\{c_i\}$. The red dashed lines indicate the left-hand cut, the blue vertical lines represent the free energies, and the blue curves depict $\Omega(E_{\rm cm})$.}
\label{fig:Omega}
\end{figure}

By minimizing $\chi^2(\{c_i\})$ using the central values of the energy levels, we obtain $c_0=-0.176698$ and $c_1=0.155040$ with a reduced chi-square of $\chi^2/{\rm dof}=0.44$ for the $^3S_1$ channel, and $c_0=-0.526277$ and $c_1=0.462596$, with $\chi^2/{\rm dof}=0.33$ for the $^1S_0$ channel. 
Fig.~\ref{fig:Omega} shows examples of $\Omega(E_{\rm cm})$, defined in Eq.~\ref{eq:Omega}, for these specific parameters $\{c_i\}$. 
The red dashed lines indicate the left-hand cut, and the blue vertical lines represent the free energies. 
The solutions $\{E_{\rm sol.}(L,{\bm P};\{c_i\})\}$ are found where the blue curves $\Omega(E_{\rm cm})$ intersect the zero lines. 
Since the Lüscher formula is not applicable below or near the left-hand cut, the lowest solutions shown in Fig.~\ref{fig:Omega} are not meaningful. 
The number of intersections agrees with the number of observed levels in the relevant energy ranges.

\begin{figure}[tbph]
\centering
\includegraphics[width=7.15cm]{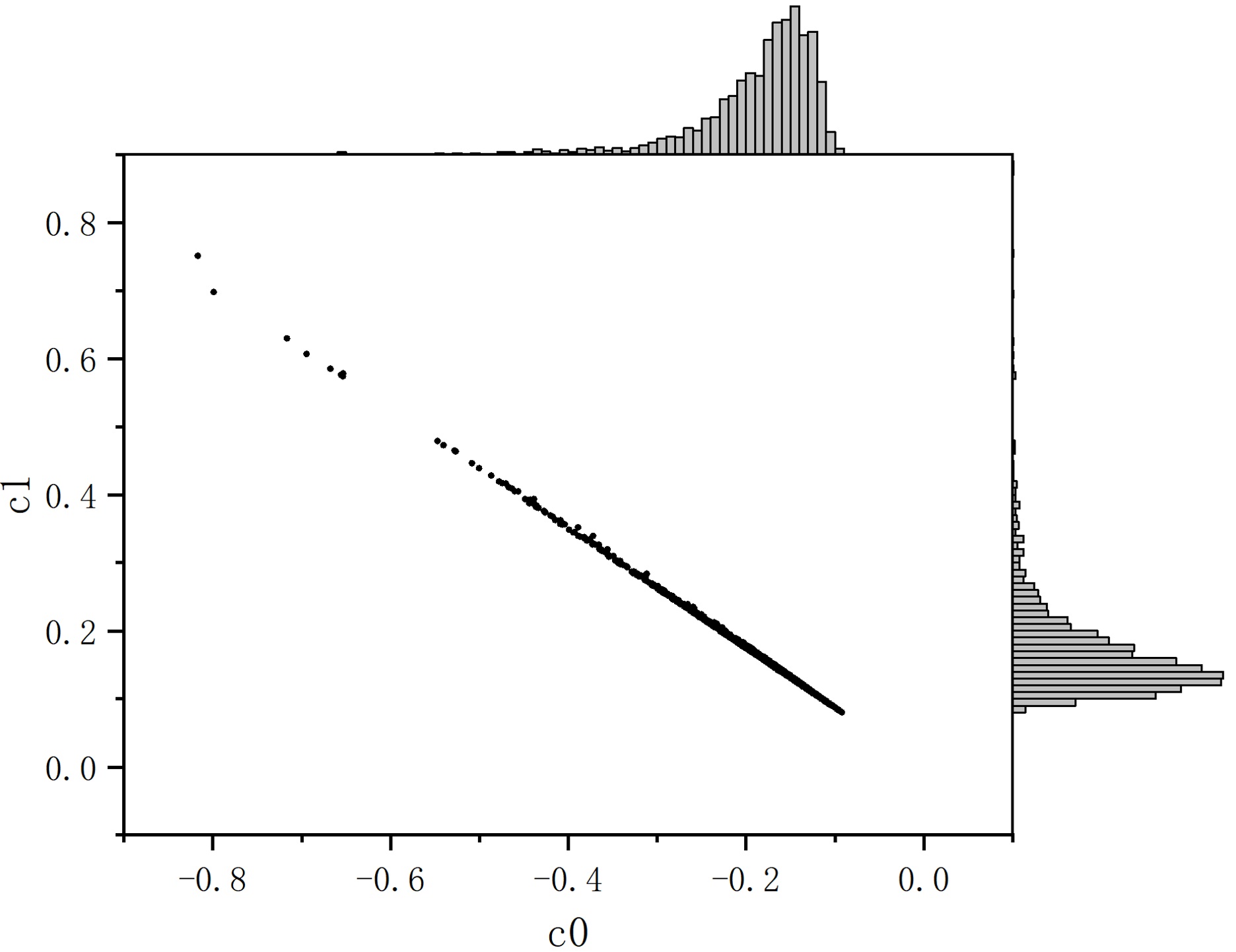}
\includegraphics[width=7cm]{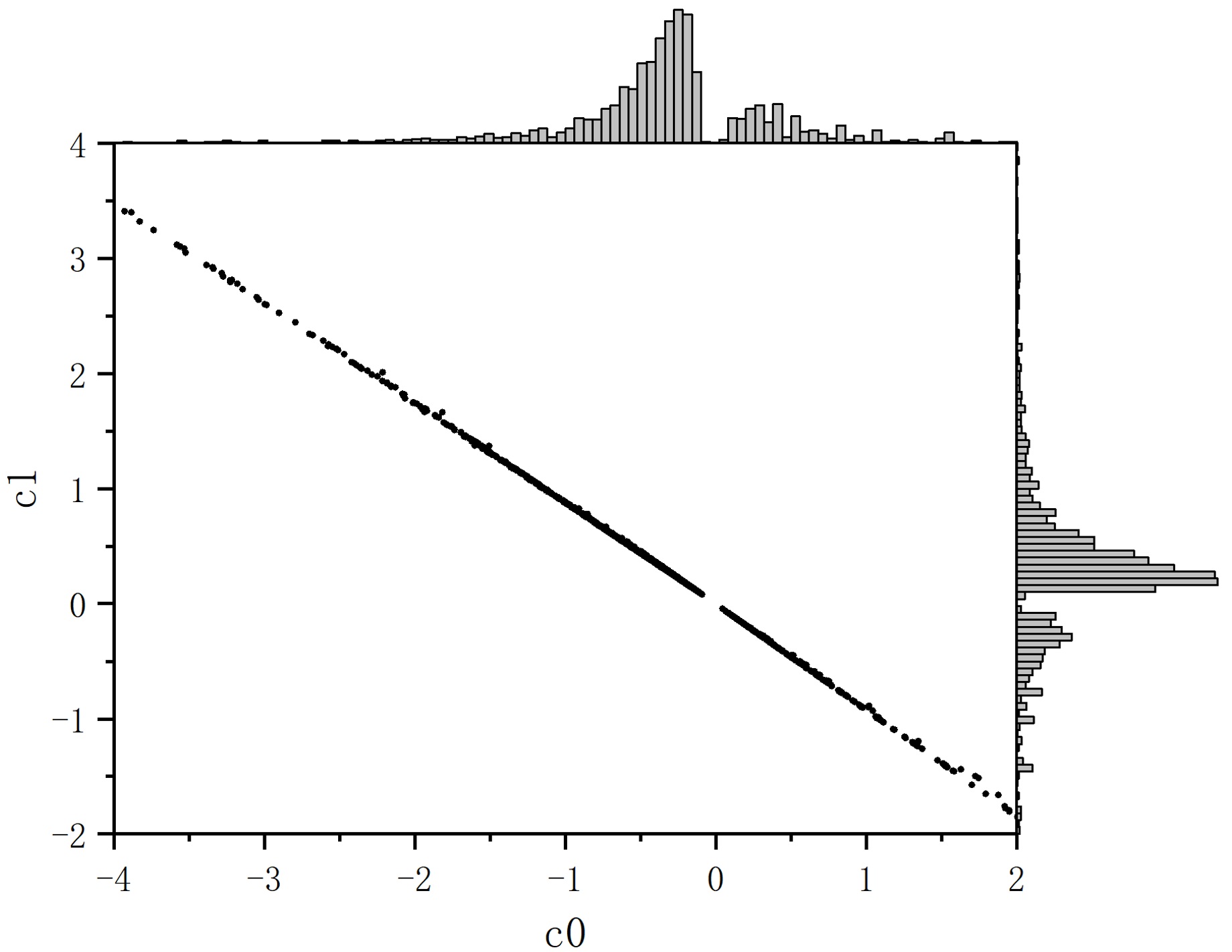}
\caption{The results for the parameters $\{c_i\}$ in each sample through minimizing $\chi^2(\{c_i\})$ in the $^3S_1$ channel (upper section) and the $^1S_0$ channel (lower section). The distributions of these parameters, shown in the marginal histograms, are asymmetrical, particularly in the $^1S_0$ channel. This asymmetry arises because the magnitude of the gradient of the Lüscher zeta function in $\Omega(E_{\rm cm})$ becomes significantly larger as $E_{\rm cm}$ approaches the free energies.}
\label{fig:para}
\end{figure}

Uncertainties are estimated by the bootstrap method. 
The parameters $\{c_i\}$ obtained for each bootstrap sample are shown in Fig.~\ref{fig:para}. 
The distributions of these parameters deviate from a Gaussian distribution and are asymmetric, so we estimate the statistical uncertainty from the central 68 percent of the samples. 
For the $^3S_1$ channel, this method yields $c_0=-0.17^{+0.04}_{-0.07}$ and $c_1=0.15^{+0.06}_{-0.03}$. 
In the $^1S_0$ channel, the distribution is not continuous. {The Lüscher zeta function is discontinuous at free energies, causing a gap at zero in the scattering phase shift ($k{\cot}\delta$) calculated from the Lüscher formula. This feature extends to the $c$ coefficients: imprecise energy levels in the $^1S_0$ intersect with the free energies, creating a similar gap at zero.}

\subsection{Pole singularities}

In Eq.~(\ref{eq:para}), the variable $k$ is present in $\rho$ and consequently in $T$. 
This results in $T$ having a branch cut when regarded as a function of the complex Mandelstam variable $s$, resulting in two Riemann sheets. 
The first, or physical sheet, is characterized by ${\rm Im}(k)>0$, while the second, or unphysical sheet, has ${\rm Im}(k)<0$. 
A pole singularity in the $T$ matrix usually indicates the existence of a state. 
Poles located on the real axis below the threshold on the physical sheet are associated with bound states, while poles off the real axis and above the threshold on the unphysical sheet correspond to resonances. 
If $T$ exhibits a pole on the real energy axis below the threshold on the unphysical sheet, it is called a virtual state. 
This type of singularity leads to an enhancement at the threshold, though it does not allow for an asymptotic state.

\begin{figure}[tbph]
\centering
\includegraphics[width=4cm]{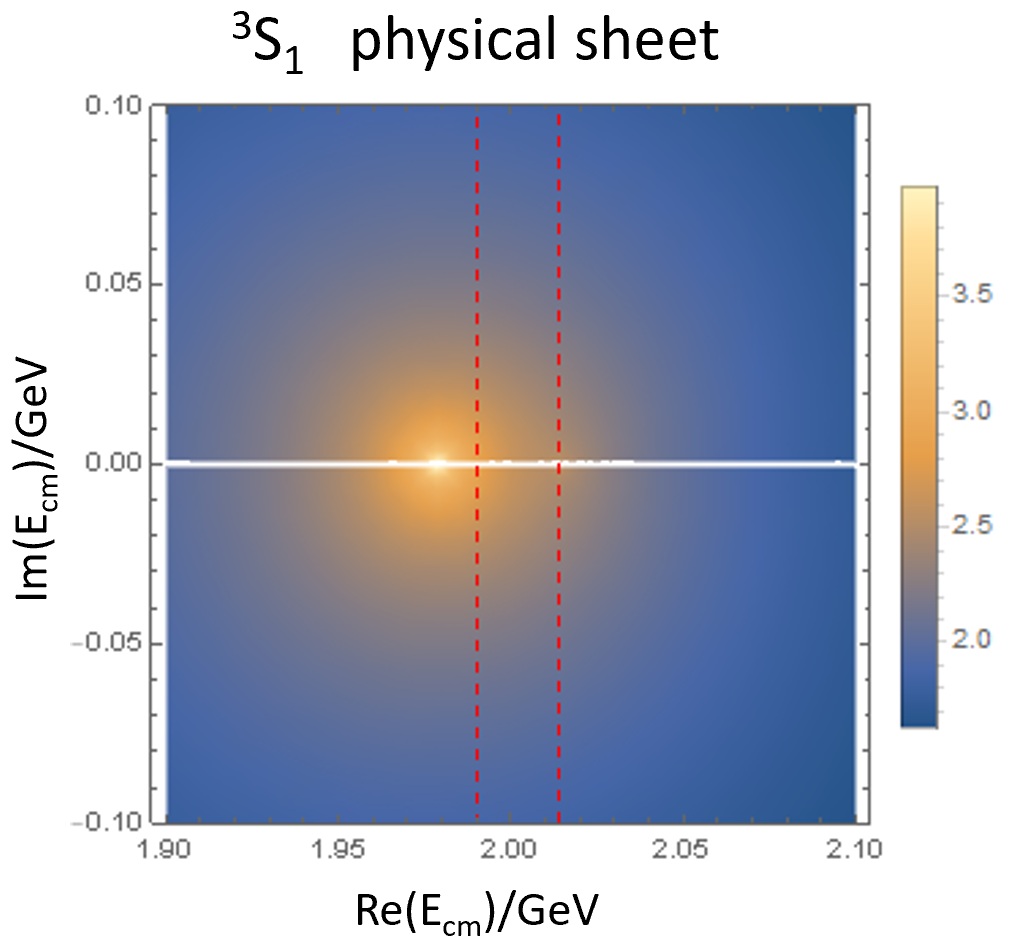}
\includegraphics[width=4cm]{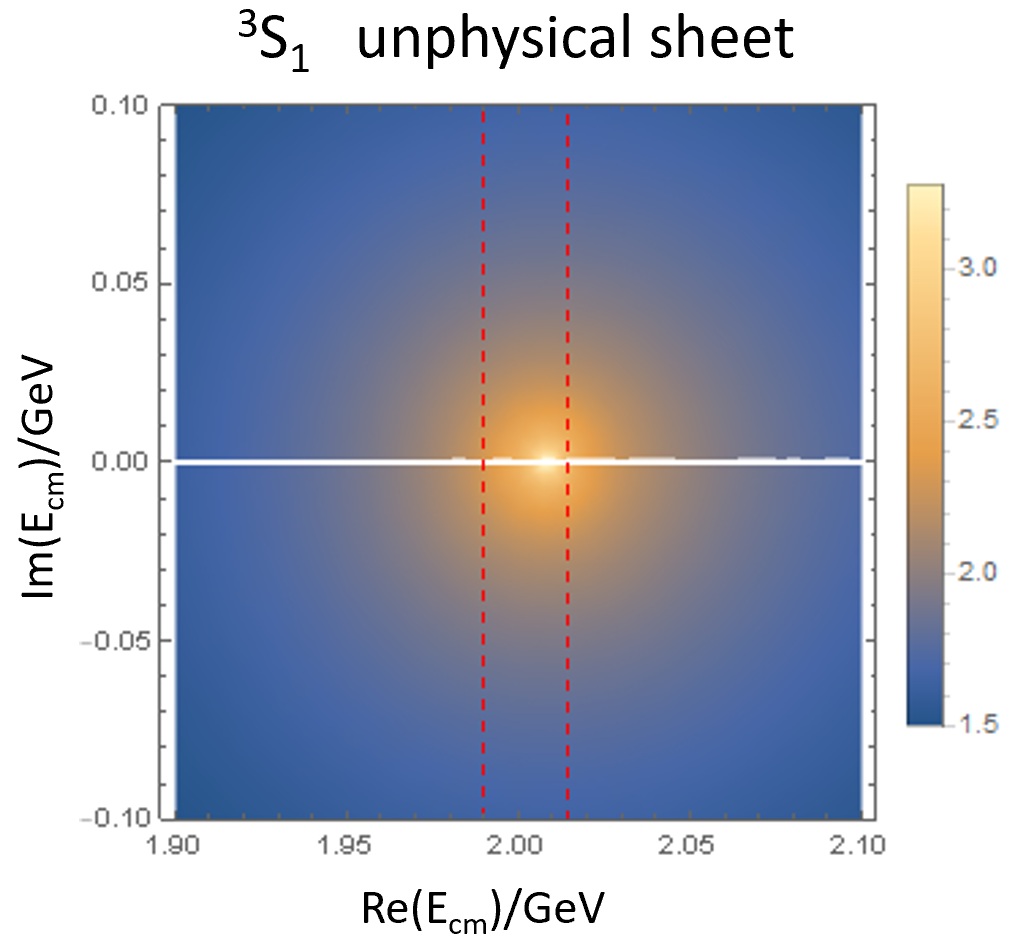}
\includegraphics[width=4cm]{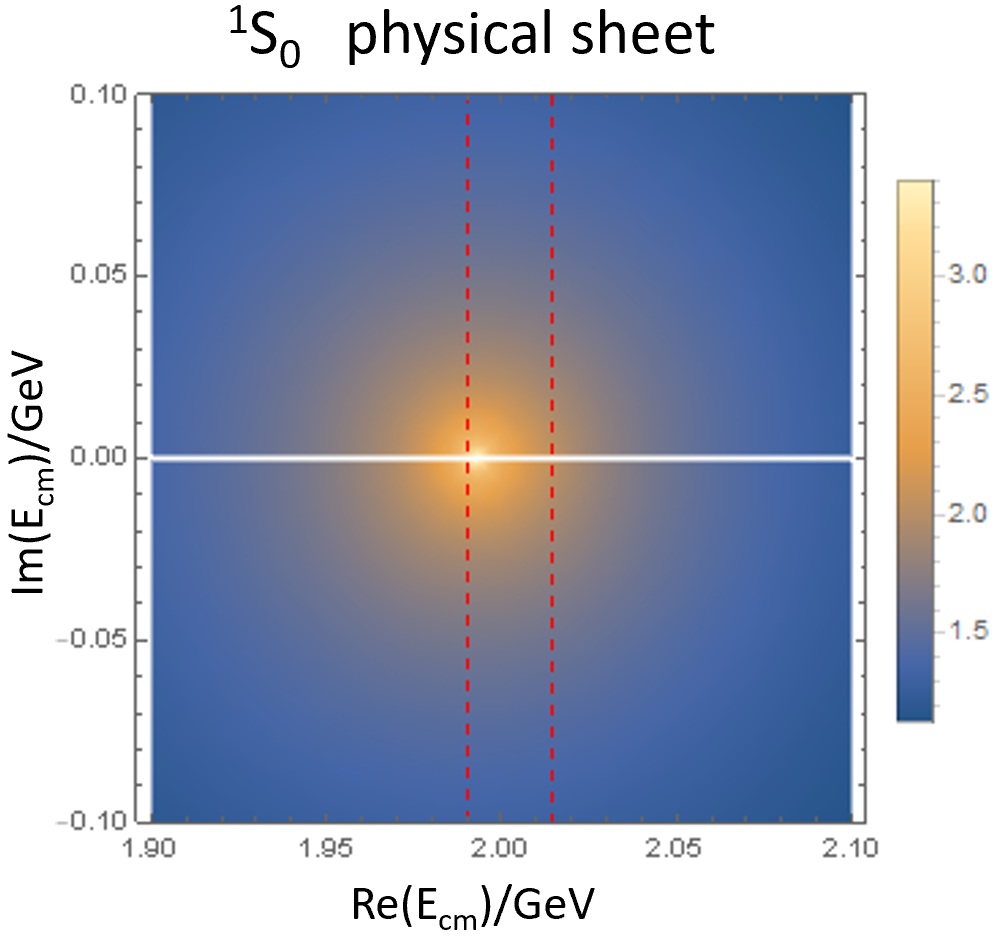}
\includegraphics[width=4cm]{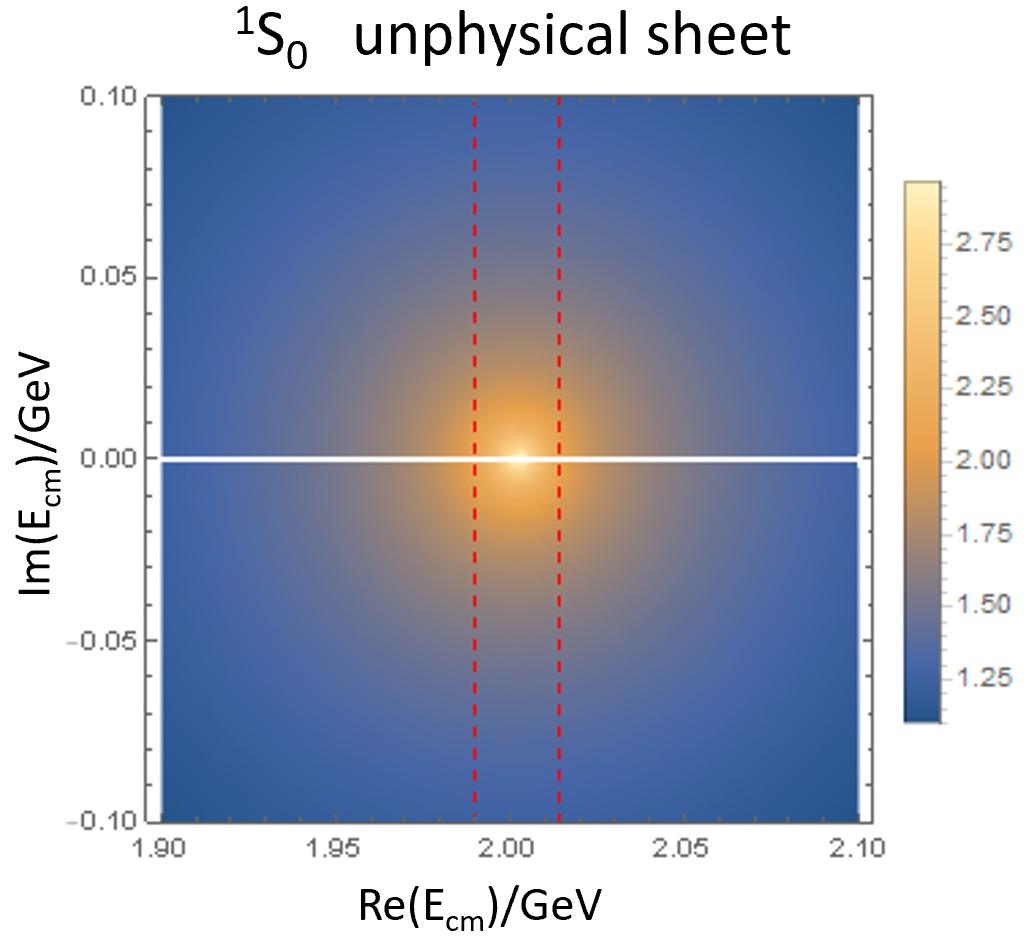}
\caption{${\rm log}|T|$ in the $^3S_1$ channel (upper row, $c_0=-0.176698$, $c_1=0.155040$) and $^1S_0$ channel (lower row, $c_0=-0.526277$ and $c_1=0.462596$), on physical sheet (left column) and unphysical sheet (right column). In each panel, the red dashed line on the left is the left-hand cut, and the line on the right marks the threshold.}
\label{fig:pole}
\end{figure}

In Fig.~\ref{fig:pole}, we plot ${\rm log}|T|$ using the parameters $\{c_i\}$ obtained from fitting the central values of the energy levels. 
%
%In the $^3S_1$ channel, there is a pole on the real axis below threshold, present on both the physical and unphysical sheets. 
%
%{\color{blue}The pole on the unphysical sheet influences the physical amplitudes on the real axis $(s+i\epsilon~{\rm with}~\epsilon\rightarrow0^+)$ in this case~\cite{Habashi:2020qgw}Therefore, the $^3S_1$ channel displays a virtual state pole. }. 
%
%{\color{red}However, the pole on the physical sheets is below the left-hand cut, where is beyond the K-matrix polynomial expansion around the threshold as shown in Eq.~(\ref{eq:para}), thus, it is neglected.The pole on the second Riemann sheet corresponds to a virtual pole, which is very close to the threshold, and will be influence the physical amplitude on the real axis~\cite{Habashi:2020qgw}.Such virtual pole demonstrates that the potential between $NN$ in $^3S_1$ channel is attractive, while it is not enough to form a bound state.}A similar situation is observed in the $^1S_0$ channel. 
In both the $^3S_1$ and $^1S_0$ channels, we find two poles on the real axis below the threshold, one on the physical sheet and the other on the unphysical sheet. The poles on the physical sheet lie below or very close to the left-hand cut, where the L\"ushcer formula and the K-matrix polynomial expansion in Eq.~(\ref{eq:para}) become invalid. Moreover, they are far away from the threshold. Therefore, we neglect these poles. In contrast, the poles on the unphysical sheet are closer to the threshold and influence the physical amplitude on the real axis. Such poles correspond to virtual poles, indicating that the interactions in both channels are attractive but not strong enough to form a bound state. 

\begin{figure}[tbph]
\centering
\includegraphics[width=7cm]{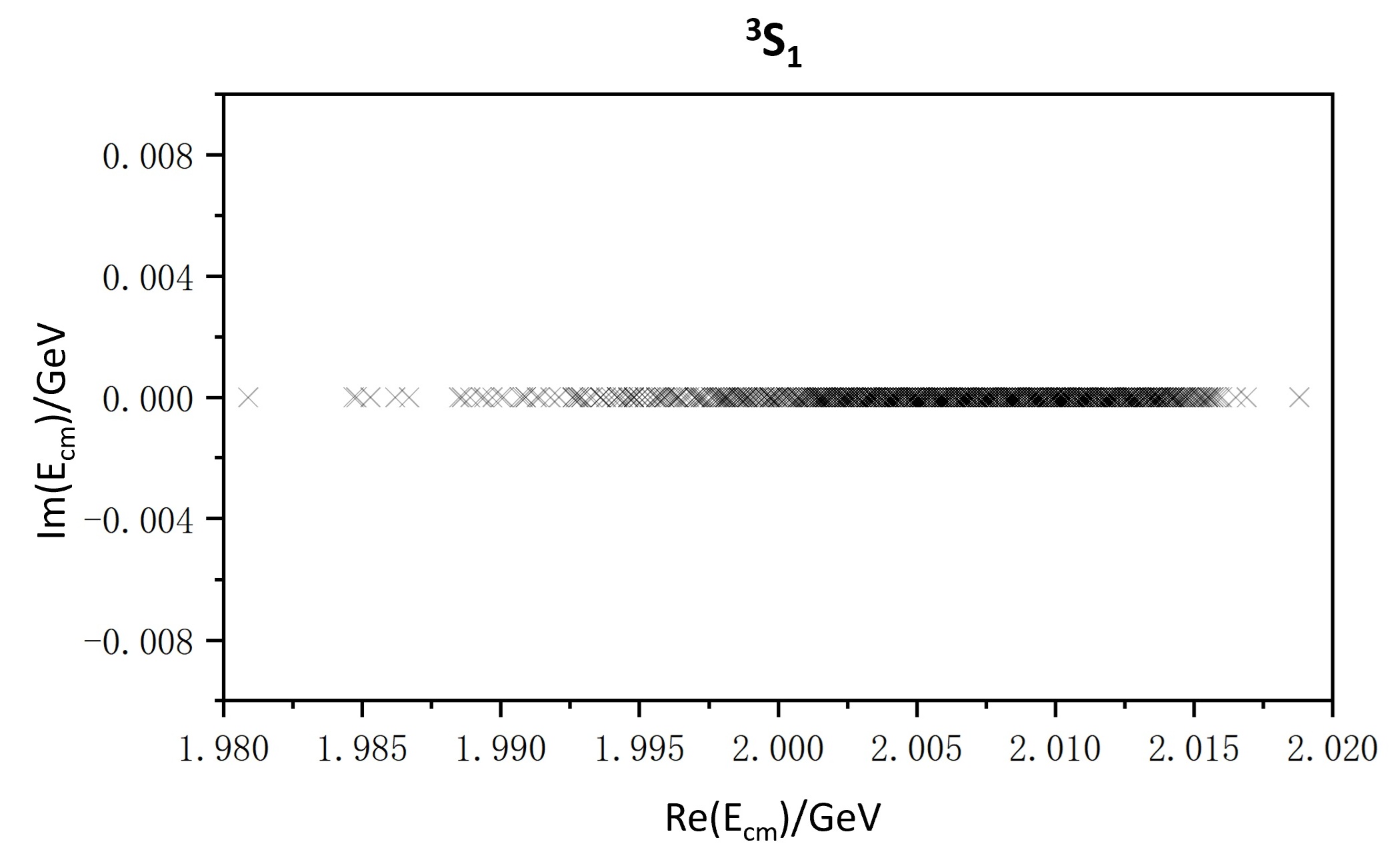}
\includegraphics[width=7cm]{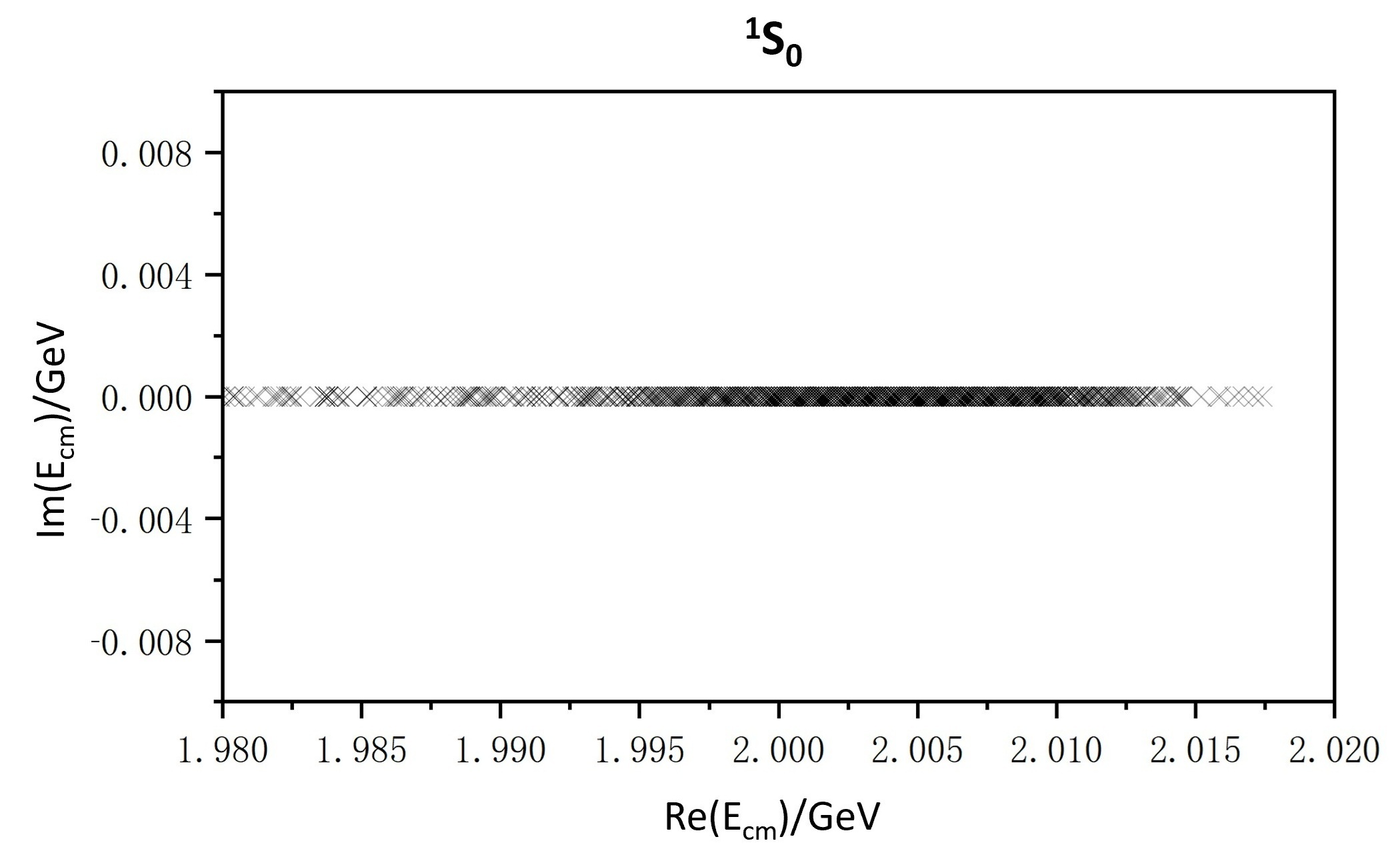}
\caption{The distributions of poles on the unphysical sheet in the $^3S_1$ channel (upper panel) and the $^1S_0$ channel (lower panel). Each sample shows a virtual state pole in the $^3S_1$ channel, and about 80 percent of the samples in the $^1S_0$ channel have a virtual state pole.}
\label{fig:poles}
\end{figure}

Fig.~\ref{fig:poles} illustrates the distribution of poles on the unphysical sheet across all bootstrap samples. 
In the $^3S_1$ channel, all the samples exhibit a virtual state pole located at $2008^{+4}_{-5}$ MeV. 
In the $^1S_0$ channel, approximately 80 percent of the samples show a virtual state pole at $2004^{+5}_{-6}$ MeV; most of the remaining samples exhibit a bound state pole, while a few samples show no pole in the studied range.

\begin{figure}[tbph]
\centering
\includegraphics[width=7cm]{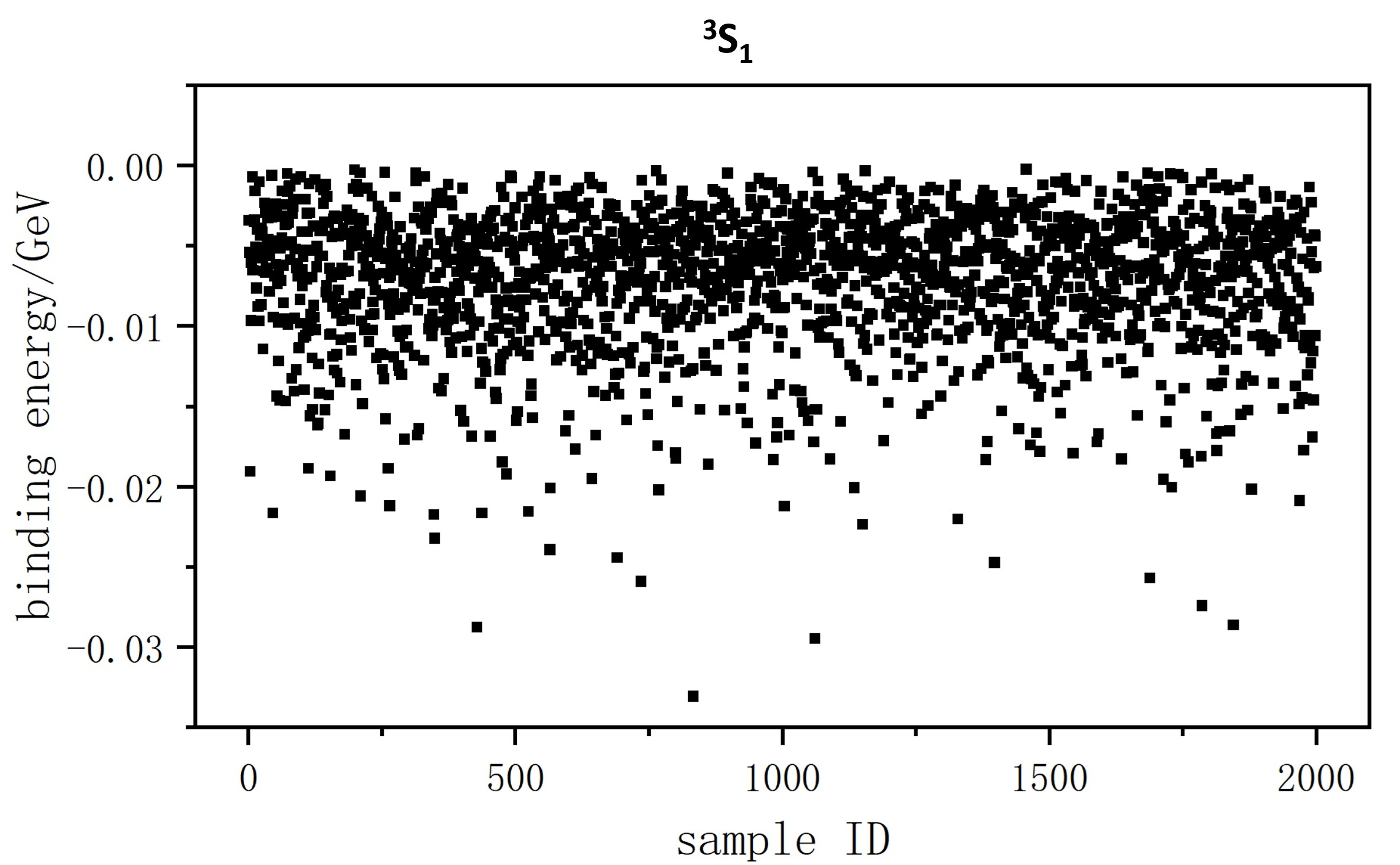}
\includegraphics[width=7cm]{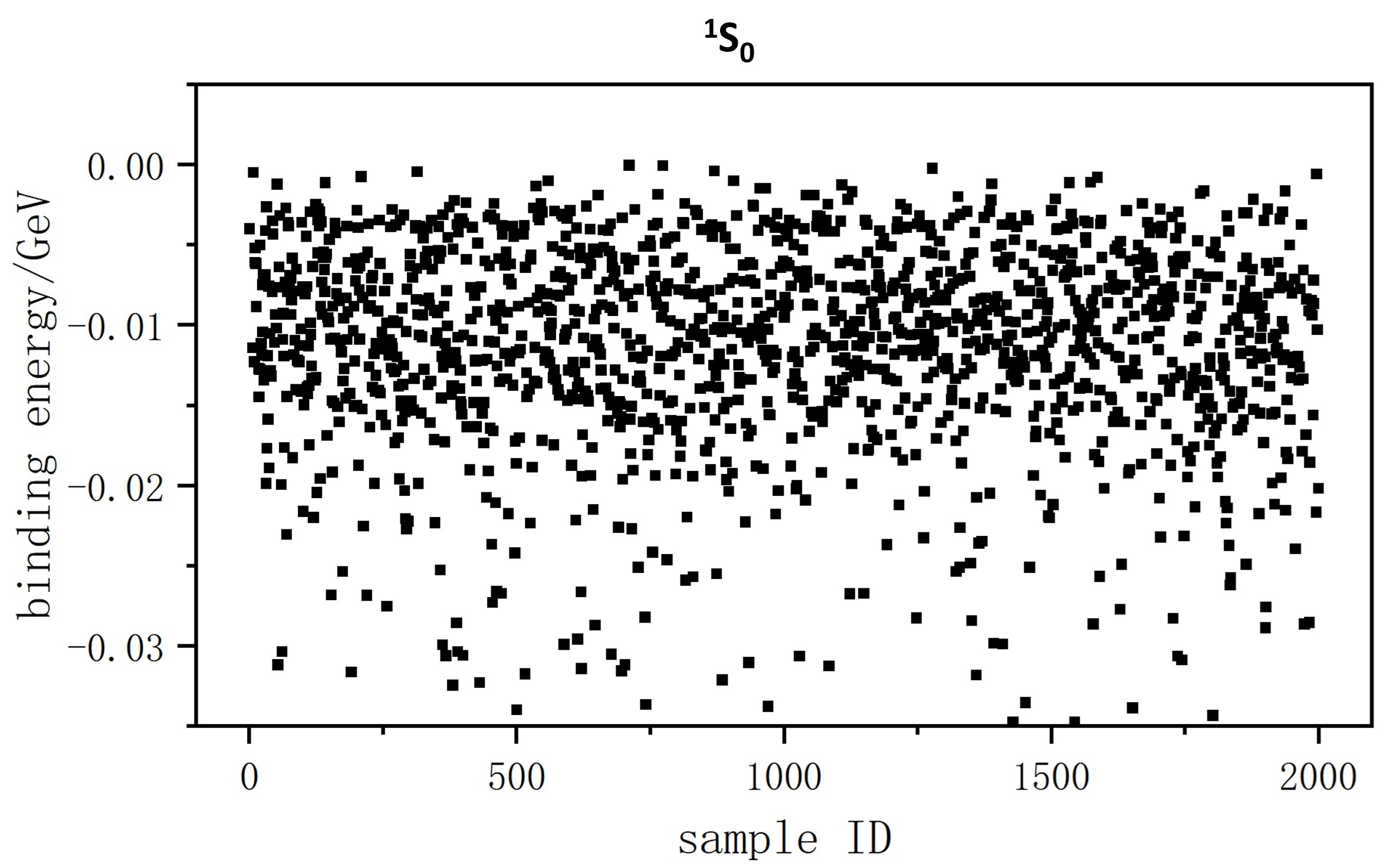}
\caption{The distributions of binding energy in the $^3S_1$ channel (upper panel) and the $^1S_0$ channel (lower panel).}
\label{fig:binding}
\end{figure}

Additionally, we present the distributions of binding energy\footnote{Here, the binding energy refers to the energy difference between the pole and the threshold.} in Fig.~\ref{fig:binding}. 
For the $^3S_1$ channel, the binding energy is $6^{+5}_{-3}$ MeV, while for the $^1S_0$ channel, it is $11^{+6}_{-5}$ MeV.

% \section{Hamiltonian Effective Field Theory}\label{sec:HEFT}
\section{Non-Perturbative Hamiltonian Framework}\label{sec:HEFT}

NPHF is an alternative framework for extracting physical observables from lattice simulations. 
This method can address the issues of the left-hand cut and slowly converging partial-waves expansion that arise from the one-pion-exchange interaction~\cite{Yu:2025gzg, Meng:2024kkp, Meng:2023bmz}, which, in principle, should not be neglected a priori for dinucleon scattering.
However, the L\"uscher formalism neglects the left-hand cut because it is designed only for short-range interactions.
%
% The workflow of NPHF is as follows.
The workflow of NPHF is as similar as Harmonic-Oscillator Based Effective Theory (HOBET)\,\cite{HOBET:1,HOBET:2,HOBET:3}:
First, we parameterize the Hamiltonian of the dinucleon system.
The finite-volume Hamiltonian matrix elements are then set up following Ref.~\cite{Yu:2025gzg}. 
The unknown parameters are determined by fitting its eigenvalues to lattice energy levels. 
For more technical details, we refer to Refs.~\cite{Yu:2025gzg, Li:2021mob}. 
In the infinite volume, the scattering amplitude $T$ is obtained by solving the partial-wave Lippmann-Schwinger equation:
\begin{align}\label{eq:LSE}
    &T^{JI}_{l^\prime l}(p,k;E) = V^{JI}_{l^\prime l}(p,k) \,+
    \notag\\
    &  \sum\limits_{l^{\prime\prime}} \int q^2 dq\, V^{JI}_{l^\prime l^{\prime\prime}}(p,q)\,G(q;E+i\epsilon)\,T^{JI}_{l^{\prime\prime} l}(q,k;E)
\end{align}
where $G(q;E)^{-1}=E-2\sqrt{q^2+m_N^2}$ and $V^{JI}$ is the partial wave potential with total angular momentum $J$ and isospin $I$.
In operator form, the solution is formally given by $T=(1-VG)^{-1}V$. 
{

In the finite volume, the potential is expanded in terms of the discrete momentum basis $|\vec{n},\sigma_1,\sigma_2\rangle$ with $\sigma$ denotes the polarization of nucleon, and the entries of matrix $V^L$ before irreducible representation projection are defined by
\begin{align}
V^L_{\sigma_1'\sigma_2',\sigma_1\sigma_2}\left(\vec{n}',\vec{n}\right)\!\! = \!\!\left(\frac{2\pi}{L}\right)^3 V_{\sigma_1'\sigma_2',\sigma_1\sigma_2}\left(\!\frac{2\pi \vec{n}'}{L},\frac{2\pi\vec{n}}{L}\!\right).
\end{align}
and the total Hamiltonian is given by 
\begin{align}
H^L_{\sigma_1'\sigma_2',\sigma_1\sigma_2}& = \left( H_0^L+V^L \right)_{\sigma_1'\sigma_2',\sigma_1\sigma_2}
\\
\hat{H}_0^L |\vec{n},\sigma_1 \sigma_2 \rangle & = 2\sqrt{(2\pi|\vec{n}|/L)^2 + m_N^2} | \vec{n},\sigma_1\sigma_2 \rangle
\end{align}
The eigenvalue of $H^L$ corresponds to the lattice spectra.
For details of irreducible representation projection, we refer to Ref.~\cite{Yu:2025gzg}
}

Considering the number of lattice spectra is limited, we parameterize the dinucleon interaction $V=V_{s} + V_\text{ope}$ at leading order based on the well-known result of chiral perturbation theory~{\cite{Machleidt:2011zz}} as follows.
For the isoscalar channel($I=0$):
\begin{align}
    &V^{I=0}_{s}(\vec{p},\vec{k}) = C_0 \,\delta_{\sigma_1^\prime \sigma_1} \delta_{\sigma_2^\prime \sigma_2} u_s(p,\Lambda_0)u_s(k,\Lambda_0) + 1\leftrightarrow 2,
    \\
    &V^{I=0}_{\text{ope}} = 3\left( \frac{g_A}{2f_\pi}\right)^2 \frac{\vec{q}_t\cdot\vec{\sigma_1} \vec{q}_t\cdot\vec{\sigma}_2}{\vec{q}_t^2 + m_\pi^2} u_\text{ope}(q_t,\Lambda_{\text{ope}}) + \vec{q}_t \to \vec{q}_u.
\end{align}
For the isovector channel($I=1$):
\begin{align}
    &V^{I=1}_{s}(\vec{p},\vec{k}) = C_1 \,\delta_{\sigma_1^\prime \sigma_1} \delta_{\sigma_2^\prime \sigma_2} u_s(p,\Lambda_0)u_s(k,\Lambda_0) - 1\leftrightarrow 2,
    \\
    &V^{I=1}_{\text{ope}} = -\left( \frac{g_A}{2f_\pi}\right)^2 \frac{\vec{q}_t\cdot\vec{\sigma_1} \vec{q}_t\cdot\vec{\sigma}_2}{\vec{q}_t^2 + m_\pi^2} u_\text{ope}(q_t,\Lambda_{\text{ope}}) - \vec{q}_t \to \vec{q}_u.
\end{align}

\begin{figure}[htb]
    \centering
    \subfigure[$\;$isoscalar(${}^3 S_1$)]{
    \includegraphics[width=0.8\linewidth]{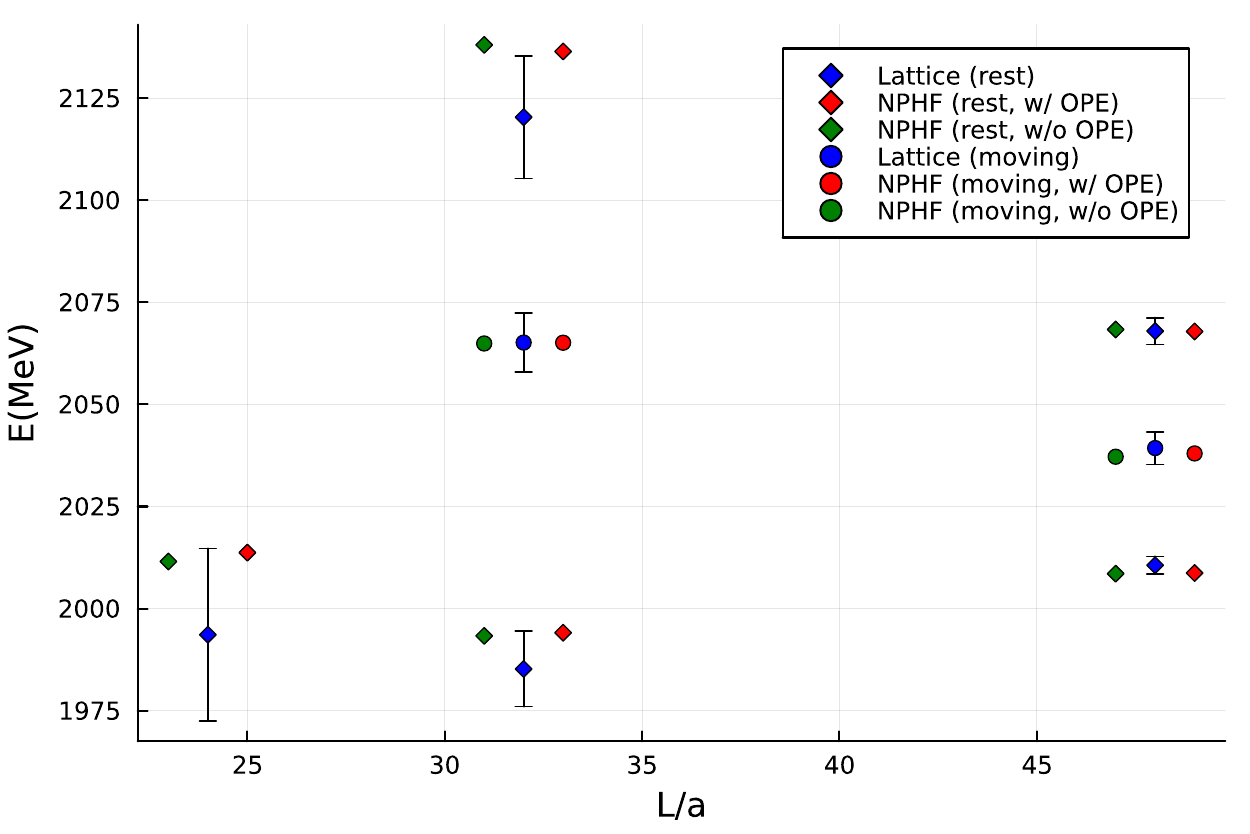}
    }
    \subfigure[$\;$isovector(${}^1 S_0$)]{    \includegraphics[width=0.8\linewidth]{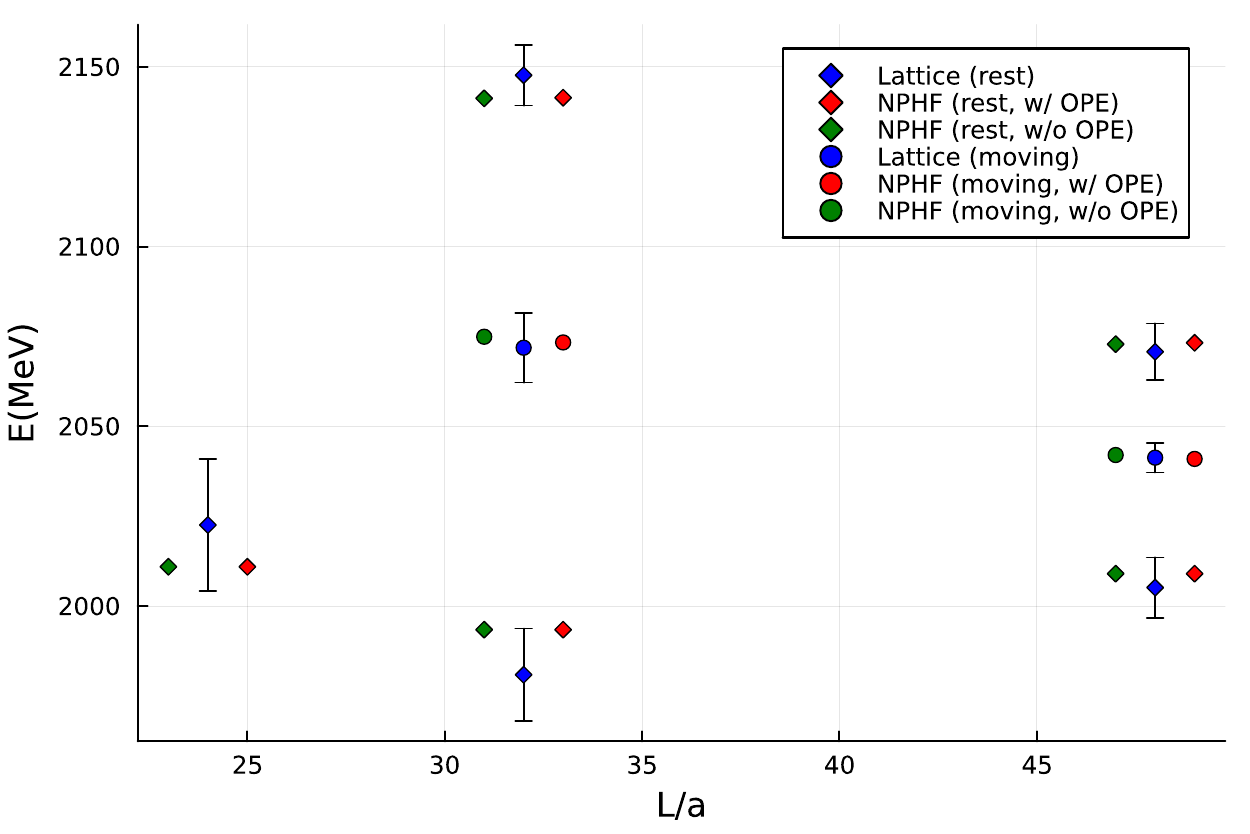}}
    \caption{ Finite volume energies from the  NPHF method compared to lattice spectra at $L/a=24,32,48$. Diamond and circle markers denote energies in the rest frame and moving frame, respectively. Blue markers with error bars are the lattice spectra. Red(green) markers without error bar, shifted right(left) for clarity, represent the energies of {NPHF} in scheme A(B). }
    \label{fig:fit:HEFT}
\end{figure}

Here, $\vec{q}_t=\vec{p}-\vec{k}$ and $\vec{q}_u=\vec{p}+\vec{k}$ denote the momentum transfer in the $t$- and $u$-channel, respectively.
$\sigma_i^{(\prime)}$ in $V_s$ explicitly denotes the nucleon spin polarization, while $\vec{\sigma}$ in $V_\text{ope}$ denotes the spin operator. 
$u_s$ and $u_\text{ope}$ are the form factors introduced phenomenologically to ensure ultraviolet convergence of the loop integral in Eq.~(\ref{eq:LSE}):
\begin{align}
    u_s(p,\Lambda) = \left( \frac{\Lambda^2}{p^2+\Lambda^2} \right)^2 \,,\, u_\text{ope}(p,\Lambda) = e^{-p^2/\Lambda^2}.
\end{align}
The coupling constants in $V_{\text{ope}}$ are given by $g_A=1.27$ and $f_\pi=92.4$ MeV\footnote{The $f_\pi$ value on the ensembles used in this work is $\sim95.5 $MeV~\cite{CLQCD:2023sdb}, which is very close to the physical value. The small difference is not expected to affect our results.}~\cite{Machleidt:2011zz}, while the other unknown parameters $C_0,\,C_1,\Lambda_0,\,\Lambda_1,\,\Lambda_\text{ope}$ are determined by jointly fitting the finite volume spectra to the isoscalar and isovector lattice spectra. 
To disentangle the effect of $V_{\text{ope}}$, we adopt two schemes: (A) $V_\text{ope}\neq0$ and (B) $V_\text{ope}=0$. 
The resulting parameters are listed in Tab.~\ref{tab:para:HEFT} and the fits are shown in Fig.~\ref{fig:fit:HEFT}.

\begin{table}[htb]
    \centering
    \begin{tabular}{|c|c|c|c|c|c|c|}
    \hline
       & $C_0$ & $\Lambda_0$ & $C_1$ & $\Lambda_1$ & $\Lambda_\text{ope}$ & $\hat{\chi}^2$ \\
       \hline
       sch. A & $-80(10)$ & $363(50)$ & $-87(12)$ & $210(50)$ & $260(100)$ & $0.67$
        \\ \hline 
        sch. B & $-85(10)$ & $323(32)$ & $-87(13)$ & $160(80)$ & $-$ & $0.64$
        \\ \hline
    \end{tabular}
    \caption{
    Fitted parameters in the two schemes. The units of $C_{0,1}$ are $\text{GeV}^{-2}$, and those of $\Lambda_{0,1,\text{ope}}$ are $\text{MeV}$. $\hat{\chi}^2=\chi^2/\text{d.o.f}$ is the reduced chi-square. 
    }
    \label{tab:para:HEFT}
\end{table}

%As shown in Based on the results, it is found that the effect of one-pion-exchange is negligible in this case, rather than which is indicated in phenomenological model at physical pion mass. 
%
%The possible reason is the pion mass in the current lattice simulation is as about twice large as the physical value and the strength of long-range interaction is approximately proportional to $1/m_\pi^2$. 

To locate the poles of the $T$ matrix, we search for zeros of $1-VG$. 
To access the second Riemann sheet, $G$ in $1-VG$ is replaced by $G^{u}(q;E) \equiv G(q;E)-i\frac{ \pi E}{ 2q_0(E)}\delta(q-q_0(E))$, where $q_0(E)=\sqrt{\frac{E^2}{4}-m_N^2}$ is the on-shell momentum with a positive imaginary part. 
Only virtual state poles are found in both schemes. 
The deviations of the pole positions from the $NN$ threshold in the isoscalar channel (${}^3S_1$) are
\begin{align}
    \text{scheme A:  } \quad E_B = 2.0(1.0) \text{ MeV},
    \\
    \text{scheme B:  } \quad E_B = 1.4(0.5) \text{ MeV},
\end{align}
and in isovector channel(${}^1 S_0$),
\begin{align}
    \text{scheme A:  } \quad E_B = 1.9(0.7) \text{ MeV},
    \\
    \text{scheme B:  } \quad E_B = 1.7(0.8) \text{ MeV}.
\end{align}
These results from the two schemes are consistent within $1\sigma$ uncertainty, and they also agree with the $K$-matrix parameterization results presented earlier within $2\sigma$.

As shown in Tab.~\ref{tab:para:HEFT}, the fitted parameters under the two mechanisms are consistent within errors.
Similarly, as shown in Fig.~\ref{fig:fit:HEFT}, the energy levels obtained from the fits for both schemes lie within the error bars of the lattice energy levels.
Nevertheless, the lattice results are not sufficiently precise to distinguish between the two schemes. 
Furthermore, the pole positions from two mechanisms are also consistent.
Therefore, in this lattice calculation, although some energy levels(see Fig.~\ref{fig:energy}) lie close to or below the left-hand cut, we do not find a  significant role of the long-range potential due to single-pion exchange. 
%
%This finding contrasts with the usual understanding that the long-range potential contributes significantly to the $NN$ interaction. 
%
%{\color{blue}A possible explanation is that the relatively large pion mass reduces the interaction strength of single-pion exchange compared to the short-range potential, rendering the peculiar effects of the long-range potential less significant.}
%
This result is consistent with the partial wave analysis of the $NN$ elastic scattering, which indicates that the long range force (one-pion exchange) is important only for the high partial waves and negligible for the $s$-wave scattering.~\cite{Machleidt:1987hj,Machleidt:2011zz}. 
Furthermore, the large pion mass also reduces the interaction strength of single-pion exchange.

At the end of this section, it is worth noting the model independence of the NPHF. 
The method may appear model-dependent because an effective potential is needed. 
However, two different models for the short range interaction would result in the same scattering amplitude at the provided energies, as long as both of them can describe the lattice spectra well~\cite{Wu:2014vma}. 
The equivalence between the NPHF and model-independent L\"uscher formalism for systems without significant long-range interaction has been examined in Ref.~\cite{Yu:2025gzg}.
On the other hand, a long-range force changes this situation because it introduces an additional pole in the potential at the left-hand cut, necessitating a proper model for the long-range force to extract resonance information or the $T$-matrix of the scattering.

\section{Summary}\label{sec:summary}

We have studied nucleon-nucleon systems using lattice QCD at a pion mass of approximately 292 MeV. 
Calculations were performed on three $N_f=2+1$ ensembles with the same pion mass and lattice spacing but different volumes, which are generated by the CLQCD Collaboration using the stout-smeared clover fermion action and Symanzik gauge action. 
Finite-volume energy levels in the $^3S_1$ and $^1S_0$ channels were extracted in both the rest frame and a moving frame. 
Using the Lüscher formalism, we determined the scattering amplitudes and found that both channels exhibit a virtual state pole. 
The corresponding binding energies are $6^{+5}_{-3}$ MeV for the $^3S_1$ channel and $11^{+6}_{-5}$ MeV for the $^1S_0$ channel.

To account for potential left-hand cut effects arising from one-pion-exchange interactions, we also analyzed the system using the NPHF method. 
The contribution of this long-range interaction was found to be insignificant at the present pion mass. This is consistent with the partial-wave analysis of $NN$ scattering. Also, the relatively heavy pion may suppress the strength of single-pion exchange relative to the short-range interaction. 
Nevertheless, the NPHF analysis also yields virtual states in both channels, with pole positions consistent with those obtained from the Lüscher approach. 
Taken together, our results indicate that at $m_\pi\approx 292$ MeV, the two-nucleon systems in both the $^3S_1$ and $^1S_0$ channels are characterized by virtual states rather than bound states.

The mixing between the $^3S_1$ and $^3D_1$ channels is ignored in this work. However, this mixing could be essential for the formation of a bound deuteron. Moreover, the pion mass may strongly influence the relative importance of long- and short-range interactions and therefore the nature of the $NN$ interaction. In future work, a controlled study at the physical pion mass with an explicit $^3S_1$–$^3D_1$ coupled-channel analysis will be a priority for establishing the bound-state nature of the deuteron on the lattice.

\section*{Acknowledgment}

We thank the CLQCD collaborations for providing us their gauge configurations with dynamical fermions, which are generated on HPC Cluster of ITP-CAS, the Southern Nuclear Science Computing Center(SNSC), the Siyuan-1 cluster supported by the Center for High Performance Computing at Shanghai Jiao Tong University and the Dongjiang Yuan Intelligent Computing Center. We are grateful to Haobo Yan and Qu-Zhi Li for valuable discussions and comments. 
This work is supported by the National Natural Science Foundation of China under Grant Nos. 12293060, 12293061, 12293063, 12175239, 12221005,  11935017, 12175279, and 12322503, 
and by the Chinese Academy of Sciences under Grant No. YSBR-101.
It was also supported by the Australian Research Council through Grant Nos.\ DP210103706 (DBL) and DP230101791 (AWT).

\bibliography{ref}

\clearpage

\begin{widetext}

\section*{Appendix}

\subsection{The quality of fitting energy levels}\label{app:ranges}

The total energy $E$ in the lab frame for a specific two-particle interacting state with total momentum $\bm{P}=\frac{2\pi}{L}\bm{d}$ is determined from the time dependence of $\lambda_n(t)$ in Eq.\ref{eq:GEVP}. The fitting process is detailed in Figs.~\ref{fig:C24P29}, ~\ref{fig:C32P29}, and ~\ref{fig:C48P29}. In our analysis, we perform a two-state fit in all cases except for the $^3S_1$ channel in the rest frame on the ensemble C48P29, where a one-state fit is used. The fitting ranges are selected by examining the fitted energies as a functions of the starting timeslice $t_{\rm min}$
, while the ending timeslice is fixed at a large value where the errors become significant. The chosen $t_{\rm min}$ corresponds to the point where the fitted energy stabilizes and the $\chi^2$ value becomes acceptably small, indicating that excited-state contaminations are under control.

\begin{figure}[tbph]
\centering
\includegraphics[width=8cm]{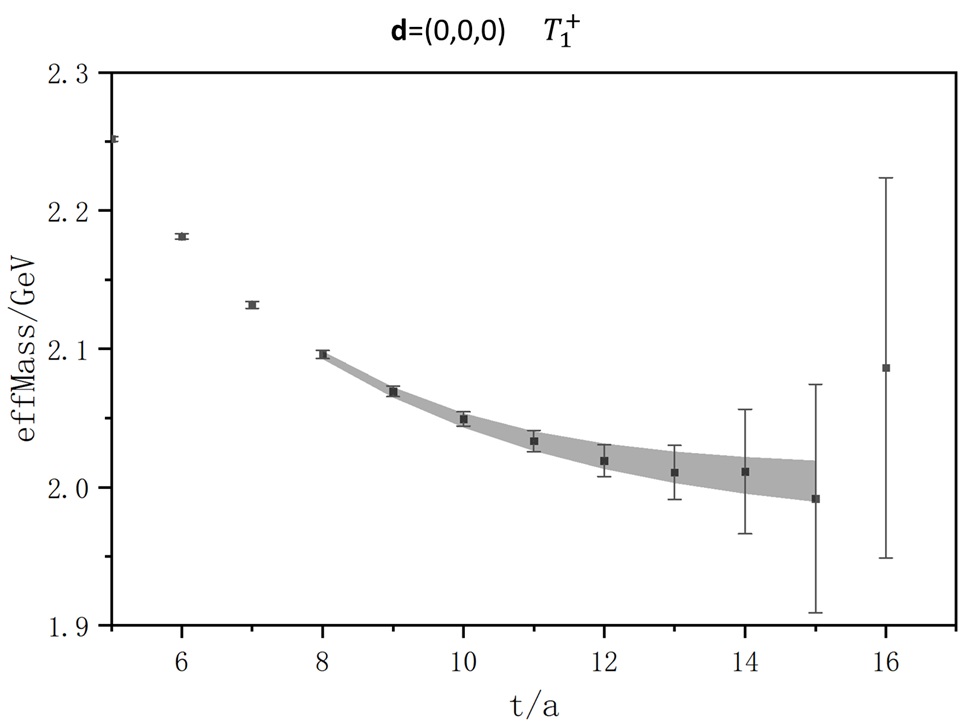}
\includegraphics[width=8cm]{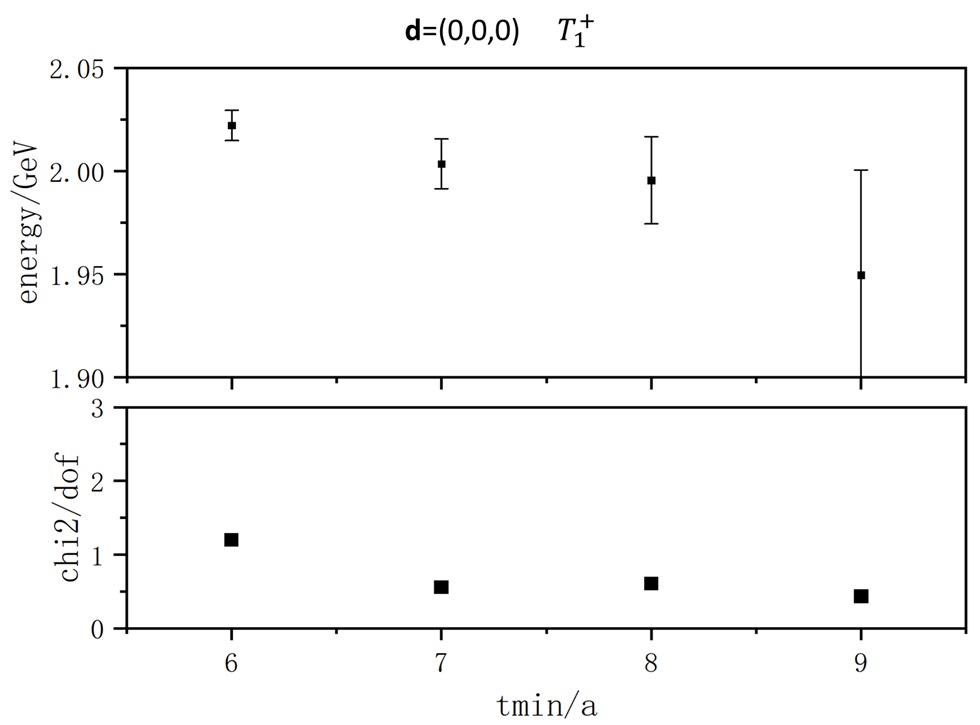}
\includegraphics[width=8cm]{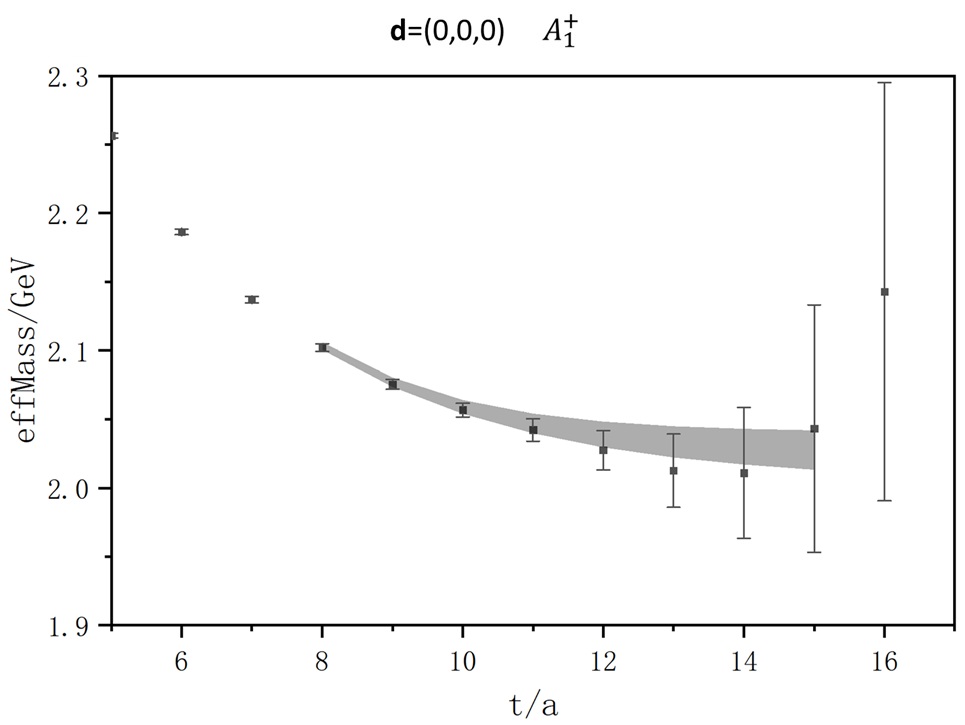}
\includegraphics[width=8cm]{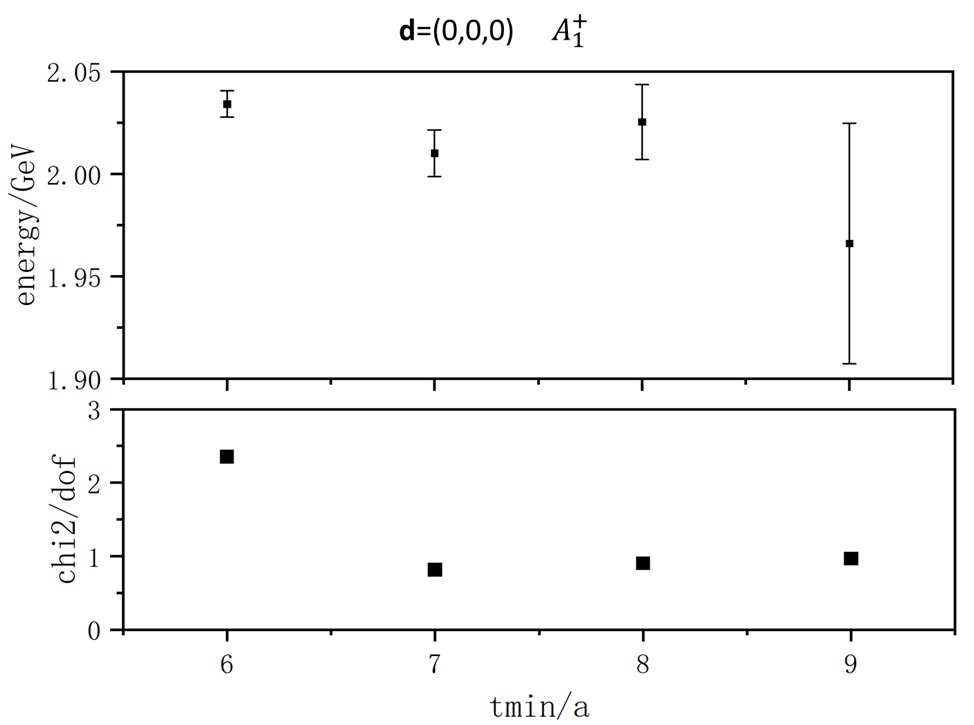}
\caption{Fitting details on the ensemble C24P29, in the $^3S_1$ channel (upper section) and the $^1S_0$ channel (lower section). The left section illustrates the effective mass of $\lambda_n(t)$, with the band representing the effective mass extracted from the fit and the fitting range used in Fig.~\ref{fig:energy}. The right section provides the fitting results across various fitting ranges ($[t_{\rm min},16a]$), including the energy levels and the $\chi^2/dof$. A two-state fit is employed for all cases. Statistical uncertainties are estimated using 2000 bootstrap samples.}
\label{fig:C24P29}
\end{figure}

\begin{figure}[tbph]
\centering
\includegraphics[width=5cm]{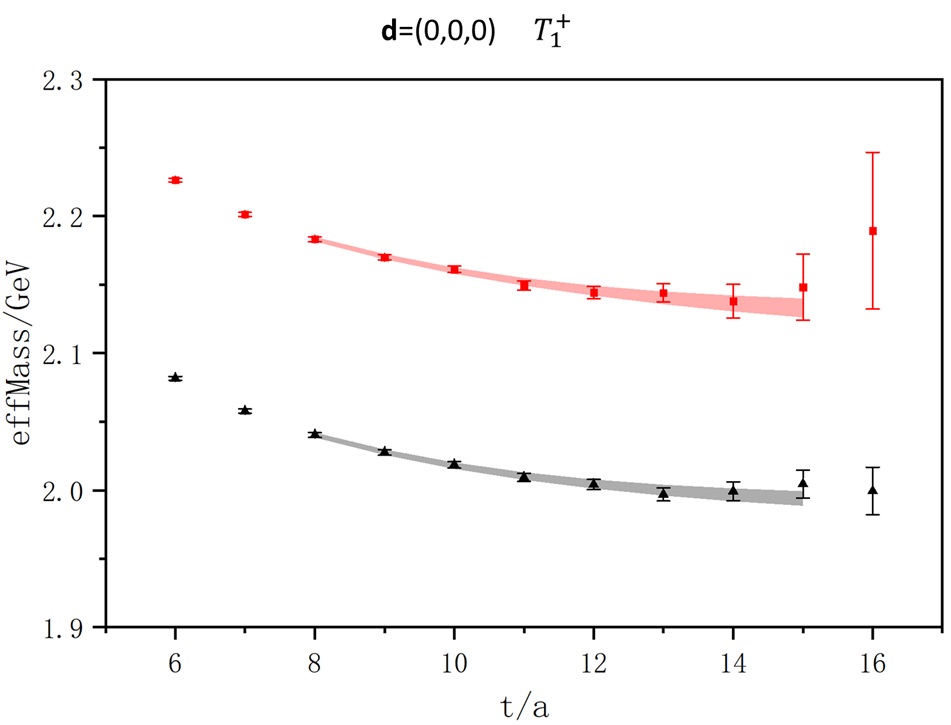}
\includegraphics[width=5cm]{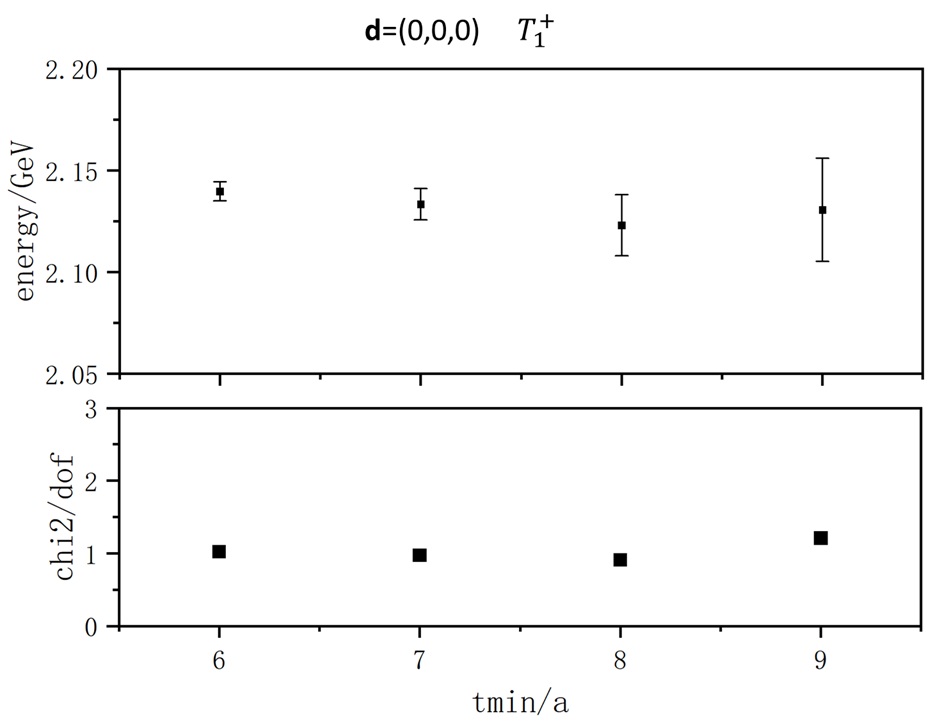}
\includegraphics[width=5cm]{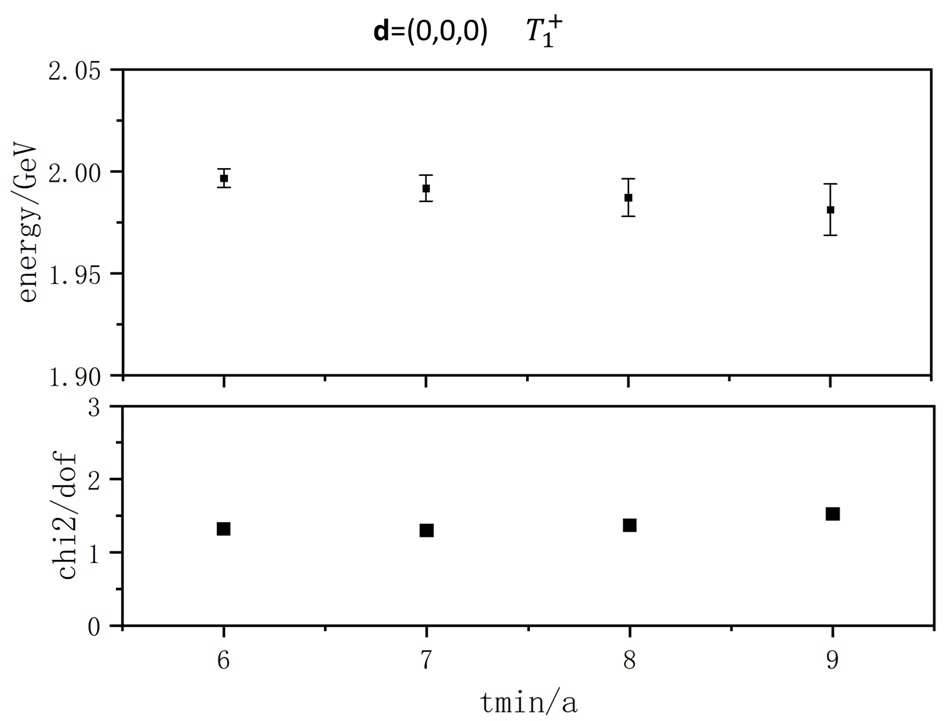}
\includegraphics[width=5cm]{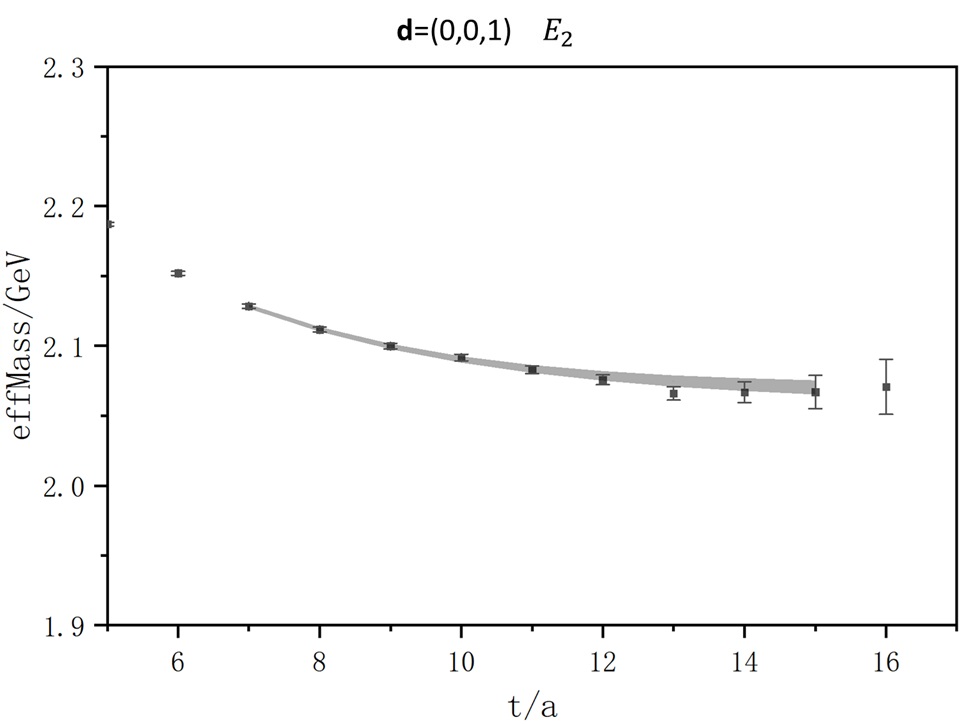}
\includegraphics[width=5cm]{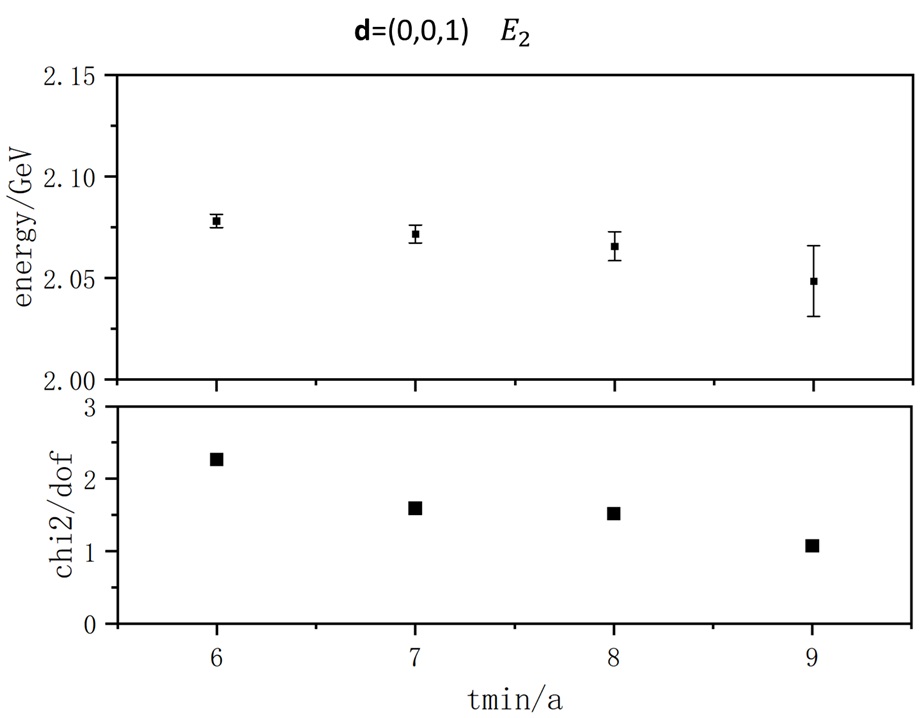}
\includegraphics[width=5cm]{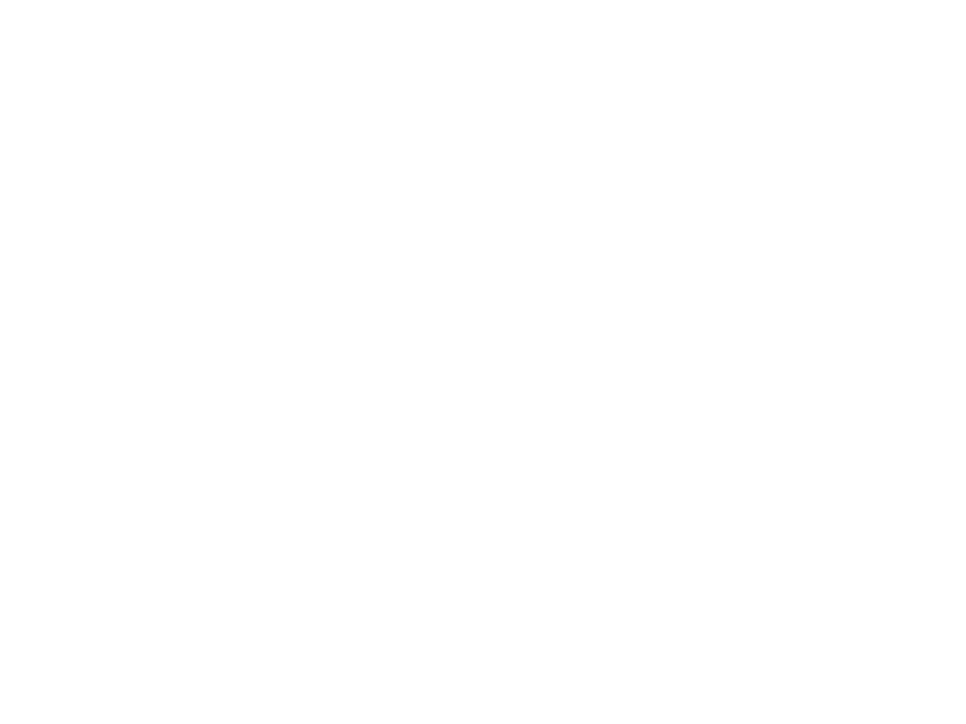}
\includegraphics[width=5cm]{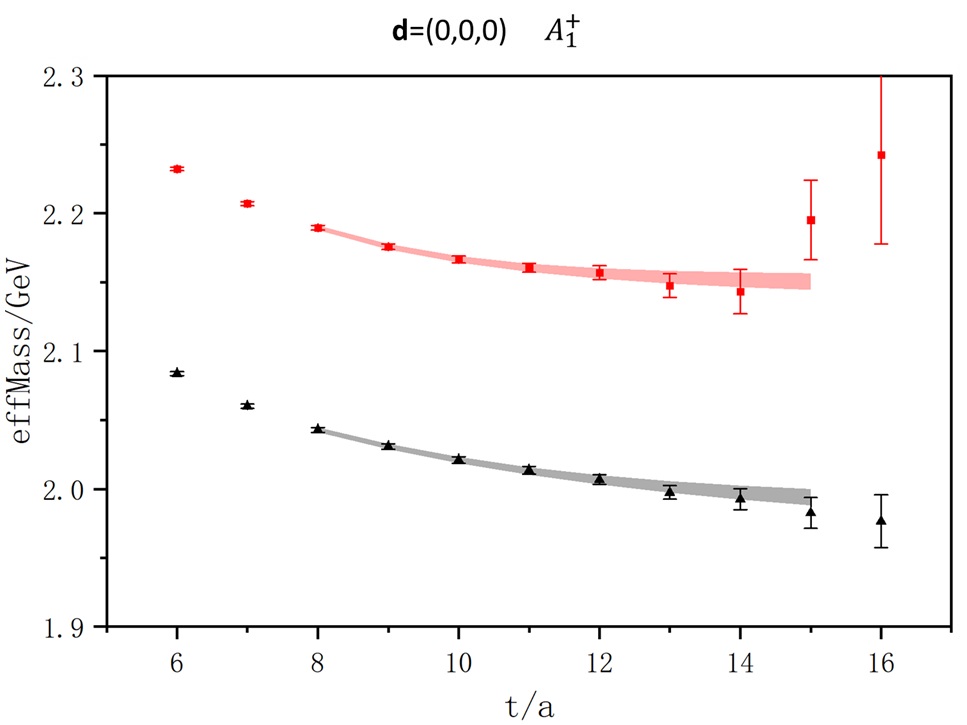}
\includegraphics[width=5cm]{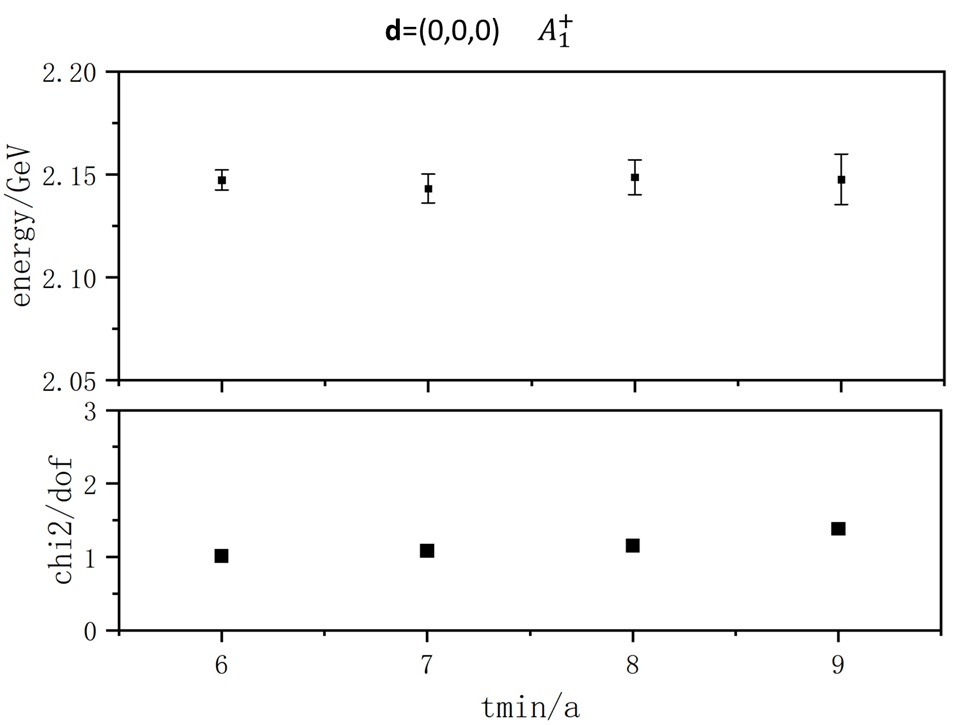}
\includegraphics[width=5cm]{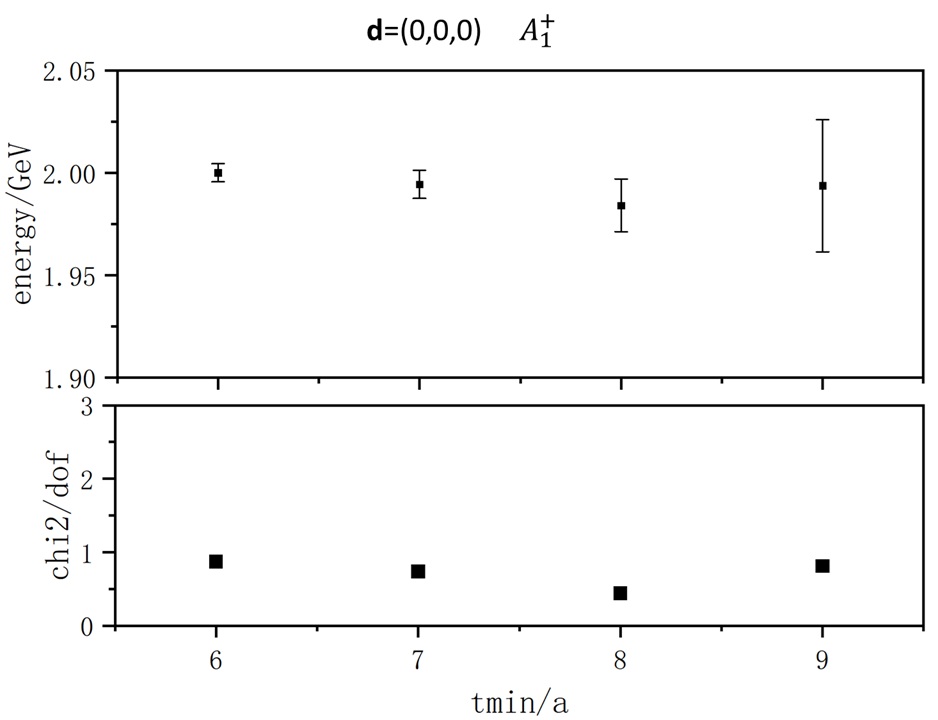}
\includegraphics[width=5cm]{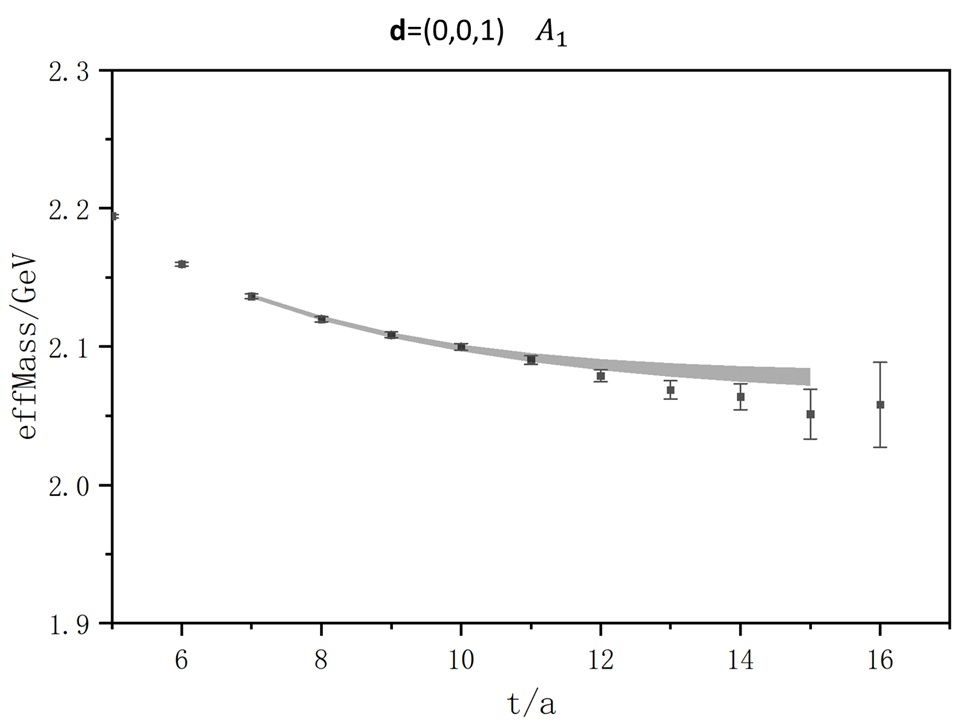}
\includegraphics[width=5cm]{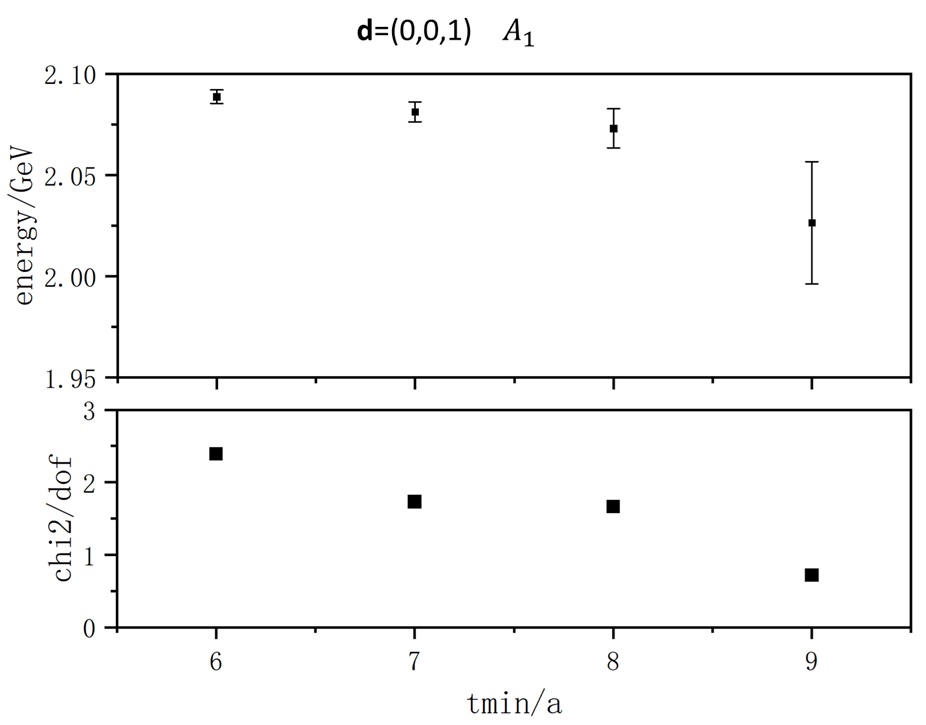}
\includegraphics[width=5cm]{spectrum/blank.jpg}
\caption{Details of the fitting for the ensemble C32P29, in the $^3S_1$ channel (first two rows) and the $^1S_0$ channel (last two rows), in the rest frame (the first and third rows) and a moving frame (the second and the forth rows). The first column indicates the effective mass of $\lambda_n(t)$. The band represents the effective mass extracted from the fit, as well as the fitting range used in Fig~\ref{fig:energy}. The other panels present fitting results across various fitting ranges ($[t_{\rm min},16a]$), showcasing both energy levels and $\chi^2/dof$. A two-state fit is applied in each case. Statistical uncertainties are assessed using 2000 bootstrap samples.}
\label{fig:C32P29}
\end{figure}

\begin{figure}[tbph]
\centering
\includegraphics[width=5cm]{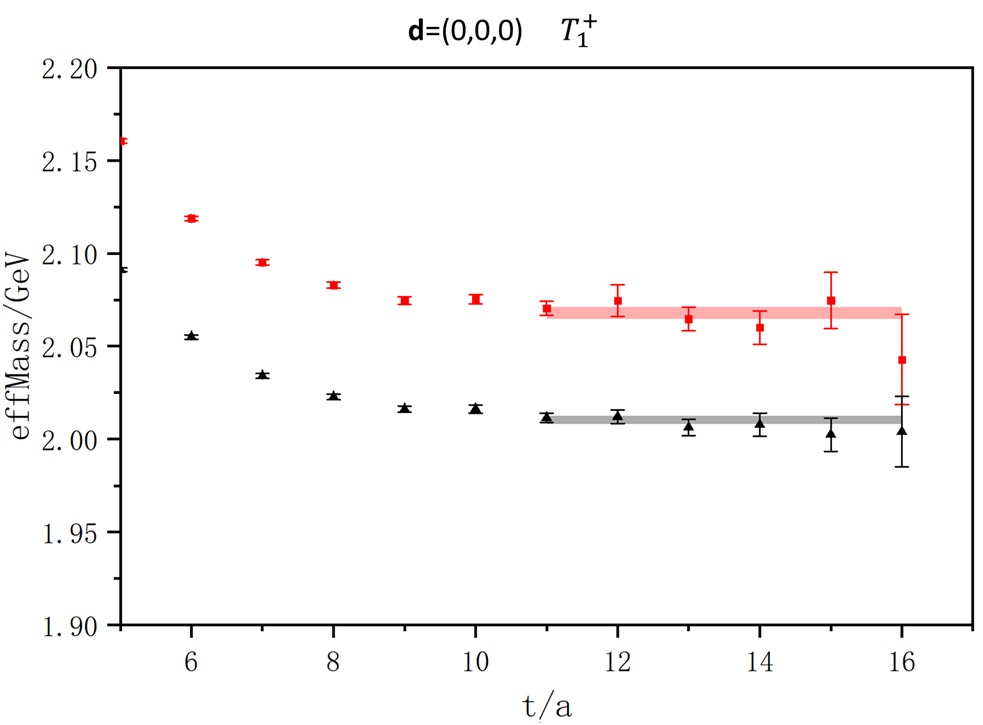}
\includegraphics[width=5cm]{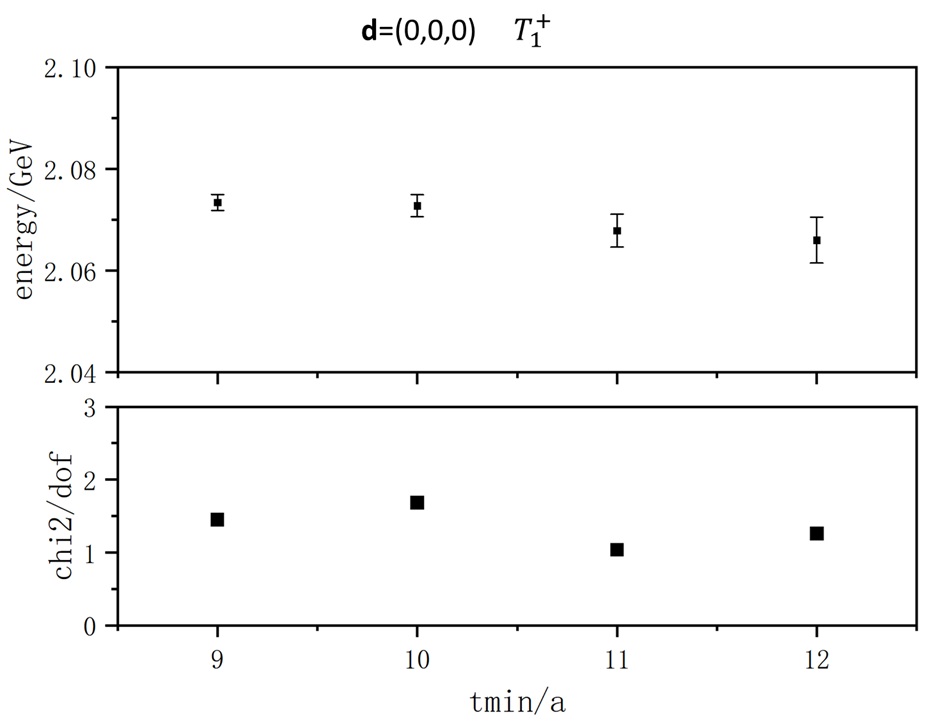}
\includegraphics[width=5cm]{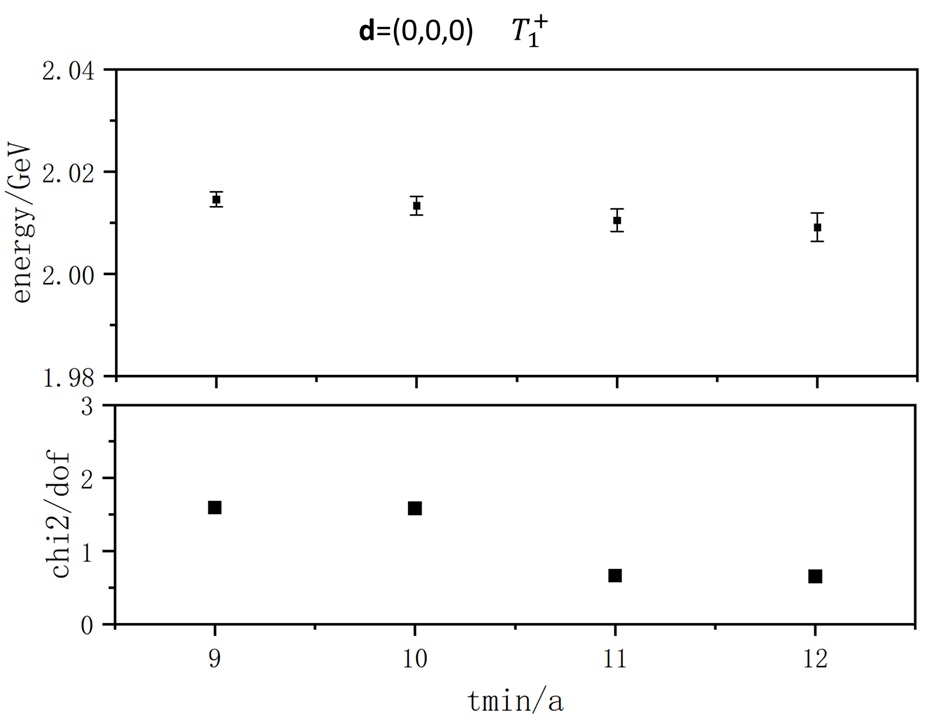}
\includegraphics[width=5cm]{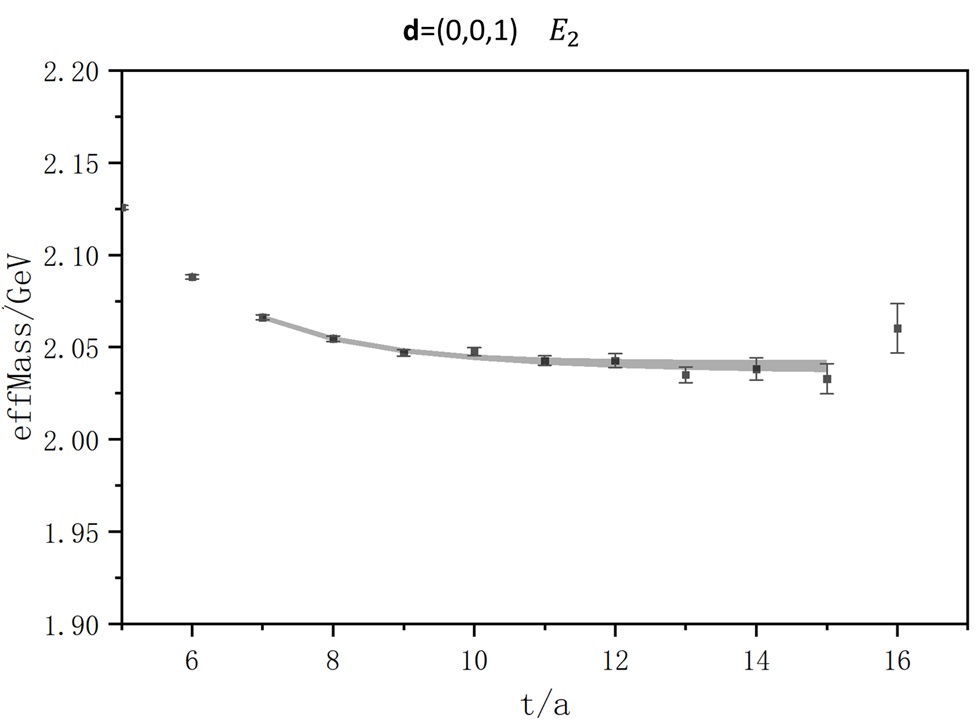}
\includegraphics[width=5cm]{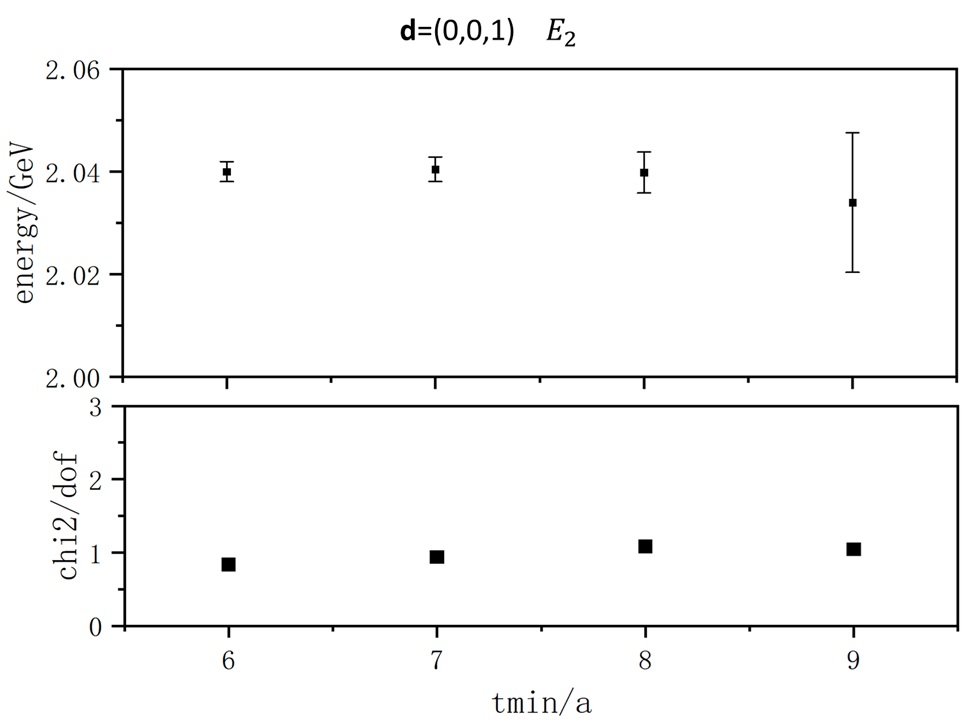}
\includegraphics[width=5cm]{spectrum/blank.jpg}
\includegraphics[width=5cm]{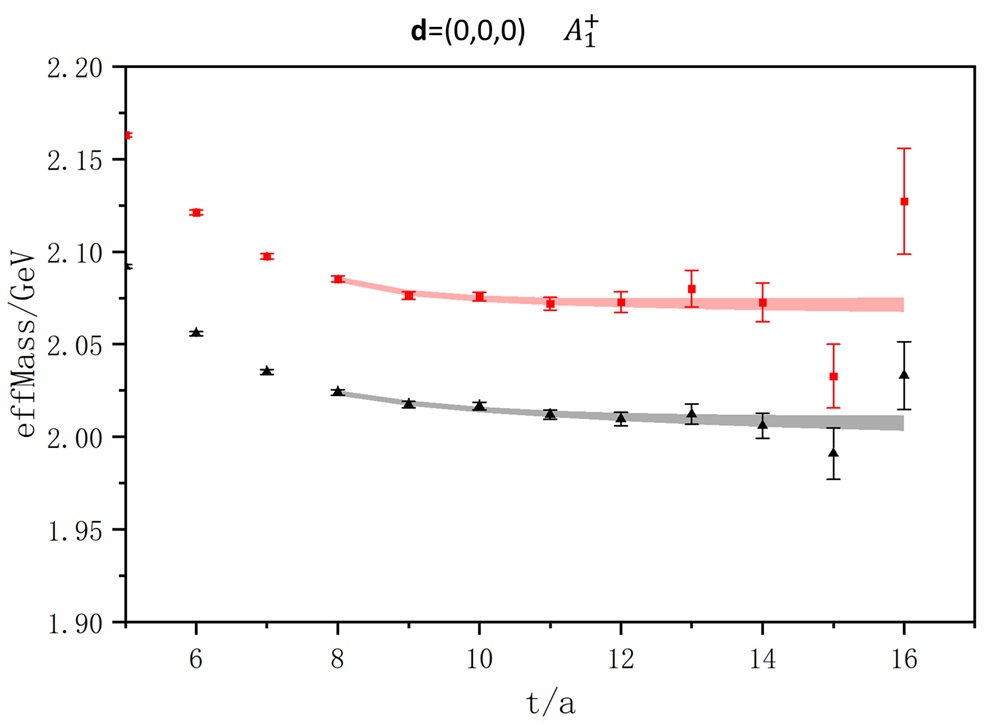}
\includegraphics[width=5cm]{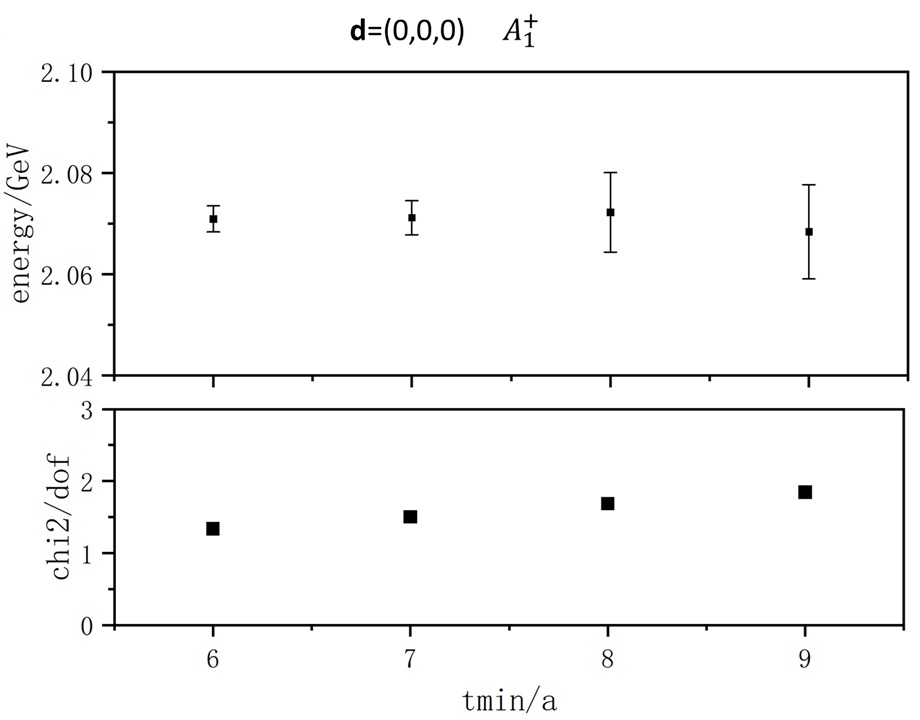}
\includegraphics[width=5cm]{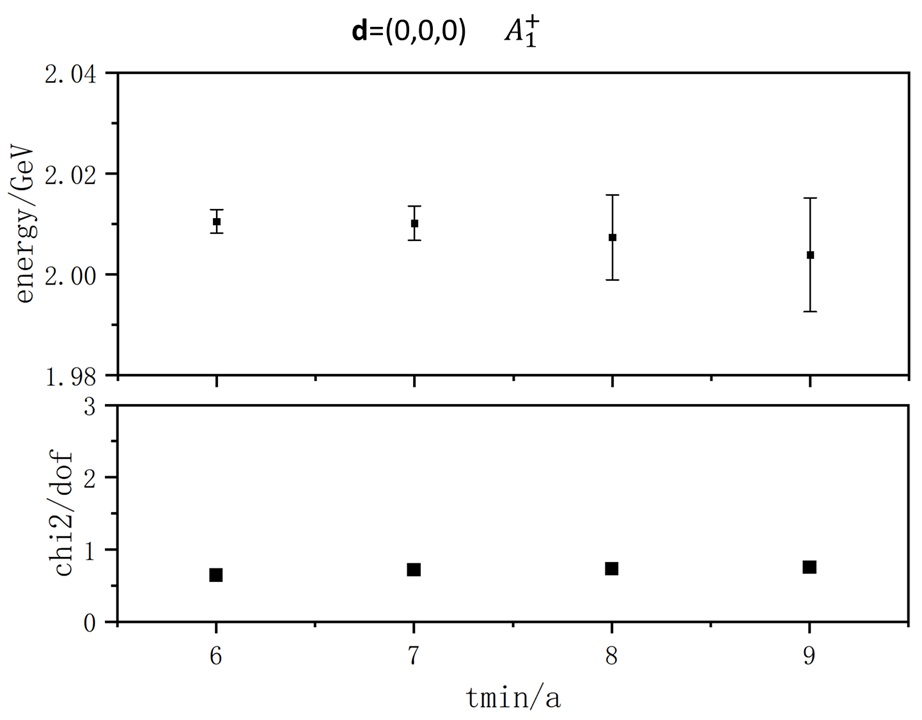}
\includegraphics[width=5cm]{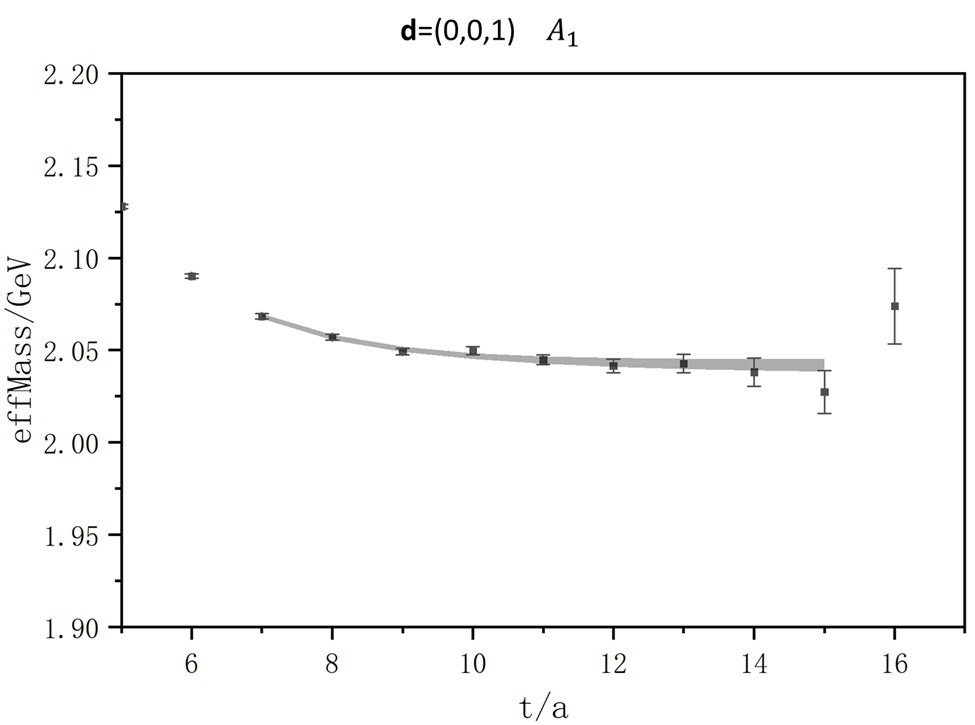}
\includegraphics[width=5cm]{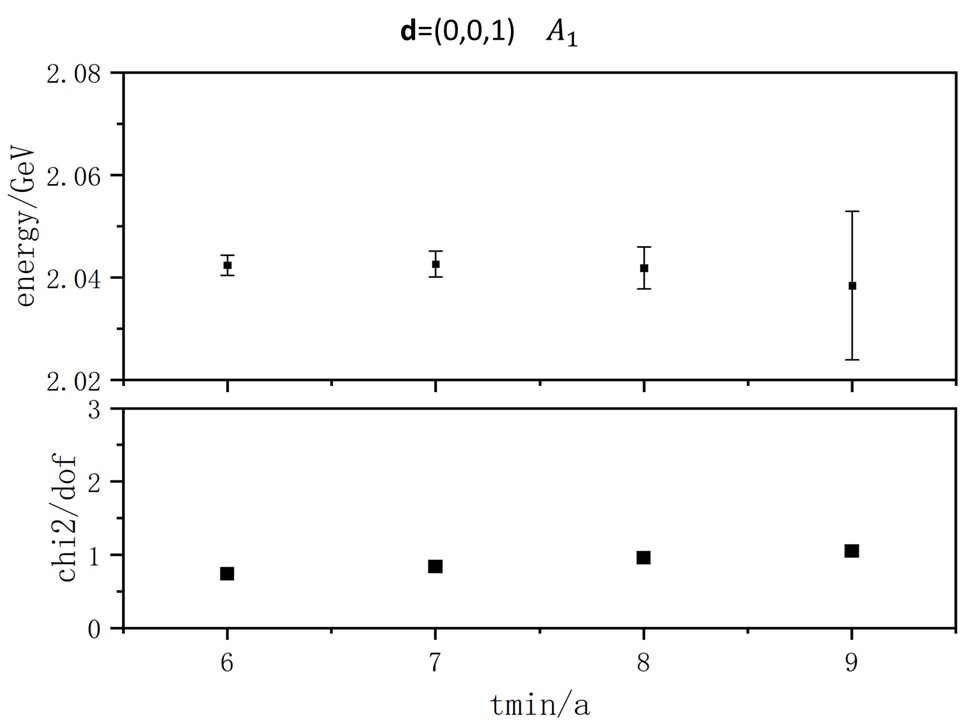}
\includegraphics[width=5cm]{spectrum/blank.jpg}
\caption{Details of the fitting process for the ensemble C48P29, in the $^3S_1$ channel (first two rows) and the $^1S_0$ channel (last two rows), in the rest frame(the first and third rows) and a moving frame (the second and the forth rows). The first column presents the effective mass of $\lambda_n(t)$. The band highlights the effective mass obtained from the fitting and the range used in Fig~\ref{fig:energy}. Other panels show energy levels and $\chi^2/dof$ for different fitting ranges ($[t_{\rm min},17a]$ in the rest frame and $[t_{\rm min},16a]$ for other cases). We use a one-state fit for the $^3S_1$ channel in the rest frame, and two-state fits for the other cases. Statistical uncertainties are determined using 2000 bootstrap samples.}
\label{fig:C48P29}
\end{figure}

\subsection{$t_0$ dependence}\label{app:t0}

{In Figs.~\ref{fig:C32P29_t0} and \ref{fig:C48P29_t0}, we compare the energies obtained with $t_0=4a, 5a$ and $6a$. We also perform the L\"uscher scattering analysis using energy levels obtained with $t_0=4a$ and $t_0=6a$. The resulting variation in pole positions due to the choice of $t_0$ is less than 0.3 MeV for both the $^3S_1$ and $^1S_0$ channels.}

\begin{figure}[tbph]
\centering
\includegraphics[width=8cm]{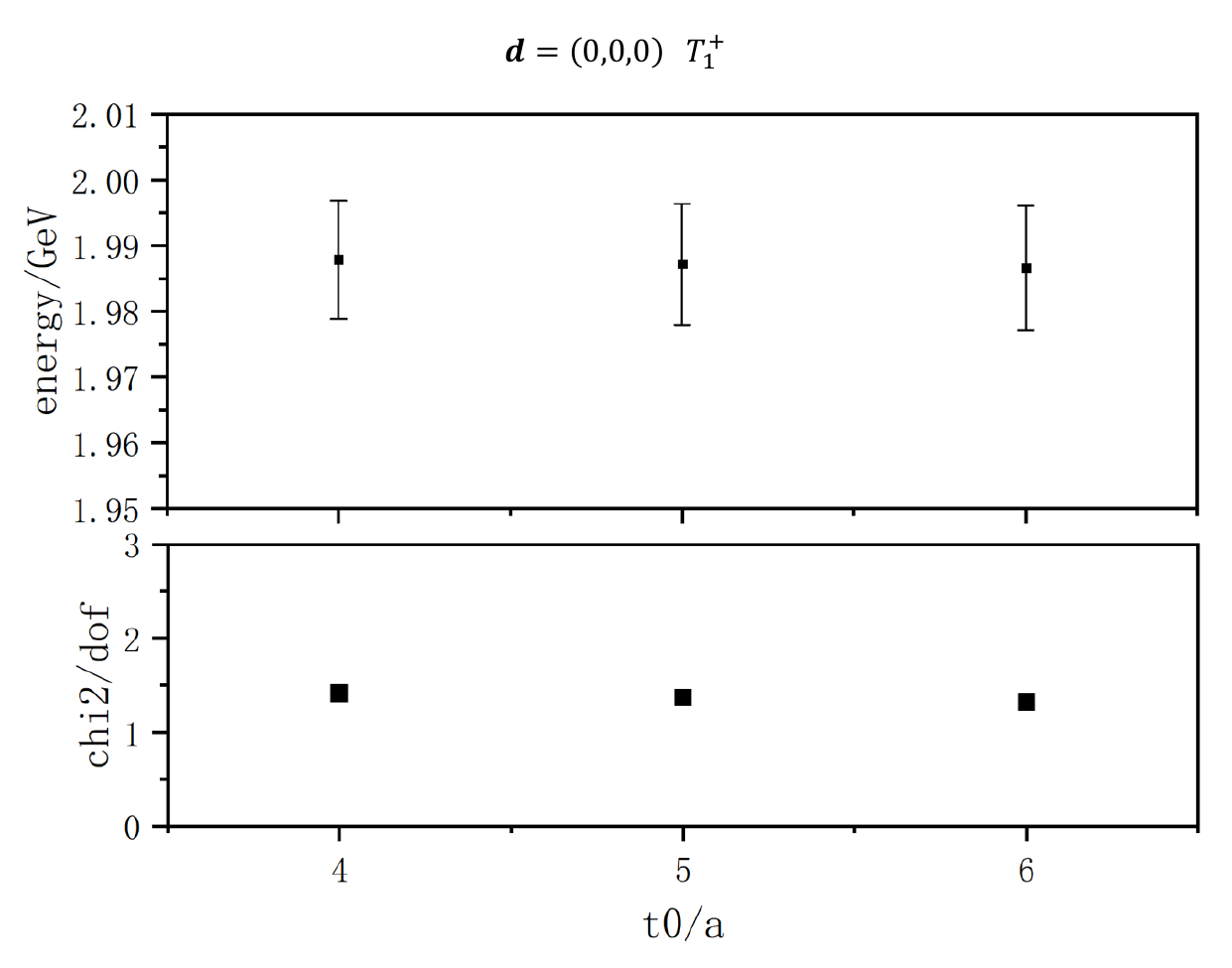}
\includegraphics[width=8cm]{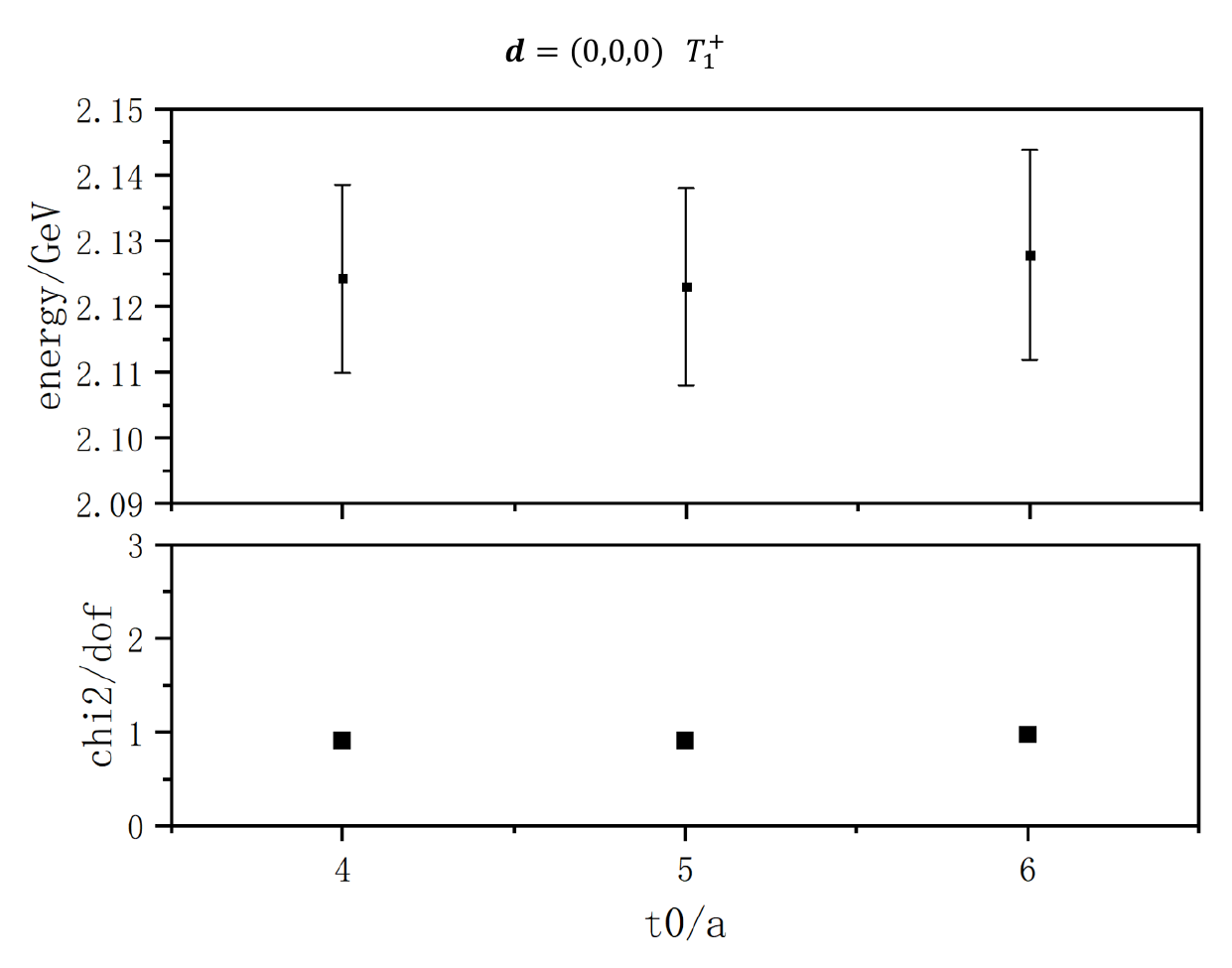}
\includegraphics[width=8cm]{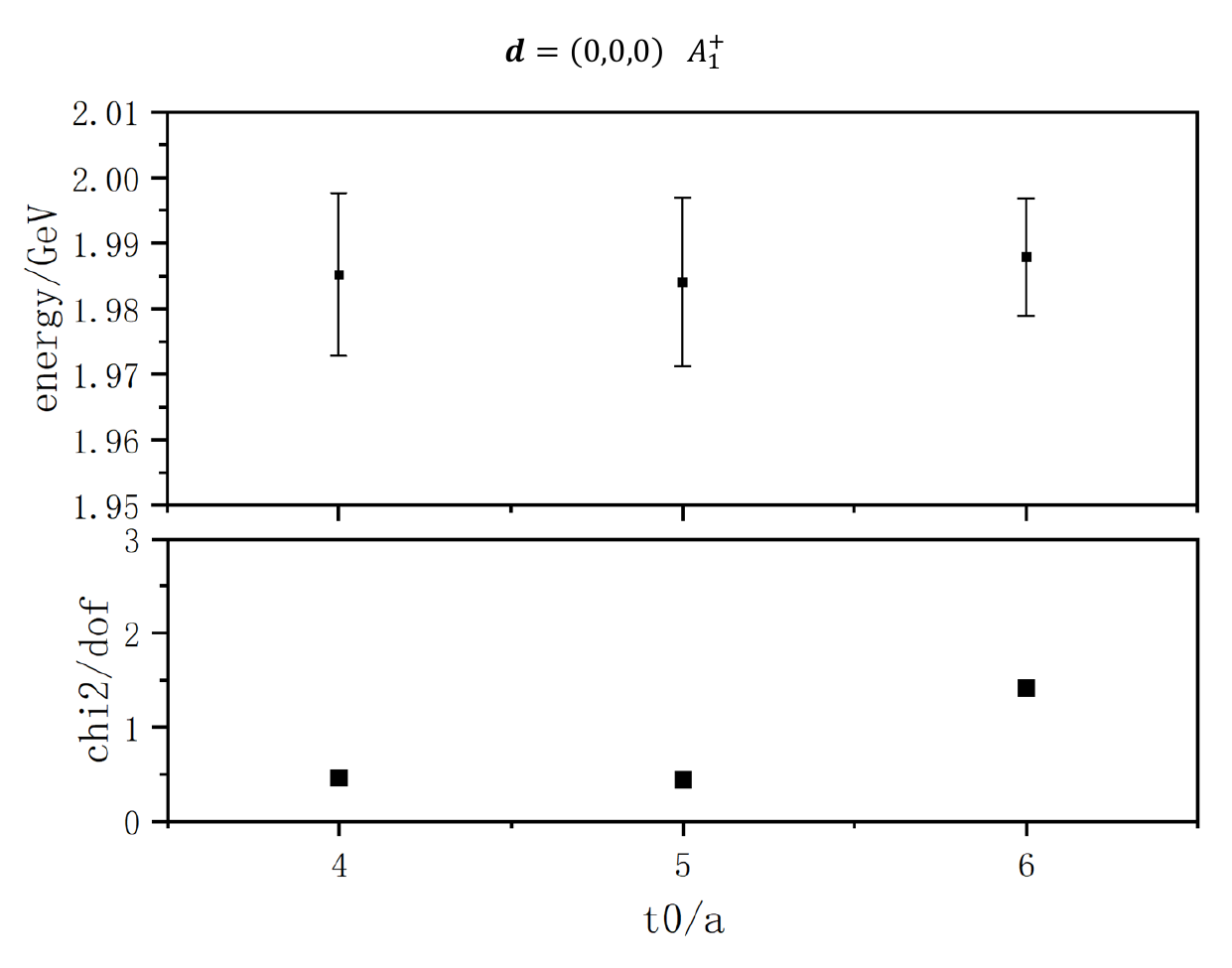}
\includegraphics[width=8cm]{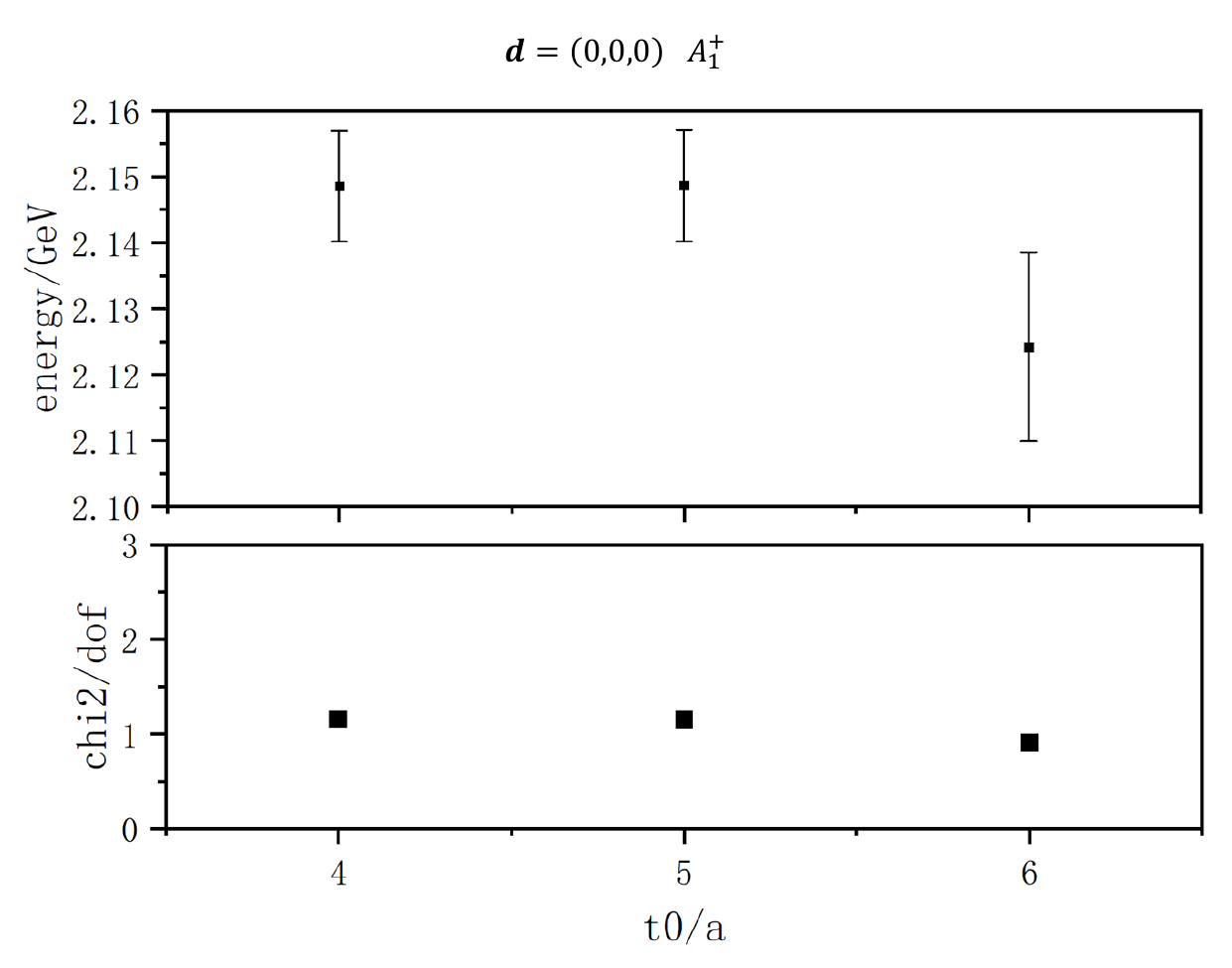}
\caption{$t_0$ dependence of the energies for the ensemble C32P29. The upper panels show the two energy levels in the $^3S_1$ channel($T_1^+$ irrep) and the lower panels show those in the $^1S_0$ channel($A_1^+$ irrep). These figures present the fitting results across various $t_0$, including the energy levels and the $\chi^2/dof$. Statistical uncertainties are estimated using 2000 bootstrap samples.}
\label{fig:C32P29_t0}
\end{figure}

\begin{figure}[tbph]
\centering
\includegraphics[width=8cm]{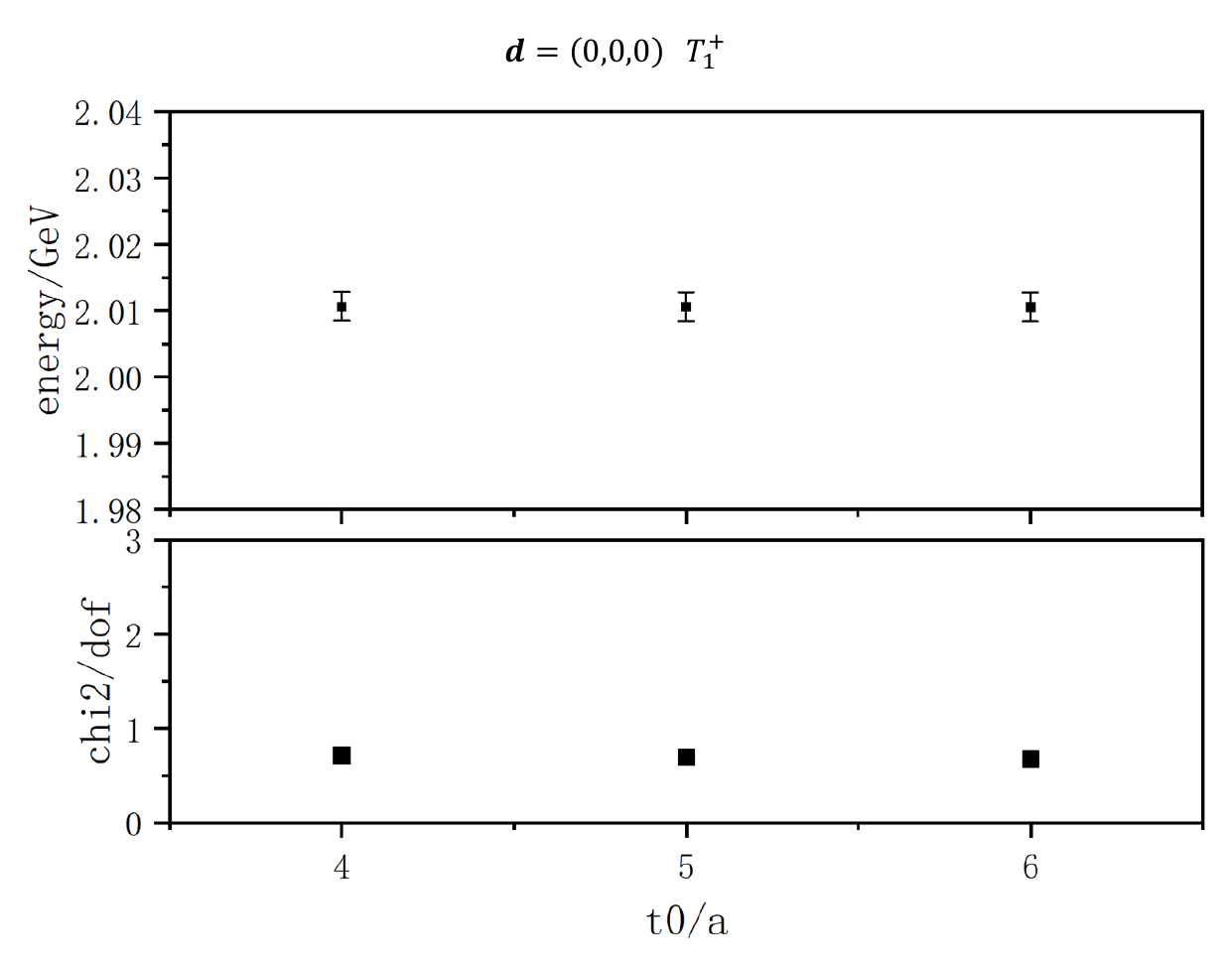}
\includegraphics[width=8cm]{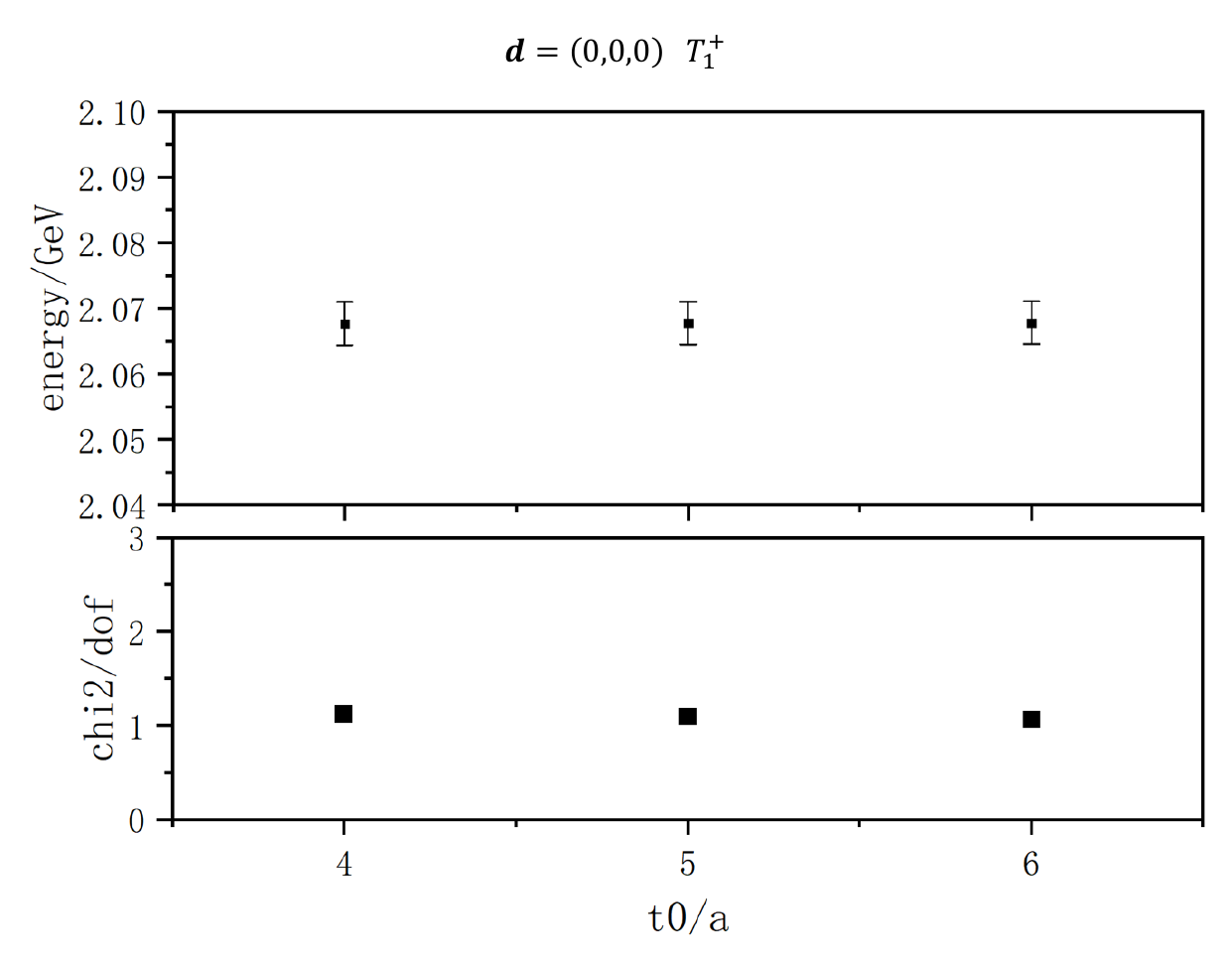}
\includegraphics[width=8cm]{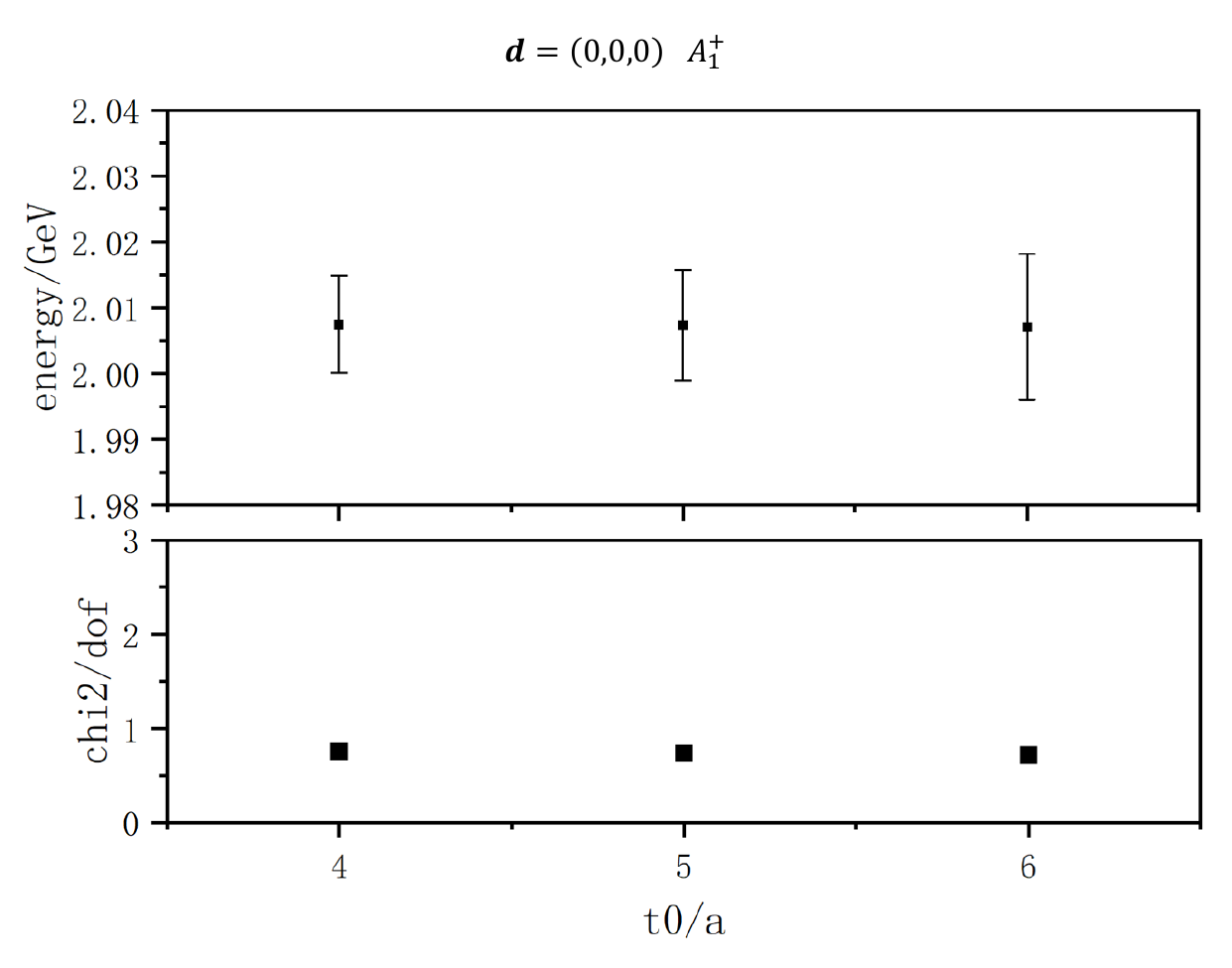}
\includegraphics[width=8cm]{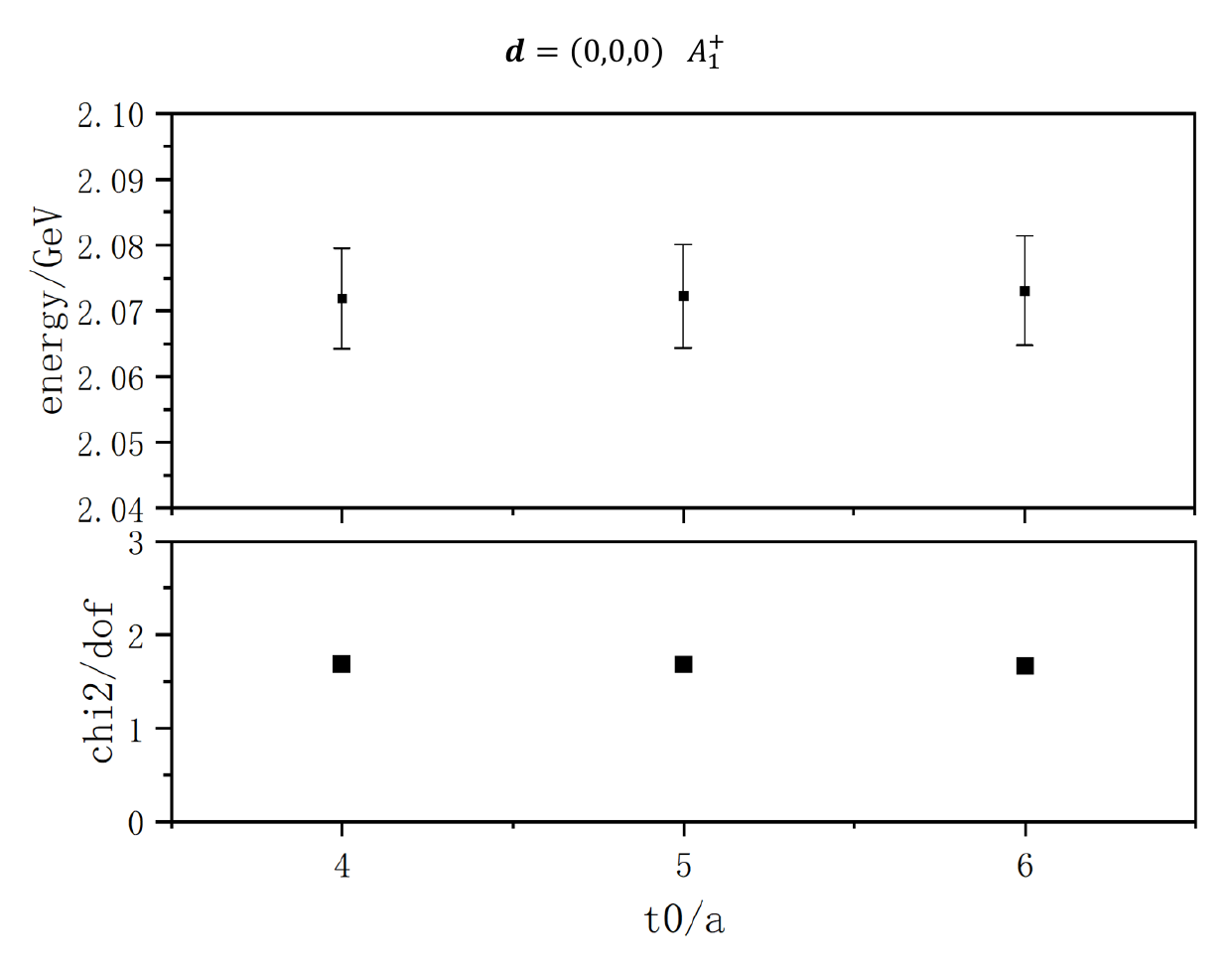}
\caption{Same as Fig.~\ref{fig:C32P29_t0}, but for the ensemble C48P29.}
\label{fig:C48P29_t0}
\end{figure}

\subsection{Explicit forms of all operators}\label{app:operators}

{The coefficients $C_{\lambda_1,\lambda_2,{\bm p_1},{\bm p_2}}$ in %Eq.~\ref{eq:NNoperator} 
Eq.(3) and Eq.(4) are listed in Table~\ref{tab:operators_T1g_0},\ref{tab:operators_T1g_1},\ref{tab:operators_E2_0},\ref{tab:operators_A1g_0},\ref{tab:operators_A1g_1},\ref{tab:operators_A1_0}.}

\begin{table}[htbp]
  \centering
  \begin{tabular}{|c|c|c|}
\hline
\hline
$T_1^+, \lambda=1$ & &\\
\hline
$\lambda_1$    & 1 & 2 \\
\hline
$\bm{p_1}$    & (0,0,0) & (0,0,0) \\
\hline
$\lambda_2$    & 1 & 2 \\
\hline
$\bm{p_2}$    & (0,0,0) & (0,0,0) \\
\hline
$C_{\lambda_1,\lambda_2,{\bm p_1},{\bm p_2}}$   & 1 & -1 \\
\hline
\hline
$T_1^+, \lambda=2$ & &\\
\hline
$\lambda_1$    & 1 & 2 \\
\hline
$\bm{p_1}$    & (0,0,0) & (0,0,0) \\
\hline
$\lambda_2$    & 1 & 2 \\
\hline
$\bm{p_2}$    & (0,0,0) & (0,0,0) \\
\hline
$C_{\lambda_1,\lambda_2,{\bm p_1},{\bm p_2}}$   & 1 & 1 \\
\hline
\hline
$T_1^+, \lambda=3$ & &\\
\hline
$\lambda_1$    & 2 & 1 \\
\hline
$\bm{p_1}$    & (0,0,0) & (0,0,0) \\
\hline
$\lambda_2$    & 1 & 2 \\
\hline
$\bm{p_2}$    & (0,0,0) & (0,0,0) \\
\hline
$C_{\lambda_1,\lambda_2,{\bm p_1},{\bm p_2}}$   & 1 & 1 \\
\hline
\hline
\end{tabular}
  \caption{The non-zero coefficients $C_{\lambda_1,\lambda_2,{\bm p_1},{\bm p_2}}$ in Eq.(3) for $T_1^+$ with $|{\bm p_1}| = 0, |{\bm p_2}| = 0$. A factor of $\frac{2\pi}{L}$ in $\bm p_1$ and $\bm p_2$ is omitted for brevity. The index $\lambda$ labels the rows of the $T_1^+$ irrep. Correlation functions are calculated for all three rows and then averaged.    }
  \label{tab:operators_T1g_0}
\end{table}

\begin{table}[htbp]
  \centering
  \begin{tabular}{|c|c|c|c|c|c|c|c|c|c|c|c|c|}
\hline
\hline
$T_1^+, \lambda=1$ & & & & & & & & & & & &\\
\hline
$\lambda_1$    & 1 & 1 & 1 & 1 & 1 & 1 & 2 & 2 & 2 & 2 & 2 & 2 \\
\hline
$\bm{p_1}$    & (1,0,0) & (-1,0,0) & (0,1,0) & (0,-1,0) & (0,0,1) & (0,0,-1) & (1,0,0) & (-1,0,0) & (0,1,0) & (0,-1,0) & (0,0,1) & (0,0,-1)  \\
\hline
$\lambda_2$    & 1 & 1 & 1 & 1 & 1 & 1 & 2 & 2 & 2 & 2 & 2 & 2 \\
\hline
$\bm{p_2}$    & (-1,0,0) & (1,0,0) & (0,-1,0) & (0,1,0) & (0,0,-1) & (0,0,1) & (-1,0,0) & (1,0,0) & (0,-1,0) & (0,1,0) & (0,0,-1) & (0,0,1) \\
\hline
$C_{\lambda_1,\lambda_2,{\bm p_1},{\bm p_2}}$   & 1 & 1 & 1 & 1 & 1 & 1 & -1 & -1 & -1 & -1 & -1 & -1 \\
\hline
\hline
$T_1^+, \lambda=2$ & & & & & & & & & & & &\\
\hline
$\lambda_1$    & 1 & 1 & 1 & 1 & 1 & 1 & 2 & 2 & 2 & 2 & 2 & 2 \\
\hline
$\bm{p_1}$    & (1,0,0) & (-1,0,0) & (0,1,0) & (0,-1,0) & (0,0,1) & (0,0,-1) & (1,0,0) & (-1,0,0) & (0,1,0) & (0,-1,0) & (0,0,1) & (0,0,-1)  \\
\hline
$\lambda_2$    & 1 & 1 & 1 & 1 & 1 & 1 & 2 & 2 & 2 & 2 & 2 & 2 \\
\hline
$\bm{p_2}$    & (-1,0,0) & (1,0,0) & (0,-1,0) & (0,1,0) & (0,0,-1) & (0,0,1) & (-1,0,0) & (1,0,0) & (0,-1,0) & (0,1,0) & (0,0,-1) & (0,0,1) \\
\hline
$C_{\lambda_1,\lambda_2,{\bm p_1},{\bm p_2}}$   & 1 & 1 & 1 & 1 & 1 & 1 & 1 & 1 & 1 & 1 & 1 & 1 \\
\hline
\hline
$T_1^+, \lambda=3$ & & & & & & & & & & & &\\
\hline
$\lambda_1$    & 2 & 2 & 2 & 2 & 2 & 2 & 1 & 1 & 1 & 1 & 1 & 1 \\
\hline
$\bm{p_1}$    & (1,0,0) & (-1,0,0) & (0,1,0) & (0,-1,0) & (0,0,1) & (0,0,-1) & (1,0,0) & (-1,0,0) & (0,1,0) & (0,-1,0) & (0,0,1) & (0,0,-1)  \\
\hline
$\lambda_2$    & 1 & 1 & 1 & 1 & 1 & 1 & 2 & 2 & 2 & 2 & 2 & 2 \\
\hline
$\bm{p_2}$    & (-1,0,0) & (1,0,0) & (0,-1,0) & (0,1,0) & (0,0,-1) & (0,0,1) & (-1,0,0) & (1,0,0) & (0,-1,0) & (0,1,0) & (0,0,-1) & (0,0,1) \\
\hline
$C_{\lambda_1,\lambda_2,{\bm p_1},{\bm p_2}}$   & 1 & 1 & 1 & 1 & 1 & 1 & 1 & 1 & 1 & 1 & 1 & 1 \\
\hline
\hline
\end{tabular}
  \caption{The non-zero coefficients $C_{\lambda_1,\lambda_2,{\bm p_1},{\bm p_2}}$ in Eq.(3) for $T_1^+$ with $|{\bm p_1}| = \frac{2\pi}{L}, |{\bm p_2}| = \frac{2\pi}{L}$. A factor of $\frac{2\pi}{L}$ in $\bm p_1$ and $\bm p_2$ is omitted for brevity. The index $\lambda$ labels the rows of the $T_1^+$ irrep. Correlation functions are calculated for all three rows and then averaged.}
  \label{tab:operators_T1g_1}
\end{table}

\begin{table}[htbp]
  \centering
  \begin{tabular}{|c|c|c|}
\hline
\hline
$E_2, \lambda=1$ & &\\
\hline
$\lambda_1$    & 1 & 2 \\
\hline
$\bm{p_1}$    & (0,0,0) & (0,0,0) \\
\hline
$\lambda_2$    & 1 & 2 \\
\hline
$\bm{p_2}$    & (0,0,1) & (0,0,1) \\
\hline
$C_{\lambda_1,\lambda_2,{\bm p_1},{\bm p_2}}$   & 1 & -1 \\
\hline
\hline
$E_2, \lambda=2$ & &\\
\hline
$\lambda_1$    & 1 & 2 \\
\hline
$\bm{p_1}$    & (0,0,0) & (0,0,0) \\
\hline
$\lambda_2$    & 1 & 2 \\
\hline
$\bm{p_2}$    & (0,0,1) & (0,0,1) \\
\hline
$C_{\lambda_1,\lambda_2,{\bm p_1},{\bm p_2}}$   & 1 & 1 \\
\hline
\hline
\end{tabular}
  \caption{The non-zero coefficients $C_{\lambda_1,\lambda_2,{\bm p_1},{\bm p_2}}$ in Eq.(3) for $E_2$ in the moving frame with $|{\bm p_1}| = 0, |{\bm p_2}| = \frac{2\pi}{L}$. A factor of $\frac{2\pi}{L}$ in $\bm p_1$ and $\bm p_2$ is omitted for brevity. The index $\lambda$ labels the rows of the $E_2$ irrep. Correlation functions are calculated for the two rows and then averaged.}
  \label{tab:operators_E2_0}
\end{table}

\begin{table}[htbp]
  \centering
  \begin{tabular}{|c|c|c|}
\hline
\hline
$A_1^+, \lambda=1$ & &\\
\hline
$\lambda_1$    & 2 & 1 \\
\hline
$\bm{p_1}$    & (0,0,0) & (0,0,0) \\
\hline
$\lambda_2$    & 1 & 2 \\
\hline
$\bm{p_2}$    & (0,0,0) & (0,0,0) \\
\hline
$C_{\lambda_1,\lambda_2,{\bm p_1},{\bm p_2}}$   & 1 & -1 \\
\hline
\hline
\end{tabular}
  \caption{The non-zero coefficients $C_{\lambda_1,\lambda_2,{\bm p_1},{\bm p_2}}$ in Eq.(4) for $A_1^+$ with $|{\bm p_1}| = 0, |{\bm p_2}| = 0$. A factor of $\frac{2\pi}{L}$ in $\bm p_1$ and $\bm p_2$ is omitted for brevity.}
  \label{tab:operators_A1g_0}
\end{table}

\begin{table}[htbp]
  \centering
  \begin{tabular}{|c|c|c|c|c|c|c|c|c|c|c|c|c|}
\hline
\hline
$A_1^+, \lambda=1$ & & & & & & & & & & & &\\
\hline
$\lambda_1$    & 2 & 2 & 2 & 2 & 2 & 2 & 1 & 1 & 1 & 1 & 1 & 1 \\
\hline
$\bm{p_1}$    & (1,0,0) & (-1,0,0) & (0,1,0) & (0,-1,0) & (0,0,1) & (0,0,-1) & (1,0,0) & (-1,0,0) & (0,1,0) & (0,-1,0) & (0,0,1) & (0,0,-1)  \\
\hline
$\lambda_2$    & 1 & 1 & 1 & 1 & 1 & 1 & 2 & 2 & 2 & 2 & 2 & 2 \\
\hline
$\bm{p_2}$    & (-1,0,0) & (1,0,0) & (0,-1,0) & (0,1,0) & (0,0,-1) & (0,0,1) & (-1,0,0) & (1,0,0) & (0,-1,0) & (0,1,0) & (0,0,-1) & (0,0,1) \\
\hline
$C_{\lambda_1,\lambda_2,{\bm p_1},{\bm p_2}}$   & 1 & 1 & 1 & 1 & 1 & 1 & -1 & -1 & -1 & -1 & -1 & -1 \\
\hline
\hline
\end{tabular}
  \caption{The non-zero coefficients $C_{\lambda_1,\lambda_2,{\bm p_1},{\bm p_2}}$ in Eq.(4) for $A_1^+$ with $|{\bm p_1}| = \frac{2\pi}{L}, |{\bm p_2}| = \frac{2\pi}{L}$. A factor of $\frac{2\pi}{L}$ in $\bm p_1$ and $\bm p_2$ is omitted for brevity.}
  \label{tab:operators_A1g_1}
\end{table}

\begin{table}[htbp]
  \centering
  \begin{tabular}{|c|c|c|}
\hline
\hline
$A_1, \lambda=1$ & &\\
\hline
$\lambda_1$    & 2 & 1 \\
\hline
$\bm{p_1}$    & (0,0,0) & (0,0,0) \\
\hline
$\lambda_2$    & 1 & 2 \\
\hline
$\bm{p_2}$    & (0,0,1) & (0,0,1) \\
\hline
$C_{\lambda_1,\lambda_2,{\bm p_1},{\bm p_2}}$   & 1 & -1 \\
\hline
\hline
\end{tabular}
  \caption{The non-zero coefficients $C_{\lambda_1,\lambda_2,{\bm p_1},{\bm p_2}}$ in Eq.(4) for $A_1$ in the moving frame with $|{\bm p_1}| = 0, |{\bm p_2}| = \frac{2\pi}{L}$. A factor of $\frac{2\pi}{L}$ in $\bm p_1$ and $\bm p_2$ is omitted for brevity.}
  \label{tab:operators_A1_0}
\end{table}

\end{widetext}

\end{document}